\documentclass[conference]{IEEEtran}
\usepackage{cite}
\ifCLASSINFOpdf
\else
\fi
\DeclareMathAlphabet{\mathpzc}{OT1}{pzc}{m}{it}

\usepackage[utf8]{inputenc}
\usepackage[T1]{fontenc}

\usepackage{amsthm}
\usepackage{sansmath}
\usepackage{mathtools}

\usepackage[bbgreekl]{mathbbol}
\usepackage{bm}
\usepackage{enumitem}
\usepackage{tikz}
\usepackage{tikzit}

\usepackage[top=2.5cm,bottom=2.5cm,left=2cm,right=2cm,marginparwidth=1.75cm]{geometry}
\usepackage{multicol}
\usepackage{proof}
\usepackage[all]{xy}

\usepackage{framed}
\usepackage{floatpag}
\usepackage{mdframed}

\usepackage{xcolor}
\definecolor[named]{ACMBlue}{cmyk}{1,0.1,0,0.1}
\definecolor[named]{ACMYellow}{cmyk}{0,0.16,1,0}
\definecolor[named]{ACMOrange}{cmyk}{0,0.42,1,0.01}
\definecolor[named]{ACMRed}{cmyk}{0,0.90,0.86,0}
\definecolor[named]{ACMLightBlue}{cmyk}{0.49,0.01,0,0}
\definecolor[named]{ACMGreen}{cmyk}{0.20,0,1,0.19}
\definecolor[named]{ACMPurple}{cmyk}{0.55,1,0,0.15}
\definecolor[named]{ACMDarkBlue}{cmyk}{1,0.58,0,0.21}
\makeatletter

\input pdf-trans
\newbox\qbox
\def\usecolor#1{\csname\string\color@#1\endcsname\space}
\newcommand\bordercolor[1]{\colsplit{1}{#1}}
\newcommand\fillcolor[1]{\colsplit{0}{#1}}
\newcommand\outline[1]{\leavevmode%
  \def\maltext{#1}%
  \setbox\qbox=\hbox{\maltext}%
  \boxgs{Q q 2 Tr \thickness\space w \fillcol\space \bordercol\space}{}%
  \copy\qbox%
}
\makeatother
\newcommand\colsplit[2]{\colorlet{tmpcolor}{#2}\edef\tmp{\usecolor{tmpcolor}}%
  \def\tmpB{}\expandafter\colsplithelp\tmp\relax%
  \ifnum0=#1\relax\edef\fillcol{\tmpB}\else\edef\bordercol{\tmpC}\fi}
\def\colsplithelp#1#2 #3\relax{%
  \edef\tmpB{\tmpB#1#2 }%
  \ifnum `#1>`9\relax\def\tmpC{#3}\else\colsplithelp#3\relax\fi
}
\bordercolor{black}
\fillcolor{white}
\def\thickness{.3}

\usepackage{amssymb}
\usepackage{color}
\usepackage{url}
\usepackage{stmaryrd}
\usepackage{xypic}

\newenvironment{varitemize}
{
\begin{list}{\labelitemi}
{\setlength{\itemsep}{0pt}
 \setlength{\topsep}{0pt}
 \setlength{\parsep}{0pt}
 \setlength{\partopsep}{0pt}
 \setlength{\leftmargin}{15pt}
 \setlength{\rightmargin}{0pt}
 \setlength{\itemindent}{0pt}
 \setlength{\labelsep}{5pt}
 \setlength{\labelwidth}{10pt}
}}
{
 \end{list}
}
\newcounter{numberone}
\newenvironment{varenumerate}
{
\begin{list}{\arabic{numberone}.}
{
  \usecounter{numberone}
  \setlength{\itemsep}{0pt}
  \setlength{\topsep}{0pt}
  \setlength{\parsep}{0pt}
  \setlength{\partopsep}{0pt}
  \setlength{\leftmargin}{15pt}
  \setlength{\rightmargin}{0pt}
  \setlength{\itemindent}{0pt}
  \setlength{\labelsep}{5pt}
  \setlength{\labelwidth}{15pt}
}}
{
\end{list}
} 

\newcommand{\hide}[1]{}
\RequirePackage{ifthen}
\newcommand{\typeof}{0}
\newcommand{\longv}[1]{\ifthenelse{\equal{\typeof}{0}}{}{#1}}
\newcommand{\shortv}[1]{\ifthenelse{\equal{\typeof}{0}}{#1}{}}

\newtheorem{theorem}{Theorem}
\newtheorem{proposition}{Proposition}
\newtheorem{lemma}{Lemma}
\newtheorem{definition}{Definition}
\newtheorem{corollary}{Corollary}

\theoremstyle{definition}
\newtheorem{example}{Example}

\newtheorem{remark}{Remark}

\newcommand{\values}{\mathcal{V}}

\newcommand{\subst}[3]{#1[#3/#2]}

\newcommand{\valone} {V}
\newcommand{\varone}{x}
\newcommand{\vartwo}{y}

\newcommand{\termone}{e}
\newcommand{\termtwo}{f}
\newcommand{\termthree}{g}

\newcommand{\termfour}{h}

\newcommand{\envone}{\env}
\newcommand{\envtwo}{\Delta}

\newcommand{\imp}{\vdash}

\newcommand{\ctxone}{\mathcal{C}}

\newcommand{\signature}{\Sigma}

\newcommand{\abs}[1]{\lambda #1.}

\newcommand{\To}{\Downarrow}

\newcommand{\lan}{\langle}
\newcommand{\ran}{\rangle}
\newcommand{\cc}{\cdots}
\newcommand{\hh}{\hdots}

\newcommand{\set}{\mathsf{Set}}
\newcommand{\powerset}{\mathcal{P}}
\newcommand{\distribution}{\mathcal{D}}

\newcommand{\preord}{\mathsf{PreOrd}}

\newcommand{\monad}{T}
\newcommand{\unit}{\eta}

\newcommand{\relthree}{Q}

\newcommand{\defeq}{\triangleq}

\newcommand{\rel}{\mathsf{Rel}}
\newcommand{\brel}{\mathsf{BRel}}
\newcommand{\howe}[1]{#1^{H}}

\newcommand{\relone}{R}
\newcommand{\reltwo}{S}
\newcommand{\vrelone}{\alpha}
\newcommand{\vreltwo}{\beta}

\newcommand{\vidrel}{1}

\newcommand{\true}{\mathsf{true}}
\newcommand{\false}{\mathsf{false}}

\newcommand{\cpoleq}{\sqsubseteq}
\newcommand{\lub}{\bigsqcup}

\newcommand{\sem}[1]{\llbracket #1 \rrbracket}

\newcommand{\kleisli}[1]{#1^{\dagger}}

\renewcommand{\valone}{v}
\newcommand{\valtwo}{w}
\newcommand{\valthree}{u}
\newcommand{\valfour}{z}

\renewcommand{\termone}{t}
\renewcommand{\termtwo}{s}
\renewcommand{\termthree}{p}
\renewcommand{\termfour}{q}

\newcommand{\TO}[1]{\Rightarrow^{n}}

\newcommand{\open}[1]{#1^{o}}
\newcommand{\closed}[1]{#1^{c}}

\newcommand{\qleq}{\leq}
\newcommand{\quantale}{\mathsf{V}}
\newcommand{\tensor}{\otimes}
\newcommand{\qunit}{k}
\newcommand{\join}{\bigvee}
\newcommand{\meet}{\bigwedge}

\newcommand{\quantaletwo}{\mathsf{W}}

\newcommand{\mmap}{\multimap}
\newcommand{\two}{\mathsf{2}}
\newcommand{\monotone}[2]{#1\Rightarrow#2}

\newcommand{\comp}{\mathbin{\circ}}
\def\tobar{\mathrel{\mkern3mu  \vcenter{\hbox{$\scriptscriptstyle+$}}%
            \mkern-12mu{\to}}}
\newcommand{\torel}{\tobar}

\newcommand{\Vrel}[1]{#1\text{-}\mathsf{Mat}}
\newcommand{\vrel}{\Vrel{\Quantale}}
\newcommand{\Wrel}[1]{#1\text{-}\mathsf{Rel}}
\newcommand{\wrelcat}{\Worlds\text{-}\mathsf{Rel}}

\renewcommand{\monad}{T}

\renewcommand{\set}{\mathsf{Set}}
\renewcommand{\powerset}{\mathcal{P}}

\newcommand{\idrel}{I}
\newcommand{\idvrel}{\mathsf{I}}
\newcommand{\Monad}{\mathbb{\monad}}

\newcommand{\setthree}{Z}
\newcommand{\setfour}{W}

\newcommand{\typeone}{\tau}
\newcommand{\typetwo}{\sigma}
\newcommand{\typethree}{\rho}
\newcommand{\typevoid}{\mathbf{Void}}
\newcommand{\typeunit}{\mathbf{Unit}}
\newcommand{\typebool}{\mathbf{Bool}}

\newcommand{\sumtype}[2]{\sum\nolimits_{#1}{#2}}

\newcommand{\typevarone}{a}

\newcommand{\rectype}[2]{\mu #1.#2}
\newcommand{\op}{\mathbf{op}}

\newcommand{\seq}[2]{\mathbf{let}\ \varone = #1\ \mathbf{in}\ #2}

\newcommand{\seqy}[2]{\mathbf{let}\ \vartwo = #1\ \mathbf{in}\ #2}

\newcommand{\fold}[1]{\mathbf{fold}\; #1}
\newcommand{\letfold}[2]{\mathbf{let}\ \fold{x} = #1\ \mathbf{in}\ #2}

\newcommand{\casebool}[3]{\mathbf{case}\ #1\ \mathbf{of}\ 
      (\mathbf{false} \to #2 \mid \mathbf{true} \to #3)}

\newcommand{\inject}[2]{\lan #1, #2 \ran}
\newcommand{\casesum}[2]{\mathbf{case}\ #1\ \mathbf{of}\ 
			\inject{i}{\varone} \to #2}

\newcommand{\unfold}[1]{\mathbf{unfold}\; #1}

\newcommand{\envOne}{\Gamma}
\newcommand{\envTwo}{\Delta}

\renewcommand{\envone}{\envOne}
\renewcommand{\envtwo}{\envTwo}

\newcommand{\emptyenv}{\cdot}

\newcommand{\substtype}[3]{#1[#3/#2]}

\renewcommand{\howe}[1]{\mathbin{#1^{\scriptscriptstyle{H}}}}
\newcommand{\closedhowe}[1]{#1_0^{\scriptscriptstyle{H}}}

\newcommand{\laxcommuterel}{\ar @{} [dr] |\subseteq}

\renewcommand{\powerset}{\mathcal{P}}
\newcommand{\divergence}{\Uparrow}

\newcommand{\baseone}{s}

\newcommand{\baseid}{\bm{1}}
\newcommand{\bang}{{!}}

\newcommand{\toval}[1]{\mathbin{#1^{\scriptscriptstyle \values}}}
\newcommand{\toterm}[1]{\mathbin{#1^{\scriptscriptstyle \Lambda}}}

\newcommand{\valimp}{\imp^{\scriptscriptstyle{\values}}}
\newcommand{\compimp}{\imp^{\scriptscriptstyle{\Lambda}}}

\newcommand{\dual}[1]{#1^{\scriptstyle{{-}}}}

\newcommand{\defiff}{
{\mathrel{\ensurestackMath{\stackon[1pt]{\iff}{\scriptstyle\triangle}}}}}
\renewcommand{\defiff}{\equiv}

\newcommand{\vsim}{\delta}

\newcommand{\refine}[1]{\widehat{#1}}
\newcommand{\substrefine}[1]{#1^{\scriptscriptstyle{\mathsf{subst}}}}

\newcommand{\mtop}{\top}
\newcommand{\qtop}{\top_{\quantale}}
\newcommand{\qbot}{\bot_{\quantale}}
\newcommand{\Fuzz}{$\mathsf{Fuzz}$}

\newcommand{\relatorsymbol}{\Gamma}

\newcommand{\semn}[2]{\sem{#1}_{#2}}

\newcommand{\eval}{\mathit{eval}}

\newcommand{\substmap}{\gamma}

\newcommand{\appsimilarity}{\similarity^{\scriptscriptstyle \mathsf{A}}}
\newcommand{\appbisimilarity}{\bisimilarity^{\scriptscriptstyle \mathsf{A}}}

\newcommand{\Maybe}{\mathbb{M}}
\newcommand{\maybe}{M}

\renewcommand{\distribution}{D}

\newcommand{\evalsymbol}{eval}

\newcommand{\Two}{\mathbb{B}}

\newcommand{\Lawvere}{\mathbb{L}}
\newcommand{\StrongLawvere}{\mathbb{S}\mathbb{L}}
\newcommand{\Quantale}{\mathbb{V}}

\makeatletter
\newcommand{\uset}[3][0ex]{%
  \mathrel{\mathop{#3}\limits_{
    \vbox to#1{\kern-7\ex@
    \hbox{$\scriptstyle#2$}\vss}}}}
\makeatother

\newcommand{\wunit}{\varepsilon}
\newcommand{\wtop}{\omega}

\newcommand{\wone}{
  \fontfamily{qcs}\selectfont \textup{w}
  }
\newcommand{\wtwo}{\fontfamily{qcs}\selectfont \textup{v}}
\newcommand{\wthree}{\fontfamily{qcs}\selectfont \textup{u}}
\newcommand{\wfour}{\fontfamily{qcs}\selectfont \textup{z}}
\newcommand{\wfive}{\fontfamily{qcs}\selectfont \textup{y}}
\newcommand{\wsix}{\fontfamily{qcs}\selectfont \textup{x}}

\renewcommand{\wone}{\mathsf{w}}
\renewcommand{\wtwo}{\mathsf{v}}
\renewcommand{\wthree}{\mathsf{u}}
\renewcommand{\wfour}{\mathsf{z}}
\renewcommand{\wfive}{\mathsf{y}}
\renewcommand{\wsix}{\mathsf{z}}

\newcommand{\wcomp}{\bullet}
\newcommand{\wleq}{\mathbin{\blue{\leq}}}
\newcommand{\wgeq}{\mathbin{\blue{\geq}}}
\newcommand{\worlds}{\mathsf{W}}
\newcommand{\Worlds}{\mathbb{W}}
\newcommand{\worldset}{\mathsf{W}}

\newcommand{\kripkequantale}{\mathbb{B}^{\mathbb{W}}}

\newcommand{\mfour}{k}
\newcommand{\mfive}{h}
\newcommand{\mzero}{\blue{0}}
\newcommand{\munit}{\blue{1}}
\newcommand{\mleq}{\mathbin{\blue{\leq}}}
\newcommand{\mgeq}{\mathbin{\blue{\geq}}}
\newcommand{\mvee}{\vee}
\newcommand{\mact}{\diamond}

\newcommand{\mstar}{\mathbin{\blue{*}}}
\newcommand{\mplus}{\mathbin{\blue{+}}}
\renewcommand{\mtop}{\infty}

\newcommand{\qone}{a}
\newcommand{\qtwo}{b}

\newcommand{\quantalescat}{\mathsf{Quant}}

\newcommand{\bbox}{\square}

\newcommand{\tbox}[1]{[#1]}

\newcommand{\letbox}[2]{\mathbf{let}\ \tbox{\varone} = #1\ \mathbf{in}\ #2}
\newcommand{\letboxy}[2]{\mathbf{let}\ \tbox{\vartwo} = #1\ \mathbf{in}\ #2}

\newcommand{\blue}{\textcolor{black}}
\newcommand{\gmap}{\multimap}

\newcommand{\wrel}[4]{#4 \Vdash #2 \mathbin{#1} #3}

\newcommand{\wrelt}[5]{#4 \Vdash #2 \mathbin{#1} #3: #5}
\renewcommand{\wrelt}[5]{#2 \mathbin{#1(#4)} #3: #5}

\newcommand{\wrelo}[6]{#1 \mid #5 \Vdash #3 \mathbin{#2} #4: #6}
\renewcommand{\wrelo}[6]{#1 \imp #3 \mathbin{#2(#5)} #4: #6}

\newcommand{\wrelcomp}[6]{#1 \midd #5 \Vdash^{\scriptscriptstyle \Lambda} #3 \mathbin{#2} #4: #6}
\renewcommand{\wrelcomp}[6]{#1 \imp^{\scriptscriptstyle \Lambda} #3 \mathbin{#2(#5)} #4: #6}

\newcommand{\wrelval}[6]{#1 \midd #5 \Vdash^{\scriptscriptstyle \values} #3 \mathbin{#2} #4: #6}
\renewcommand{\wrelval}[6]{#1 \imp^{\scriptscriptstyle \values} #3 \mathbin{#2(#5)} #4: #6}

\newcommand{\wrelcompclosed}[5]{#4 \Vdash^{\scriptscriptstyle \Lambda} #2 \mathbin{#1} #3: #5}
\renewcommand{\wrelcompclosed}[5]{#2 \mathbin{\toterm{#1}(#4)} #3: #5
}

\newcommand{\wrelvalclosed}[5]{#4 \Vdash^{\scriptscriptstyle \values} #2 \mathbin{#1} #3: #5}
\renewcommand{\wrelvalclosed}[5]{#2 \mathbin{\toval{#1}(#4)} #3: #5}

\renewcommand{\appsimilarity}{\preceq}
\renewcommand{\appbisimilarity}{\simeq}

\newcommand{\duplicate}{\mathtt{d}}

\newcommand{\graded}[3]{#1 :_{#3} #2}

\newcommand{\diagramrel}[8]{
  \xymatrix{
    \laxcommuterel
    #3
    \ar[r]^-{#1} 
    \ar[d]_{#5}|@{|}  
    & #6   
    \ar[d]^{#8}|@{|}
    \\
    #4
    \ar[r]_-{#2}   
    & #7    
    }
}

\newcommand{\corelator}[2]{\Delta_{#1}(#2)}
\newcommand{\corelatorsymbol}{\Delta}
\newcommand{\mask}[2]{{\uparrow_{#1}}{#2}}

\newcommand{\public}{\mathtt{low}}
\newcommand{\secret}{\mathtt{high}}

\newcommand{\opposite}[1]{#1^{\mathsf{op}}}

\newcommand{\howen}[2]{#1^{\scriptscriptstyle H}_{#2}}

\renewcommand{\eval}[1]{\sem{#1}}
\renewcommand{\evalsymbol}{\sem{-}}
\newcommand{\evaln}[2]{\sem{#2}_{#1}}
\newcommand{\evalsymboln}[1]{\sem{-}_{#1}}
\renewcommand{\divergence}{\bot}

\newcommand{\metalambda}{%
    \rlap{$\lambda$}%
    \mkern2mu
    \raisebox{.2ex}{$\lambda$}
}
\newcommand{\metafun}[2]{\metalambda{#1}.#2}

\newcommand{\wpredone}{\varphi}
\newcommand{\wpredtwo}{\psi}

\newcommand{\gradealg}{\mathcal{J}}
\newcommand{\gplus}{+}

\newcommand{\gstar}{\mathbin{*}}
\newcommand{\gzero}{0}
\newcommand{\gunit}{1}
\newcommand{\gtop}{\infty}
\newcommand{\gleq}{\leq}
\newcommand{\ggeq}{\geq}
\newcommand{\gradeone}{j}
\newcommand{\gradetwo}{i}
\newcommand{\gradethree}{g}
\newcommand{\gradefour}{h}
\newcommand{\gradefive}{l}

\newcommand{\gradealgset}{J}

\newcommand{\relcomp}[2]{#1;#2}

\newcommand{\mactop}[2]{\widehat{#1}(#2)}

\newcommand{\substarrow}{\mathtt{subst}}
\renewcommand{\wrel}[4]{#2 \mathbin{#1(#4)} #3}

\newcommand{\vect}[1]{\mathbf{#1}}

\renewcommand{\distribution}{\mathcal{D}}

\renewcommand{\ctxone}{C}

\renewcommand{\maybe}{\mathsf{M}}

\renewcommand{\distribution}{\mathcal{D}}
\renewcommand{\powerset}{\mathcal{P}}

\newcommand{\TODO}[1]{\textcolor{red}{#1}}

\newcommand\sbullet[1][.7]{\mathbin{\vcenter{\hbox{\scalebox{#1}{$\bullet$}}}}}

\renewcommand{\wcomp}{\sbullet}

\begin{document}
%
\title{Modal Reasoning = Metric Reasoning\\ 
 via Lawvere\\
 \vspace{0.2cm}
 \LARGE{(Extended Version)}
 }

 \author{\IEEEauthorblockN{Ugo Dal Lago}
 \IEEEauthorblockA{University of Bologna \\
 Inria Sophia Antipolis}
 \and
 \IEEEauthorblockN{Francesco Gavazzo}
 \IEEEauthorblockA{University of Bologna \\
 Inria Sophia Antipolis}
 }

%


\maketitle

\begin{abstract}
Graded modal types systems and coeffects are becoming a standard 
formalism to deal with context-dependent computations where 
code usage plays a central role. The theory of program equivalence 
for modal and coeffectful languages, however, is considerably 
underdeveloped if compared to the denotational and 
operational semantics of such languages. This raises the 
question of how much of the theory of ordinary program equivalence 
can be given in a modal scenario. In this work, 
we show that \emph{coinductive} equivalences can 
be extended to a modal setting, and we do so by generalising 
Abramsky's applicative bisimilarity to coeffectful behaviours. 
To achieve this goal, we develop a general theory of 
ternary program relations based on the novel notion of a 
\emph{comonadic lax extension}, on top of which we define
a modal extension of 
Abramsky's applicative bisimilarity (which we dub 
\emph{modal applicative bisimilarity}). 
We prove such a relation to be a congruence, this way 
obtaining a compositional technique for reasoning about 
modal and coeffectful behaviours. 
But this is not the end of the story:  
we also establish a correspondence 
between modal program relations and program distances. 
This correspondence shows that modal applicative bisimilarity 
and (a properly extended) applicative bisimilarity distance 
coincide, this way revealing that
modal program equivalences and program distances 
are just two sides of the same coin.
\end{abstract}

%
\IEEEpeerreviewmaketitle

\section{Introduction}

Program equivalence, the study of notions of equality between programs, 
is a central topic in programming language semantics since the very birth of the discipline. For a fixed programming language, a notion of program equivalence 
is usually given in terms of an equivalence relation between program phrases 
relating pairs of programs exhibiting the same behaviour. 
Obviously, if the programming language at hand is equipped with a 
\emph{denotational semantics}, a notion of program equivalence is given by 
mathematical equality in the semantic model. However, interesting notions 
of program equivalence can be given even \emph{without} a denotational semantics, 
but 
relying on the operational behaviour of programs, only. 
Examples of such equivalences are Morris' contextual equivalence 
\cite{Morris/PhDThesis}, CIU equivalence \cite{MasonTalcott/1991},  
logical relations \cite{Plotkin-Lambda-definability-logical-relations,Reynolds/Logical-relations/1983}, and
applicative bisimilarity \cite{Abramsky/RTFP/1990}.

Although different, all the aforementioned equivalences share some 
common properties, the main one being 
\emph{compositionality}, whereby equivalent programs can be safely replaced for
one another inside larger software systems. 
More precisely, compositionality states that program equality 
is compatible with the syntactical constructs of the language, 
so that putting equivalent programs 
in arbitrary contexts always produces equivalent programs. 
If we think about contexts of the language as \emph{tests} an 
external observer can perform on programs, then compositionality states that 
\emph{no} test can distinguish 
between two equivalent programs. Therefore, program equivalence captures the idea of 
program indistinguishability in a black-box testing
scenario.
Compositionality then entails that program equality is \emph{context insensitive}: 
to be equal, 
two programs must exhibit the same behaviour in \emph{any} possible 
context, i.e. under \emph{any} possible test.

Oftentimes, however, we would like to reason about programs 
precisely \emph{in virtue} of the environment in which such programs are used. 
This ability is crucial in scenarios where, e.g., resource consumption, 
data security, and information leakage play a central role, as it often happens 
in today's software systems. 
Let us think, for instance, to a software manipulating 
sensitive data. In such a scenario, we would like users with unprivileged permissions 
to identify \emph{all} programs critically relying on classified data 
(such as medical data or passwords), even if an idealised 
observer could tell them apart. Why? Because 
if a user with unprivileged permissions can discriminate between programs 
that differ only for the \emph{classified} data they manipulate, 
then the user can infer some information about such data, 
meaning that the program causes an \emph{information leakage}. 
This is precisely the idea behind \emph{non-interference}, one 
of the main properties studied in 
the field of information flow \cite{DBLP:conf/popl/AbadiBHR99}.
A similar story can be told for, e.g., resource consumption, where 
one may want to identify programs that can only be discriminated 
by means of expensive computations, or only on certain hardwares, 
which may not be available to an external observer.

What does the literature offers to cope with these scenarios? 
From a programming language perspective, several new type systems 
 \cite{Gaboradi-et-al/ICFP/2016,Brunel-et-al/ESOP/2014,Orchard:2019:QPR:3352468.3341714,DBLP:conf/lics/Atkey18,Mycroft-et-al/ICFP/2014,DBLP:journals/pacmpl/BernardyBNJS18,DBLP:journals/pacmpl/AbelB20}
disciplining code usage have been recently developed. 
All these systems share two common features: 
\emph{(i)} they are resource-sensitive, relying 
on some sort of linearity 
\cite{DBLP:journals/tcs/Girard87};
\emph{(ii)} they have \emph{modal} type constructors 
indexed by grades, resources, or capabilities that specify how code can be manipulated. 
Such type constructors
generalise bounded exponential modalities \cite{DBLP:journals/tcs/GirardSS92}
and can be instantiated to recover modalities for resource analysis 
\cite{Orchard:2019:QPR:3352468.3341714},
program sensitivity \cite{Pierce/DistanceMakesTypesGrowStronger/2010},
and information flow \cite{DBLP:conf/popl/AbadiBHR99,DBLP:journals/jcs/VolpanoIS96} 
For these reasons, we generically speak of \emph{graded modal types}
or \emph{coeffects}.\footnote{The word \emph{coeffect} is usually employed 
to denote the family of program behaviours related to code usage and 
context-dependency.}

But what about program equivalence? 
Having good notions of program equivalence is crucial when dealing 
with modal types, as context-sensitive 
reasoning 
is usually modelled as a form of program equivalence. 
This is witnessed by a number of important theorems --- such as 
\emph{non-interference} \cite{DBLP:conf/popl/AbadiBHR99}, 
\emph{metric preservation}, and \emph{proof irrelevance} \cite{DBLP:conf/lics/Pfenning01} --- 
which are cornerstones of programming language techniques  
in fields like information flow, program security, and differential privacy.
Non-interference, for instance, relies on a notion of program equality 
parametrised by users' permissions (say, \emph{high} and \emph{low}),
and states that 
a function has the non-interference property if for any pair of 
\emph{secret} inputs, 
the resulting outputs are \emph{low equivalent}. 
All the aforementioned results, however, 
are tailored for specific modal type systems and coeffects. 
Hence our research question: \emph{is 
there a general notion of modal and coeffectful program equivalence 
of which non-interference, metric preservation, and the like 
are instances of?}
As there are many notions of non-modal program equivalence, 
our question can be declined in several ways (obviously, one can 
define a notion of contextual equivalence 
for a language with modal types, hence obtaining \emph{a} 
desired notion of modal equivalence). Therefore, a better question is: 
\emph{how much of the theory of non-modal program equivalence can be extended 
to a modal and coeffectful setting?}

Recently, Abel and Bernardy \cite{DBLP:journals/pacmpl/AbelB20} 
gave a positive answer to this question in the case of logical 
relations for a strongly normalising polymorphic $\lambda$-calculus 
with modal types. 
Logical relations, however, are just \emph{one} 
of the many ways to define program equivalence. Additionally, 
logical relations have the major drawback of not being able to 
readily handle infinitary behaviours, such as nontermination or 
recursive types. This raises an important question: 
is it possible to deal with programs exhibiting 
both \emph{coeffectful} and \emph{infinitary} behaviours? 

Abel and Bernardy's logical relations cannot provide an answer to the above question,
as they are defined on a strongly normalising
calculus. 
A promising way to obtain a positive answer is to move  
from inductive to \emph{coinductive} notions of program equivalence.
Among such equivalences, Abramsky's applicative bisimilarity 
\cite{Abramsky/RTFP/1990} is arguably the most well-known one 
in the context of higher-order sequential languages. 
From an operational perspective, applicative bisimilarity 
tests programs as argument-passing interactive process;
from a mathematical perspective, applicative bisimilarity equates functions according to 
the function extensionality principle in a coinductive fashion, 
this way obtaining powerful proof techniques (the coinduction proof principle, 
in first place) to handle infinitary computational behaviours. 
So the main question now becomes: can we extend coinductive
equivalences
to a modal and coeffectful setting? 

In this work, we answer this question in the affirmative
by extending Abramsky's applicative bisimilarity 
\cite{Abramsky/RTFP/1990} to coeffectful behaviours.  
Such an extension is obtained through the development of 
a general \emph{relational} theory of modal program equivalence 
based on monoidal Kripke relations and lax extensions of comonads. 
Let us expand on that.

\paragraph{Ternary Relations}
As witnessed by, e.g., non-interference, 
in a modal setting equivalences should relate programs 
with respect to objects reflecting 
usage constraints of (modal) programs. Following standard 
practice in logic \cite{DBLP:journals/mst/Lambek68,ROUTLEY1973199,DBLP:journals/jsyml/Urquhart72,DBLP:books/daglib/0006566},
we model this feature by working in a category of \emph{ternary relations}, 
which we call monoidal 
Kripke relations,
relating programs with respect to possible worlds 
living in monoidal preorders, 
i.e. preorders endowed with a monoid structure. 

\paragraph{Comonadic Lax Extensions}
The main novelty of our approach is the interpretation of modal types  
in terms of monoidal Kripke relations: to define 
such an interpretation we introduce the novel notion of a 
\emph{lax extension of a comonad}. 
Lax extensions of \emph{monads} \cite{Kurz/Tutorial-relation-lifting/2016,Hoffman-Seal-Tholem/monoidal-topology/2014,Hoffman/Cottage-industry/2015} 
are maps extending the action of monads on functions to (binary) relations, 
which play a a central role in topology \cite{Barr/LMM/1970,DBLP:journals/fss/ClementinoH17,Hoffman-Seal-Tholem/monoidal-topology/2014,Hoffman/Cottage-industry/2015}, 
coalgebra \cite{Thijs/PhDThesis/1996,Venema-Marti,Katsumata-Sato/FOSSACS/20S13}, 
and programming 
language semantics \cite{DalLagoGavazzoLevy/LICS/2017,DBLP:conf/esop/LagoG19,dal-lago/gavazzo-mfps-2019,GoubaultLasotaNowak/MSCS/2008,Simpson-Niels/Modalities/2018}.
Here, we show that modal types and coeffectful behaviour can be uniformly understood in terms of 
lax extensions of \emph{comonads} to the category of monoidal Kripke relations. 
Actually, we do not even need to extend arbitrary comonads: we can consider 
lax extension of the identity comonad 
(which we call \emph{comonadic lax extensions}), only. 
This reflects the 
intuition that the action of a modal type on a program 
does not modify the program, but its usage. 

Comonadic lax extensions 
give an abstract axiomatisation of 
the action of changing possible worlds, the consequence being 
that comparing programs via a comonadic lax extension forces 
one to compare the very same programs, but in a different 
possible world (where, for instance, the observer has more or less resources 
at her disposal).
Compared with other relational interpretations of modal types 
\cite{DBLP:journals/pacmpl/AbelB20,DBLP:conf/csl/BreuvartP15}, 
comonadic lax extensions build upon the rich theory of relation 
lifting 
\cite{Kurz/Tutorial-relation-lifting/2016,Hoffman/Cottage-industry/2015}: 
it is precisely this level of abstraction that allows us to relate 
(monadic) effectful and (comonadic) coeffectful behaviours 
relationally, and thus to account for both \emph{infinitary}
(and, as we will see, generic monadic) and 
\emph{coeffectful} phenomena at the same time, something not readily 
possible in other relational frameworks. 

\paragraph{Modal Applicative Bisimilarity}
On top of this relational framework, we define modal applicative 
bisimilarity --- the extension of applicative bisimilarity to a modal 
setting --- for a call-by-value $\lambda$-calculus with (graded) modal and 
recursive types. 
Our first main result (Theorem~\ref{theorem:congruence}) states 
that modal applicative bisimilarity is a
\emph{congruence}, from which a general 
compositionality theorem subsuming the aforementioned non-interference, 
proof irrelevance, and metric preservation theorems follows. 
Proving such a congruence result, however, is nontrivial. In fact, 
proving applicative bisimilarity to be a congruence 
is difficult already in a 
non-modal setting, 
the proof being based on a non-elementary relational construction known as 
Howe's method \cite{Howe/IC/1996,Pitts/ATBC/2011}. In a modal setting, 
the situation is even worse, as possible worlds have to be taken into 
account, with the consequence that  
routine lemmas (notably, substitutivity) 
now require nontrivial proofs based on the axiomatics of a 
comonadic lax extension. 

\paragraph{Modalities as Metrics}
But this is not the end of the story. 
Our second main result (Theorem~\ref{theorem:modal-reasoning-equal-metric-reasoning} and 
Theorem~\ref{theorem:abstract-metric-preservation}) states that 
modal program equivalence and program distance are one and the same. 
More precisely, our notion of a comonadic lax extension 
can be used to improve Gavazzo's theory of abstract program distance 
\cite{Gavazzo/LICS/2018} 
(the latter being a theory of coinductively-defined distances based on 
Lawvere analysis of metric spaces as enriched categories \cite{Lawvere/GeneralizedMetricSpaces/1973}) with the consequent result 
that modal applicative bisimilarity and applicative bisimilarity distance are just two sides of 
the same coin. 
This equivalence (that can be easily extended to other 
notions of equivalence and metric) 
allows the modal and the metric worlds to improve one another. 
For instance, it is an easy exercise to show that the many results on 
program distance for languages with 
\emph{monadic effects} can be now imported into the modal world, this way 
obtaining a large family of operationally-based techniques for 
\emph{effectful} and \emph{coeffectful} languages that, to the best of the 
authors' knowledge, are not available in any of the operational theories
of coeffects present in the literature.

\textbf{Summary and Outline}\; 
Summing up, our contributions are: 
\emph{(i)} the definition of a relational theory of modal program
equivalence
based on the novel notion of a comonadic lax extension (Section~\ref{section:relational-reasoning} and 
Section~\ref{section:corelators-and-relators}); 
\emph{(ii)} the definition of modal applicative bisimilarity
and a compositionality theorem for it 
based on a nontrivial extension 
of Howe's method (Section~\ref{section:modal-applicative-bisimilarity}); 
\emph{(iii)} a correspondence between modal equivalence and 
program distance  (Section~\ref{section:kripke-meets-lawvere}).

\section{Modal Calculi}
\label{section:modal-calculi}

The vehicle calculus of this work is a call-by-value \emph{affine} 
$\lambda$-calculus 
with \emph{modal necessity types} graded in an algebra $\gradealg$, 
which we are going to formally introduce.
We call our calculus $\Lambda_{\gradealg}$. 
The raw syntax of $\Lambda_{\gradealg}$ is given by the following grammars, 
respectively for types, values, and terms 
(where $\gradeone$ ranges over elements in $\gradealg$).
\begin{align*}
\typeone  
  &::= \typevarone 
  \mid \typeone \gmap \typeone 
  \mid \rectype{\typevarone}{\typeone} 
  \mid \bbox_{\gradeone} \typeone
\\
\valone 
  &::= \varone 
  \mid \abs{\varone}{\termone}
  \mid \fold{\valone}
  \mid \tbox{\valone} 
\\
\termone 
  &::= \valone 
  \mid \valone \valone
  \mid \unfold{\valone}
  \mid \seq{\termone}{\termone}
  \mid \letbox{\valone}{\termone}.
\end{align*}

For ease of exposition, we consider a minimal set of types consisting of 
recursive types $\rectype{\typevarone}{\typeone}$, 
\emph{affine} arrow types $\typeone \gmap \typetwo$, and 
\emph{graded modal} necessity types $\bbox_{\gradeone} \typeone$: this is enough 
to study modal, higher-order, and infinitary properties of programming languages. 
Nonetheless, our results have been developed for an extension of  
$\Lambda_{\gradealg}$ including product, tensor, and finitary 
sum types. 

We adopt standard notational and terminological conventions \cite{Barendregt/Book/1984}. 
In particular, we work with types and expressions modulo $\alpha$-conversion and
write $\subst{\termone}{\varone}{\termtwo}$ for the 
capture-avoiding substitution of $\termtwo$ for all the free occurrences of 
the variable $\varone$ in $\termone$. We employ a similar notation for types.
Before going any further, it is convenient to formally introduce 
grade algebras. 

\begin{definition}
A \emph{grade algebra} 
$\gradealg = (\gradealgset, \mleq, \mplus, \mstar, \mzero, \munit, \mtop)$ is 
a \emph{preordered semiring with top element}, i.e 
a semiring $(\gradealgset, \gplus, \gstar, \gzero, \gunit)$ 
together with a preorder $\gleq$ for which both $\gplus$ and 
$\gstar$ are monotone, and $\gtop$ is the top element.
\end{definition}

Elements of a grade algebra $\gradealg$
are called \emph{grades} (but also \emph{resources} or \emph{capabilities}
\cite{Orchard:2019:QPR:3352468.3341714}) and are used to constrain the way 
code can be manipulated. The following examples will clarify the concept. 

\begin{example} 
\label{ex:modal-signatures}
\begin{varenumerate}
  \item The one element semiring $\{\infty\}$ is used to model 
    the exponential modality of linear logic 
    \cite{DBLP:conf/lics/BentonW96,DBLP:journals/tcs/Girard87}. 
    An expression of type $\bbox_{\infty} \typeone$ represents a 
    piece of code that can be freely duplicated and discharged. 
    This property comes from idempotency of the 
    semiring addition, which gives $\infty + \infty = \infty$. 
    Notice that here the zero, unit, and top elements of the semiring coincide.
  \item The semiring of natural numbers extended with infinity
    $(\blue{\mathbb{N}^{\infty}}, =, \blue{+}, \blue{\cdot}, 
    \blue{0}, \blue{1}, \infty)$ is used to model 
    the 
    \emph{exact usage} modality from bounded linear logic 
    \cite{DBLP:journals/tcs/GirardSS92}. Accordingly, 
    a term of type $\bbox_{\blue{n}} \typeone$ represents a 
    piece of code that has to be used exactly $n$ times. 
    Notice that here the semiring addition is not idempotent.
  \item The semiring of extended non-negative real numbers\footnote{
    Recall that $\infty + r = r + \infty = \infty$ and 
    that $\infty \cdot r = r \cdot \infty = \infty$, if $r \neq 0$, 
    and $\infty \cdot 0 = 0 \cdot \infty = 0$. The latter equality 
    captures our intuitive understanding that unused programs should be always 
    regarded as equivalent.
  }  
    $([0,\infty], \leq, \blue{+}, \blue{\cdot}, 
    \blue{0}, \blue{1}, \infty)$ is used to model type systems 
    for program sensitivity \cite{Pierce/DistanceMakesTypesGrowStronger/2010}. 
    Accordingly, we think about expressions of type $\bbox_{\gradeone} \typeone \to \typetwo$ as representing functions that are $\gradeone$-Lipschitz continuous.
  \item 
    Distributive lattices 
    $(\mathcal{L}, \geq, \wedge, \vee, \top, \bot)$ 
    \longv{
    Notice that the order of 
    the lattice is reversed, so that the semiring addition is given by 
    the meet of the lattice (its cartesian product), 
    whereas multiplication is given by the join of the lattice.
    }
    are used to model  
    information flow modalities, such as modalities describing
    security levels \cite{DBLP:journals/cacm/Denning76,DBLP:journals/jcs/VolpanoIS96,DBLP:conf/popl/AbadiBHR99}. 
    As a running example, we consider the two-element lattice 
    $\{\public \leq \secret\}$. A term of type 
    $\bbox_{\secret} \typeone$ represents a piece of code accessible only 
    by users with high confidentiality level. Dually, a term 
    of type $\bbox_{\public} \typeone$ can be used by users with at least 
    low confidentiality level, and thus by any user. 
\end{varenumerate}
\end{example}

Notice that grade algebras are not required to satisfy $\mzero \neq \munit$. However, 
it is convenient to require $\mzero$ to be the bottom element 
of the algebra (i.e. $\mzero \mleq \gradeone$, for any $\gradeone)$. 
Moreover, we also require 
the join of two elements to exist, so that for all elements 
$\gradeone, \gradetwo \in \gradealgset$, the element $\gradeone \vee \gradetwo$ 
exists and belongs to $\gradealgset$.\footnote{Actually, we 
do not need $\gradeone \vee \gradetwo$ to exist for any $\gradetwo$, but just 
for $\gradetwo = \munit$.} Such a condition 
is necessary to guarantee soundness of our type system \cite{Gavazzo/LICS/2018}. 

\subsection{Statics and Dynamics}
\label{section:typing}
Let us now fix a grade algebra 
$\gradealg = (\gradealgset, \gleq, \gplus, \gstar, \gzero, \gunit, \gtop)$. 
We endow $\Lambda_{\gradealg}$ with a typing system 
relying on the judgments 
$\envone \valimp \valone: \typeone$ and $\envone \compimp \termone: \typeone$,
where $\typeone$ is a \emph{closed} type and $\envone$ is an environment.
Environments are sets of \emph{graded variables}, i.e. 
expressions of the form $\graded{\varone}{\typeone}{\gradeone}$ acting as placeholders for 
code of type $\typeone$ that has to be manipulated according to $\gradeone$. 
For instance, if $\gradeone$ belongs to the algebra of 
natural numbers extended with infinity, then 
$\graded{\varone}{\typeone}{\gradeone}$ is a placeholder for a piece of code that
has to be used $\gradeone$ times. We write $\emptyenv$ for the empty environment, 
and extend the semiring structure of $\gradealg$ to environments 
in the natural way. 

We use judgments to distinguish between arbitrary terms and 
values \cite{Levy/InfComp/2003}, 
the former representing programs to be evaluated, and the latter representing the 
result of the evaluation of such programs.\footnote{Notice that any value can be 
regarded as a term that simply evaluates to itself.} 
Accordingly, a judgment 
of the form $\envone \valimp \valone: \typeone$ asserts that 
$\valone$ is a \emph{value} of type $\typeone$ in 
environment $\envone$, whereas judgments of the form $\envone \compimp \termone: \typeone$
assert that 
$\termone$ is a term of type $\typeone$ in environment $\envone$.

Finally, we endow $\Lambda_{\gradealg}$ with the type system 
defined in Figure~\ref{fig:simply-typed}. 
We write $\Lambda^{\envone}_{\typeone}$ and $\values^{\envone}_{\typeone}$ 
for the collections of terms and values having type $\typeone$ in 
the environment $\envone$, simply writing $\Lambda_{\typeone}$ and $\values_{\typeone}$ if the environment is empty.

\begin{figure*}[htbp]
\hrule 
\vspace{0.2cm}
\[
\infer
  {\envone, \graded{\varone}{\typeone}{\gradeone} \valimp x: \typeone}{\gradeone \mgeq \munit}
\quad
\infer
  {\envone \valimp \abs{\varone}{\termone}: \typeone \gmap \typetwo}
  {\envone, \graded{\varone}{\typeone}{\munit} \compimp \termone: \typetwo}
\quad 
\infer{\envone \mplus 
\envtwo \compimp \valone \valtwo: \typetwo}
{\envone \valimp \valone: \typeone \gmap \typetwo
& 
\envtwo \valimp \valtwo: \typeone}
\quad
\infer{\envone \compimp \valone: \typeone}{\envone \valimp \valone:\typeone}
\quad
\infer{
  (\gradeone \vee \munit) \mstar \envone \mplus \envtwo \compimp \seq{\termone}{\termtwo}: 
  \typetwo
}
{\envone \compimp \termone: \typeone
&
\envtwo, \graded{\varone}{\typeone}{\gradeone} \compimp \termtwo: \typetwo
}
\]
\vspace{-0.2cm}
\[
\infer{\envone \valimp \fold{\valone}: \rectype{\typevarone}{\typeone}}
{\envone \valimp \valone:
\substtype{\typeone}{\typevarone}{\rectype{\typevarone}{\typeone}}}
\quad 
\infer{\envone \compimp \unfold{\valone}: \substtype{\typeone}{\typevarone}{\rectype{\typevarone}{\typeone}} }
{\envone \valimp \valone: \rectype{\typevarone}{\typeone}}
\quad
\infer{\gradeone \mstar \envone \valimp \tbox{\valone}: \bbox_{\gradeone} \typeone}
{\envone \valimp \valone: \typeone}
\quad
\infer{\gradetwo \mstar \envone \mplus \envtwo \compimp \letbox{\valone}{\termone}: \typeone'}
{\envone \valimp \valone: \bbox_{\gradeone} \typeone
&
\envtwo, \graded{\varone}{\typeone}{\gradetwo \mstar \gradeone} \compimp \termone: \typeone'
}
\]
\hrule
\caption{$\Lambda_{\gradealg}$: Core Fragment}
\label{fig:simply-typed}
\end{figure*}

Let us now comment on the rules in Figure~\ref{fig:simply-typed}, 
starting with the role played by semiring operations on environments. 
Following the intuition that terms of
type $\bbox_{\gradeone} \typeone$ are pieces of code of type 
$\typeone$ that can be manipulated according to the modal label $\gradeone$,
we see that whenever we have a value $\valone$ with free variables in $\envone$ 
and we want to use $\valone$ in place of a variable used by another program 
according to $\gradeone$, then 
we need variables in $\envone$ to be themselves usable according $\gradeone$: 
this is formalised by the environment $\gradeone \gstar \envone$, where we 
use semiring multiplication to give conditions on code to be 
used inside other code. 
For instance, if a variable $\varone$ is used by $\valone$ twice and 
we want to replace $\valone$ for a variable $\vartwo$ used $3$ times in 
a term $\termone$, then $\varone$ will be used $6$ times in 
$\subst{\termone}{\vartwo}{\valone}$.

\longv{
      Semiring addition is used to put together independent pieces of code. 
      Consider the case of application, and suppose to have two 
      values $\valone$ and $\valtwo$ using a free variable $x$ 
      according to $\gradeone$ and $\gradetwo$, respectively (e.g., 
      $\gradeone$ and $\gradetwo$ are the number of times $\valone$ and 
      $\valtwo$ use $\varone$, respectively). Then 
      $x$ will be used according to $\gradeone \gplus \gradetwo$ by $\valone\valtwo$. 
}
We can now look at some technical features of the 
rules in Figure~\ref{fig:simply-typed}. 
Most of such rules are standard in the context of 
graded calculi, although there are minor differences with other presentations. 
For instance, in the first rule in Figure~\ref{fig:simply-typed}
it is often required $\gradeone=\munit$ and the environment 
$\mzero \mstar \envone$ is used in place of $\envone$, this way staying 
closer to linear calculi. 
Our choice of allowing $\gradeone$ to be greater than or equal to $\munit$ 
is in line with examples coming from differential privacy and information flow 
\cite{Pierce/DistanceMakesTypesGrowStronger/2010,GaboardiEtAl/POPL/2017,DBLP:conf/popl/AbadiBHR99}. 
Nonetheless, all our results hold for calculi where one requires $\gradeone = \munit$.

Another important difference is given by the typing rule for sequencing, which 
comes from type systems for abstract program metrics \cite{Gavazzo/LICS/2018}
and that has been used in the context of modal types 
\cite{DBLP:journals/pacmpl/AbelB20} more recently. The rationale behind such a rule 
can be easily understood in terms of resource usage. Suppose to have a term 
$\termtwo$ using a variable $\varone$ zero times, and a term 
$\termone$ using a variable $\varone$ two times. How many times 
$\seq{\termone}{\termtwo}$ uses $\vartwo$? One may be tempted to say that 
$\vartwo$ is not used in $\seq{\termone}{\termtwo}$ at all, as $\termone$ is 
simply thrown away in $\termtwo$. However, in a call-by-value scenario the term
$\seq{\termone}{\termtwo}$ first evaluates $\termone$, and then it throws it away. 
As a consequence, the variable $\vartwo$ is still used twice in $\seq{\termone}{\termtwo}$. 
If we were to replace $\gradeone \vee \gunit$ with $\gradeone$ in 
the conclusion of the sequencing rule,
then we would obtain that $\vartwo$ is not used in $\seq{\termone}{\termtwo}$, which 
is just unsound. Our choice of using $(\gradeone \mvee \munit) \mstar \envone$ in place 
$\gradeone \mstar \envone$ models the fact that $\termone$ is evaluated in 
$\seq{\termone}{\termtwo}$, and thus it is used at least once.

\shortv{
  \begin{example}[Modal Calculi]
  \label{ex:modal-calculi}
  We now instantiate $\Lambda_{\gradealg}$ with suitable 
  grade algebras to recover several examples of modal calculi
  that have been studied in the literature both in isolation 
  \cite{DBLP:conf/lics/BentonW96,DBLP:conf/popl/AbadiBHR99,Pierce/DistanceMakesTypesGrowStronger/2010,GaboardiEtAl/POPL/2017,DBLP:journals/mscs/PfenningD01,DBLP:journals/jacm/DaviesP01,DBLP:journals/tcs/GirardSS92} 
  and in the context of general modal, quantitative, and coeffectful calculi \cite{Mycroft-et-al/ICFP/2014,DBLP:conf/esop/GhicaS14,Gaboradi-et-al/ICFP/2016,Brunel-et-al/ESOP/2014,Orchard:2019:QPR:3352468.3341714,DBLP:conf/lics/Atkey18,DBLP:journals/pacmpl/AbelB20,DBLP:journals/corr/abs-2005-02247}. 
  \begin{enumerate}
    \item
      Let us instantiate $\gradealg$ as the 
      one-element semiring
      $\{\infty\}$. Obviously, addition and multiplication on $\{\infty\}$ 
      are idempotent operations so that 
      in a judgment of the form $\envone \imp \termone: \typeone$, all variables in 
      $\envone$ can be freely erased and duplicated. Therefore, 
      $\Lambda_{\gradealg}$ is nothing but a standard, non-modal 
      call-by-value $\lambda$-calculus.
    \item Instantiating $\gradealg$ as the three-element chain 
      $\{0 \leq 1 \leq \infty\}$, we recover a standard \emph{affine} 
      call-by-value $\lambda$-calculus.
    \item 
      Instantiating $\gradealg$ 
      as
      $(\mathbb{N}^{\infty}, \leq, +, \cdot, 0,1, \infty)$, we obtain a type 
      system for approximate usage analysis. Accordingly, a term of $\bbox_{n}\typeone$ 
      can be used at most $n$ times and a judgment of the form 
      $\graded{\varone}{\typeone}{n} \compimp \termone$ states that 
      to produce (a single unit) $\termone$ we need to use at most 
      $n$ copies of $\varone$. 
      \item 
      Instantiating $\gradealg$ 
      as
      $([0,\infty], \leq, +, \cdot, 0,1, \infty)$, we obtain 
      a variation of \textsf{Fuzz} \cite{Pierce/DistanceMakesTypesGrowStronger/2010}, 
      a call-by-value $\lambda$-calculus with a type system tracking program sensitivity
      \cite{Pierce/DistanceMakesTypesGrowStronger/2010,DBLP:conf/lics/AmorimGHK19,GaboardiEtAl/POPL/2017,DBLP:conf/icfp/DAntoniGAHP13}. 
      We read 
      judgments of the form 
      $\graded{\varone}{\typeone}{\gradeone} \compimp \termone: \typetwo$
      as stating that $\termone$ is a $\gradeone$-continuous function, in the sense of 
      Lipschitz-continuity, and refer to $\gradeone$ as the sensitivity of 
      of $\varone$ in $\termone$. 
      \item
      Instantiating $\gradealg$ as the opposite of a security lattice 
      \cite{DBLP:journals/cacm/Denning76}
      $(\mathcal{L}, \leq, \wedge, \vee, \bot, \top)$,
      we obtain a call-by-value $\lambda$-calculus for information flow 
      \cite{DBLP:conf/popl/AbadiBHR99,DBLP:journals/jcs/VolpanoIS96}.
      Here, we regard elements in $\mathcal{L}$ as security levels 
      and read $\ell_1 \leq \ell_2$ as stating that data labelled with 
      $\ell_2$ are more secure (or confidential) than those labelled with 
      $\ell_1$. We take the two-element 
      lattice $\{\secret \geq \public\}$ as a running example.
      A term of type $\bbox_{\gradetwo}\typeone$ is a piece of code that can 
      be used in any security level below $\gradetwo$, so that if 
      $\termone$ is a high confidential term (so that it 
      has type $\bbox_{\secret} \typeone$), then it can be used 
      also as a low confidential term, but the vice versa does not hold.
      More generally, we regard the opposite
      $(\mathcal{L}, \geq, \wedge, \vee, \top, \bot)$
      of any security lattice 
      as a grade algebra.
    \end{enumerate}
    \end{example}

    We conclude this section with a remark on our design choices. 

      \begin{remark}[On Language Extensions and Comparison]
      \label{rem:extneded-calculus}
      $\Lambda_{\gradealg}$ is a minimal calculus whose type system supports
      higher-order, modal, and infinitary features. However, all our results 
      are robust with respect to language extensions, such as the addition of 
      finitary product, tensor, and sum types.
      Several calculi for graded modal 
      types have been recently proposed in the literature, most of them 
      differing (one another) for small details, such as side conditions on typing rules 
      or the axioms of grade algebras. 
      $\Lambda_{\gradealg}$ is meant to be a simple vehicle calculus
      to illustrate our relational techniques, rather 
      than as a calculus giving a comprehensive account of modal types. 
      In fact, it is easy to realise that our techniques and results can be adapted to 
      all the main (graded) modal calculi in the literature.
      \end{remark}
}

\longv{
      We conclude this section with a remark on our design choices. 

      \begin{remark}[On Language Extensions and Comparison]
      \label{rem:extneded-calculus}
      $\Lambda_{\gradealg}$ is a minimal calculus whose type system supports
      higher-order, modal, and infinitary features. However, all our results 
      are robust with respect to language extensions, such as the addition of 
      finitary product, tensor, and sum types.
      Another important feature that $\Lambda_{\gradealg}$ lacks is modal subtyping, 
      which can be added via the following rule
      (notice that modal subtyping is contravariant):
      \[
      \infer{\envone \imp \termone: \bbox_{\gradetwo}\typeone}
      {\envone \imp \termone: \bbox_{\gradeone} \typeone 
      & \gradetwo \mleq \gradeone}
      \]
      Again, our techniques can be extended to calculi with modal subtyping with 
      minor efforts.  

      It is also important to stress that several calculi for graded modal 
      types have been recently proposed in the literature, most of them 
      differing one another for small details, such as side conditions on typing rules 
      or the axioms of grade algebras. For instance, 
      several authors \cite{Gaboradi-et-al/ICFP/2016,DBLP:conf/esop/GhicaS14,Orchard:2019:QPR:3352468.3341714} do not
      require grade algebras to have a top element, as they deal with 
      strongly normalising calculi only; or they prefer dual contexts formalism 
      \cite{DBLP:journals/jacm/DaviesP01,DBLP:journals/mscs/PfenningD01,DBLP:conf/lics/Plotkin93,DBLP:conf/mfcs/Wadler93,DBLP:conf/mfps/Wadler93,DBLP:journals/apal/Girard93,DBLP:journals/logcom/Andreoli92,Barber96} for defining typing judgments distinguishing 
      between linear and graded variables.
      Again, $\Lambda_{\gradealg}$ is meant to be a simple vehicle calculus
      to illustrate our operational techniques, rather 
      than as a calculus giving a comprehensive account of modal types. 
      In fact, it is easy to realise that our techniques and results can be easily adapted to 
      all the main (graded) modal calculi in the literature.
      \end{remark}

        \section{Modal Calculi: Examples}
        \label{section:examples-modal-calculi}

        In this section we introduce some examples of modal calculi. 
        Such calculi have been extensively studied both in isolation 
        \cite{DBLP:conf/lics/BentonW96,DBLP:conf/popl/AbadiBHR99,Pierce/DistanceMakesTypesGrowStronger/2010,GaboardiEtAl/POPL/2017,DBLP:journals/mscs/PfenningD01,DBLP:journals/jacm/DaviesP01,DBLP:journals/tcs/GirardSS92} 
        and in the context of general coeffectful calculi \cite{Mycroft-et-al/ICFP/2014,DBLP:conf/esop/GhicaS14,Gaboradi-et-al/ICFP/2016,Brunel-et-al/ESOP/2014,Orchard:2019:QPR:3352468.3341714,DBLP:conf/lics/Atkey18,abel,DBLP:journals/corr/abs-2005-02247}.

        \begin{example}[Non-modal Calculi]
        Our first example is given by taking as modal signature $\gradealg$ the 
        one-element semiring
        $\{\infty\}$. Obviously, addition and multiplication on $\{\infty\}$ 
        are idempotent operations 
        ($\infty \mplus \infty = \infty \mstar \infty = \infty$) and 
        $\mzero = \munit = \infty$. As a consequence, 
        in a judgment of the form $\envone \imp \termone: \typeone$ all variables in 
        $\envone$ are graded by $\infty$, meaning that they can be freely erased and 
        duplicated in $\termone$. That is, $\Lambda_{\gradealg}$ is nothing but 
        a standard, non-modal call-by-value $\lambda$-calculus.
        \end{example}

        \begin{example}[Usage Analysis]
        Our second example shows how to use modal types to 
        perform approximate usage analysis. For that, we consider 
        the modal signature
        $(\mathbb{N}^{\infty}, \leq, +, \cdot, 0,1)$.
        The type $\bbox_{n}\typeone$ is inhabitated by 
        those pieces of code that can be used at most $n$ times. 
        Accordingly, we read judgments of the form 
        $\graded{\varone}{\typeone}{n} \imp \termone$ as stating that 
        to produce (a single unit) $\termone$ we need to use at most 
        $n$ copies of $\varone$. Here are some examples of typing rule:
        \[
        \infer{\graded{\varone}{\typeone}{n + m} \compimp \valone \valtwo: \typethree}
        {\graded{\varone}{\typeone}{n} \valimp \valone: \typetwo \to \typethree
        &
        \graded{\varone}{\typeone}{m} \valimp \valtwo: \typetwo}
        \qquad
        \infer
        {\graded{\varone}{\typeone}{n \cdot m} \valimp \tbox{\valone}: \bbox_{m}\typetwo}
        {\graded{\varone}{\typeone}{n} \valimp \valone: \typetwo}
        \]
        Let us comment some of their main features. 
        The application rule states that if we need (at most) 
        $n$ copies of $\varone$ to produce $\valone$, and (at most)
        $m$ copies of $\varone$ to produce $\valtwo$, then we need 
        (at most) $n + m$ copies of $\varone$ to produce the application 
        $\valone\valtwo$.
        Similarly, we read the introduction rule for the box modality 
        as follows: if we need (at most) $n$ copies of $\varone$ to produce $\valone$, 
        then to produce $m$ copies of $\valone$ we need $m \cdot n$ 
        copies of $\varone$. 
        \end{example}

        \begin{example}[Program Sensitivity]

        Replacing $\mathbb{N}^{\infty}$ with $[0,\infty]$ in previous example, we obtain 
        a variation of \textsf{Fuzz} \cite{Pierce/DistanceMakesTypesGrowStronger/2010}, 
        a call-by-value $\lambda$-calculus with a type system tracking program sensitivity
        used to statically guarantee differential privacy properties of programs \cite{Pierce/DistanceMakesTypesGrowStronger/2010,DBLP:conf/lics/AmorimGHK19,GaboardiEtAl/POPL/2017,DBLP:conf/icfp/DAntoniGAHP13}. 
        Formally, we consider the modal 
        signature $([0,\infty], \leq, +, \cdot, 0,1)$ and read 
        judgments of the form $\graded{\varone}{\typeone}{s} \imp \termone: \typetwo$
        as stating that $\termone$ is a $s$-continuous function, in the sense of 
        Lipschitz-continuity, and refer to $s$ as the sensitivity of 
        of $\varone$ in $\termone$. 
        Inhabitants of types $\bbox_{s}\typeone$ 
        are thus expressions that can be used as inputs to programs that are $s$-sensitive 
        to their input. 
        \end{example}

        \begin{example}[Information Flow]
        So far we focused on \emph{monoidal} modal signatures. 
        Starting with the seminal work by Abadi et al. \cite{DBLP:conf/popl/AbadiBHR99},
        \emph{cartesian} modal signatures have been extensively 
        used to define programming-language approaches to security and 
        information flow. The idea behind these approaches is to keep track of how information 
        flows throughout programs by decorating types with elements 
        of a lattice 
        $(\mathcal{L}, \leq, \wedge, \vee, \bot, \top)$
        of security levels \cite{DBLP:journals/cacm/Denning76}, with the informal 
        reading that if $\ell_1 \leq \ell_2$, then data labelled with 
        $\ell_2$ are more secure (or confidential) than those labelled with 
        $\ell_1$. As a prototypical example, we consider the following two-element 
        lattice.
        \[
        \xymatrix{
          \secret \ar@{-}[d] 
          \\
          \public
        }
        \]
        An expression of type $\bbox_{\gradetwo}\typeone$ is a piece of code that can 
        be used in any security level below $\gradetwo$, so that if 
        $\termone$ is a secret/confidential/private expression (so that it 
        has type $\bbox_{\secret} \typeone$), then it can be used 
        also as a public code, but the vice versa does not hold.
        More generally, we regard any lattice $(\mathcal{L}, \geq, \wedge, \vee, \top, \bot)$
        as a modal signature
        $(\gradealg, \mleq, \mplus, \mstar, \mzero, \munit)$. 
        Notice that, contrary to standard presentations of lattices, 
        here we use the order $\geq$. As a consequence, the bottom element of the semiring 
        is $\top$, whereas the top one is $\bot$. Moreover, 
        semiring addition and multiplication are given by (binary) meet and join, respectively.
        We summarise this correspondence in the following table.

        \begin{center}
        \begin{tabular}{|cccccc|}
        \hline
        $\mathcal{L}$ & $\geq$ & $\wedge$ & $\vee$ & $\top$ &$\bot$
        \\
        \hline
        $\gradealg$ & $\mleq$ & $\mplus$ & $\mstar$ & $\mzero$ & $\munit$
        \\
        \hline
        \end{tabular}
        \end{center}

        For the two-element lattice, we thus obtain typing rules such as 
        the following ones.
        \[
        \infer
        {\secret \vee \envone, \graded{\varone}{\typeone}{\public} 
        \valimp \varone: \typeone}
        {}
        \quad
        \infer{\graded{\varone}{\typeone}{\secret  \wedge \public} \compimp 
        \valone \valtwo: \typetwo}
        {\graded{\varone}{\typeone}{\secret} \valimp 
        \valone: \typeone \to \typetwo
        &
        \graded{\varone}{\typeone}{\public} \compimp 
        \valtwo: \typeone
        }
        \]
        \[
        \infer{\graded{\varone}{\typeone}{\secret \vee \public} \valimp \tbox{\valone}: 
        \bbox_{\secret} \typetwo}
        {\graded{\varone}{\typeone}{\public} \valimp \valone: 
        \typetwo}
        \quad
        \infer{\graded{\varone}{\typeone}{\public \vee \secret} \compimp 
        \letboxy{\valone}{\termone}: \typethree}
        {\graded{\varone}{\typeone}{\secret} \valimp \valone: 
        \bbox_{\secret} \typetwo
        &
        \graded{\vartwo}{\typetwo}{\public \vee \secret} \compimp \termone: \typethree}
        \]
        The variable rule requires the used variable $\varone$ to be public, whereas 
        unused variables can be secret. The rule for application shows that
        although $\valtwo$ uses a secret variable 
        $\varone$, the same variable must be public in $\valone$, and thus we 
        need it to be public in $\valone\valtwo$ 
        (in fact, $\secret \wedge \public = \public$). 
        Dually, the introduction rule for the box modality  shows that 
        if we make a value $\valone$ secret, then we must classify all its variables (in fact, 
        $\secret \vee \public = \secret$).
        Although a similar reading explains the elimination rule for the box modality, 
        it is instructive to look at an example of how the type system prevents 
        undesired behaviours. 

        By undesired behaviours, we mean 
        secret information flow out of programs in the following sense.
        Let us consider an expression 
        $\graded{\varone}{\typeone}{\secret} \imp \termone$. Then, a user with public 
        permission cannot infer any information on the input of $\termone$ 
        from the behaviour of $\termone$. Stated otherwise, no (secret) information
        about the input $\varone$ flows out of $\termone$. This idea is formalised 
        by the so-called \emph{non-interference}, which we will discuss in 
        Section~\ref{section:compositionality-metric-preservation-and-non-interference}.
        For the moment, let us approach non-interference by stating that, 
        given $\graded{\varone}{\typeone}{\secret} \imp \termone$ and two input values 
        $\valone$ and $\valtwo$, if we have public permission only, then 
        we cannot discriminate between $\subst{\termone}{\varone}{\valone}$ 
        and $\subst{\termone}{\varone}{\valtwo}$. 

        For instance, we may want to infer information on two secret boolean values 
        by considering an expression $\termone$ of the form 
        $\textbf{if}\ \varone\ \textbf{then}\ I\ \textbf{else}\ \Omega$, where 
        $\varone$ is a secret boolean variable, $I$ is the identity combinator 
        $\abs{\varone}{\varone}$ and $\Omega$ is the purely divergent expression\footnote{
          As usual, $\Omega$ is defined using recursive types. Notice that we 
          need $\gradealg$ to have a top element for that.
        }
        Then, if $\subst{\termone}{\varone}{\valone}$ converges, then we can infer that 
        the secret value $\valone$ is $\textbf{true}$, and if 
        $\subst{\termone}{\varone}{\valone}$ diverges, then we infer $\valone$ to 
        be $\textbf{false}$. Luckily, the type system prevents these situations. 
        To see why, let us consider an extension of $\Lambda_{\gradealg}$ with finite 
        sum types $\sumtype{i \in I}{\typeone_i}$, where $I$ is a (possibly) empty 
        finite set. The grammar of values is extended with expressions of the form 
        $\inject{\hat{\imath}}{\valone}$, whereas computations are extended with 
        expressions of the form $\casesum{\valone}{\termone_i}$. We also extend the 
        typing rules of Figure~\ref{fig:simply-typed} with the following rules:
        \[
        \infer
        {\envone \valimp \inject{\hat{\imath}}{\valone}: \sumtype{i \in I}{\typeone_i}}
        {\envone \valimp \valone: \typeone_{\hat{\imath}}}
        \qquad
        \infer{\gradeone \mstar \envone \mplus \envtwo \compimp \casesum{\valone}{\termone_i}: \typetwo}
        {\envone \valimp \valone: \sumtype{i \in I}{\typeone_i}
        &
        \envtwo, \graded{\varone}{\typeone_i}{\gradeone} \compimp \termone_i : \typeone 
        & (\forall i \in I)}
        \] 
        We define the type of booleans $\typebool$ as $\typeunit + \typeunit$, where 
        $\typeunit$ is defined as $\typevoid \to \typevoid$, $\typevoid$ being the 
        empty sum type. 

        Let us now come back to our example. First, we notice that we cannot derive 
        a judgment of the form $\graded{\varone}{\typebool}{\secret} \valimp \varone: \typebool$.
        If we want to have a secret variable, we have to make it such using modalities. 
        Therefore, what we can infer is 
        $\graded{\varone}{\typebool}{\secret} \valimp \tbox{\varone}: \bbox_{\secret}\typebool$. 
        Next, we encode the if-then-else constructor as 
        $\casebool{-}{-}{-}$ (we employ some syntactic sugar in place of $\inject{1}{-}$ and 
        $\inject{2}{-}$).
        Putting pieces together, we would like to complete the following derivation:
        \[
        \infer{\graded{\varone}{\typebool}{p \vee \secret} \compimp 
        \letboxy{\tbox{\varone}}{(\casebool{\vartwo}{I}{\Omega})}: \typetwo \to \typetwo}
        {
          \infer{\graded{\varone}{\typebool}{\secret} 
          \valimp \tbox{\varone}: \bbox_{\secret}\typebool}
          {\graded{\varone}{\typebool}{\public} \valimp \varone: \typebool}
          & 
          \deduce[ ]
          {\graded{\vartwo}{\typebool}{p \vee \secret} \compimp \casebool{\vartwo}{I}{\Omega}: 
          \typetwo \to \typetwo}
          {\vdots}
        }
        \]
        in such a way that $\graded{\varone}{\typebool}{p \vee \secret} = \secret$. 
        Since the latter is always the case, what we need is to derive 
        $$
        \graded{\vartwo}{\typebool}{\secret} \compimp \casebool{\vartwo}{I}{\Omega}: 
          \typetwo \to \typetwo
        $$
        which in turn requires us to derive the judgment
        $\graded{\vartwo}{\typebool}{\secret} \valimp \vartwo: \typebool$. 
        That, however, is not possible and thus the type system 
        prevents one to write programs whose (branching) behaviour 
        depends on secret parameters.
        We will say more about that in 
        Section~\ref{section:compositionality-metric-preservation-and-non-interference}.
        \end{example}

        Our last two examples introduce the construction of new modal signatures 
        from old ones. The constructions we analysed are the so-called \emph{product} 
        of modal signatures 
        \cite{Orchard:2019:QPR:3352468.3341714}, \emph{intervals} over a a modal 
        signature 
        \cite{Orchard:2019:QPR:3352468.3341714}, and \emph{endomorphisms} over a modal signature
        \cite{Gavazzo/LICS/2018}.

        \begin{example}[Product]
        Without much of a surprise, the product \cite{Orchard:2019:QPR:3352468.3341714} 
        of two modal 
        signature $\gradealg_1$ and $\gradealg_2$ is defined by taking 
        the product of their carrier with ordering and operations defined 
        pointwise. This construction is useful for combining different 
        modal analysis, such as resource usage and information flow.
        \end{example}

        \begin{example}[Substructural Systems via Intervals] 
        Our last example deals with (commutative\footnote{The analysis of non-commutative 
        systems would require several modifications to our type system. A promising 
        way to develop a uniform account of substructural systems 
        may be the extension of the techniques developed in this work to 
        calculi akin to Licata et al.'s mode theories \cite{DBLP:conf/rta/LicataSR17}. 
        We leave the study of such extensions for future wok.}) 
        substructural type systems modelled using the so-called interval modal
        signatures by Orchard et al. \cite{Orchard:2019:QPR:3352468.3341714}. 
        One possible way to deal with linearity is given by the three-element 
        monad signature $\{0, 1, \infty\}$, with 
        $0$ standing for unused variables, $1$ for linear variables, and 
        $\infty$ for intuitionistic variables. 
        Another option is to use the interval semiring: given a modal signature 
        $(\gradealg, \mleq, \mplus, \mstar, \mzero, \munit)$, we form the 
        modal signature of $\gradealg$-intervals, whose 
        carrier is $\{[\gradeone, \gradetwo] \mid \gradeone \mleq \gradetwo\}$ 
        and whose operations and order are defined pointwise. 
        In particular, if we start withe the modal signature 
        $(\mathbb{N}^\infty, \leq, +, \cdot, 0,1)$, then we can recover 
        \emph{affine}, \emph{relevant}, \emph{linear}, and \emph{intuitionistic} 
        types as indicated in the following table.

        \begin{center}
        \begin{tabular}{cccccc}
        \hline
        $\vspace{-0.2cm}$
        \\
        \textbf{Affine}
        & \textbf{Relevant}
        & \textbf{Linear}
        & \textbf{Intuitionistic}
        &\textbf{Permissive Affine}
        &\textbf{Restricted Relevant}
        \\
        $\bbox_{[0,1]} \typeone$
        &
        $\bbox_{[1,\infty]} \typeone$
        &
        $\bbox_{[1,1]} \typeone$
        &
        $\bbox_{[0,\infty]} \typeone$
        &
        $\bbox_{[0,n]} \typeone$
        &
        $\bbox_{[1,n]} \typeone$
        \\
        $\vspace{-0.2cm}$
        \\
        \hline
        \end{tabular}
        \end{center}
        For instance, an expression of a relevant type $\bbox_{[1,\infty]} \typeone$
        is a piece of code that can be used $n \in [1, \infty]$ times, 
        and thus \emph{at least} once. Using intervals, we can also define 
        new substructural modalities, giving, e.g., \emph{permissive affinity} 
        (whereby an expression of type $\bbox_{[0,n]} \typeone$ 
        can be used at most $n$ times) and \emph{restricted relevance}
        (whereby an expression of type $\bbox_{[1,n]} \typeone$ 
        must be used at least once, but at most $n$ times).
        \end{example}

        \begin{example}[Semiring Endomorphisms]
        \label{example:semiring-endomorphism}
        Our last example has been studied (in a more general setting, though) 
        in the context of abstract program distances \cite{Gavazzo/LICS/2018}. 
        Any modal signature $\gradealg$ induces 
        a new modal signature (denoted by $\mathsf{End}(\gradealg)$) 
        given by endomorphisms on $\gradealg$ 
        with all operations defined pointwise with the exception of 
        unit and multiplication which are instead defined as the identity function
        and function composition, respectively. 
        That is, elements of $\mathsf{End}(\gradealg)$ 
        are monotone semiring homomorphisms $h: \gradealg \to \gradealg$ 
        ordered pointwise. The addition and zero element of the semiring 
        are defined pointwise (so that the 
        zero element of $\mathsf{End}(\gradealg)$ is given by $\mzero$-constant 
        function). The semiring multiplication, instead, 
        is defined as function composition, meaning that 
        $h \mstar k$ is defined as $h \comp k$). As a consequence, its neutral element 
        is given by the identity function. 
        Actually, one can even take any sub-semiring of $\mathsf{End}(\gradealg)$, i.e. 
        any subset of $\mathsf{End}(\gradealg)$ closed 
        by addition and composition, and
        containing the always $\mzero$ and identity functions 
        (as well as the always $\mtop$ function, if one requires $\gradealg$ 
        to come with a top element).
        \end{example}
}

Last but not least, we define the dynamic semantics of 
$\Lambda_{\gradealg}$ as an ordinary operational semantics 
for a call-by-value $\lambda$-calculus with recursive types extended 
with the reduction rule $\letbox{\tbox \valone}{\termone} \mapsto 
\subst{\termone}{\varone}{\valone}$. 
Notice that modal types essentially play no role here. 
We write $\termone \Downarrow \valone$ if $\termone$ converges. 
The relation $\Downarrow$ being deterministic, we obtain 
a (type-indexed family of) map(s) $\sem{-}: \Lambda_{\typeone} \to (\values_{\typeone})_{\bot}$
defined by $\sem{\termone} \defeq \valone$ if $\termone \Downarrow \valone$, and 
$\bot$ otherwise.

\section{Relational Reasoning}
\label{section:relational-reasoning}

{
\newcommand{\rone}{\wone}
\newcommand{\rtwo}{\wtwo}
\newcommand{\rthree}{\wthree}
\newcommand{\rfour}{\wfour}
\newcommand{\rfive}{\wfive}
\newcommand{\rzero}{\wunit}
\newcommand{\runit}{\blue{1}}
\newcommand{\rleq}{\wleq}
\newcommand{\rgeq}{\wgeq}

\newcommand{\rstar}{\TODO{?}}
\newcommand{\rplus}{\wcomp}

In the previous section, we introduced the syntax and semantics of $\Lambda_{\gradealg}$, 
a call-by-value $\lambda$-calculus 
parametric with respect to grade algebras. Now we want to 
define notions of program equivalence for such a calculus. 
As a first observation, we notice that according to our operational semantics,
modal types \emph{do not} influence the operational behaviour of programs, as the
presence of modalities does not affect 
program execution. 
According to our information reading, what modalities do is to act on 
the \emph{external observer} by modifying the way she can use and test (and thus 
ultimately \emph{discriminate between}) programs.
We formalise this intuition by defining notions of program equivalence not 
as relations, but as relations indexed by possible worlds (or, equivalently, 
as \emph{ternary} relations), the latter capturing 
suitable observer's features.

Formally, instead of associating to any type $\typeone$ a relation
between terms of type $\typeone$, we associate to it a function 
$\sim_{\typeone}: \worlds \to \powerset(\Lambda_{\typeone} \times \Lambda_{\typeone})$ 
from possible worlds to relations mapping each world $\wone$ to a relation 
$\sim_{\typeone}^{\wone}$ meant to capture operational indistinguishability 
for programs of type $\typeone$ at world $\wone$.   
We thus use possible worlds to represent possible states in which 
an external observer compares programs.

\begin{definition}
A \emph{monoidal Kripke frame} (MKF, for short) is a 
  \emph{symmetric monoidal preorder}
$\Worlds = (\worlds, \wleq, \wcomp, \wunit)$ such that $\wunit$
is the bottom element. 
A \emph{$\Worlds$-relation} is a monotone map 
$\relone: (\worlds, \wleq) \to (\powerset(X \times Y), \subseteq)$. 
\end{definition}

We call elements $\wone, \wtwo, \hh$ of a MKF possible worlds 
and refer to generic $\Worlds$-relations as monoidal Kripke relations.
Fixed a MKP $\Worlds$, 
a $\Worlds$-relation is thus a
monotone map $\relone: \worlds \to \powerset(X \times Y)$ --- 
equivalently described as a monotone ternary relation $\relone \subseteq 
X \times Y \times \worlds$ ---
in the sense that
$\wone \wleq \wtwo$ implies $\relone(\wone) \subseteq \relone(\wtwo)$.
Following the resource-semantics reading \cite{DBLP:journals/mst/Lambek68,ROUTLEY1973199,DBLP:journals/jsyml/Urquhart72,DBLP:books/daglib/0006566}, 
we think about possible worlds as consumable resources, 
this way regarding program comparison as a resource consuming process. 
$\Worlds$-relations share most of the algebra of the usual 
(binary) relations, so that we can transfer several notions used 
in relational reasoning (such as the notion of an equivalence or 
of a preorder) to the realm $\Worlds$-relations. We now recall 
the basic algebraic construction needed in this work.

\longv{
    Requiring $\wunit$ to be the bottom element 
    reflects our reading of worlds 
    as resources. 
    In fact, since $\wunit \wleq \wtwo$, for any world $\wtwo$, 
    monotonicity of $\wcomp$ gives $\wone = \wone \wcomp \wunit \wleq \wone \wcomp \wtwo$, 
    meaning that we do not loose resource by joining resources together. 
}

We denote by $\wrelcat(X,Y)$ the collection of 
$\Worlds$-relations over sets
$X$ and $Y$, and 
write $\relone: X \torel Y$ in place of $\relone \in \wrelcat(X,Y)$, 
provided that $\Worlds$ is clear from the context. 
Given a $\Worlds$-relation $\relone$, 
we use the notations 
$\wrel{\relone}{x}{y}{\wone}$, $\relone(x,y,\wone)$, 
and $(x,y,\wone) \in \relone$ interchangeably.
\longv{
  Notice that since both $\worlds$ and $\rel(X,Y)$ are ordered sets, then 
  $\wrelcat(X,Y)$ is nothing but the hom-set 
  $\preord(\worlds, \rel(X,Y))$, where $\preord$ is the 
  category of (pre)ordered sets and monotone maps.
}
$\Worlds$-relations form a category, denoted by $\wrelcat$,
 whose objects are sets and whose arrows 
are $\Worlds$-relations. For any set $X$, 
the identity $\Worlds$-relation $\idrel$ over $X$ is
defined as
$\idrel = \{(x,x,\wone) \mid x \in X, \wone \in \worlds\}$, 
whereas
the composition $\relone;\reltwo$ of   
$\Worlds$-relations $\relone, \reltwo$ (of the appropriate type), 
is defined as\footnote{In this work, we use the symbol 
$\equiv$ for the ``if and only if'' connective.}
$$
\wrel{(\relcomp{\relone}{\reltwo})}{x}{z}{\wone} 
\equiv 
\exists y,{\wtwo},{\wthree}.\
\wone \wgeq \wtwo \wcomp \wthree\; \&\;
\wrel{\relone}{x}{y}{\wtwo}\; \&\; \wrel{\reltwo}{y}{z}{\wthree}.
$$
Notice that $\idrel$ and $\relcomp{\relone}{\reltwo}$ are 
monotone
(provided that $\relone$ and $\reltwo$ are), and thus indeed $\Worlds$-relations.
We define the converse of a 
$\Worlds$-relation $\relone: X \torel Y$ as 
the $\Worlds$-relation $\dual{\relone}: Y \torel X$ 
defined by $\wrel{\dual{\relone}}{y}{x}{\wone} \equiv 
\wrel{\relone}{x}{y}{\wone}$.
As a consequence, we say that a $\Worlds$-relation $\relone \in \wrelcat(X,X)$ is 
reflexive if $\idrel \subseteq \relone$, symmetric if $\dual{\relone} \subseteq \relone$,
and transitive if $\relcomp{\relone}{\relone} \subseteq {\relone}$. 
Altogether, we obtain the notions of a $\Worlds$-preorder and 
$\Worlds$-equivalence.
Additionally, each set $\wrelcat(X,Y)$ forms a complete lattice when endowed 
with (the pointwise extension of) subset inclusion, so that we can 
define $\Worlds$-relations both inductively and coinductively.
Finally, we observe that any function $f: X \to Y$ can 
be regarded as an arrow in $\wrelcat$ via the 
$\Worlds$-relation $\{(x,f(x),\rone) \mid x \in X, \wone \in \worlds\}$. 
For simplicity, we use the notation $f: X \to Y$ 
even when regarding $f$ as an arrow in $\wrelcat$.

An important construction on $\Worlds$-relations we will extensively use
in this work, is the tensor product of $\Worlds$-relations: given $\Worlds$-relations
$\relone: X \torel Y$, $\reltwo: X' \torel Y'$, 
define $\relone \otimes \reltwo: X \times X' \torel Y \times Y'$ 
by $\wrel{(\relone \otimes \reltwo)}{(x,x')}{(y,y')}{\wone}$ iff 
there exist $\wtwo,\wthree$ such that $\wone \wgeq \wtwo \wcomp \wthree$ 
and $\wrel{\relone}{x}{y}{\wtwo}$, $\wrel{\reltwo}{x'}{y'}{\wthree}$. 
Notice that $\relone \otimes \reltwo$ is the monoidal counterpart 
of the usual (cartesian) product $\relone \times \reltwo$, 
defined as $\wrel{(\relone \times \reltwo)}{(x,x')}{(y,y')}{\wone}$ iff 
$\wrel{\relone}{x}{y}{\wone}$ and $\wrel{\reltwo}{x'}{y'}{\wone}$.

Now that we have introduced some basic notions on $\Worlds$-relations, 
we give some examples of MKFs and their corresponding $\Worlds$-relations 
one finds in the literature on programming language semantics.

\longv{
          It is useful to keep in mind the pointwise reading of $\quantale$-relations  
          of the form $\dual{g} \comp \reltwo \comp f$,
          for a relation $\reltwo: Z \torel W$ and functions 
          $f: X \to Z$, $g: Y \to W$:
          $$
          (\dual{g} \comp \reltwo \comp f)(x,y) = 
          \reltwo(f(x),g(y)).
          $$
          Given $\relone: X \torel Y$ we can thus express a generalised  
          monotonicity condition in pointfree fashion as:
          $$
          \relone \subseteq \dual{g} \comp \reltwo \comp f.
          $$
          Indeed, taking $f = g$, we obtain standard monotonicity of $f$. 
          We will make extensively use of the following \emph{adjunction rules} 
          (which can be derived by their metric-based counterparts 
          we will see in Section~\ref{section:kripke-meets-lawvere}), 
          for $f: X \to Y$, $g: Y \to Z$, $\relone: X \torel Y$, 
          $\reltwo: Y \torel Z$, and $\relthree: X \torel Z$:
          \begin{align*}
            g \comp \relone \subseteq \relthree 
            &\iff \relone \subseteq \dual{g} \comp \relthree
            \label{adj-1}\tag{adj 1}
            \\
            \relthree \comp \dual{f} \subseteq \reltwo 
            &\iff
            \relthree \subseteq \reltwo \comp f.
            \label{adj-2}\tag{adj 2}
          \end{align*}
          Using \eqref{adj-1} and \eqref{adj-2} we see that generalised monotonicity 
          $\relone \subseteq \dual{g} \comp \reltwo \comp f$ can be equivalently 
          expressed via the following lax commutative diagram: 
          \begin{center}
          \(
          \vcenter{
          \xymatrix{
          \laxcommuterel
          X     
          \ar[r]^{f} 
          \ar[d]_{\relone}|@{|}  &  
          \setthree   
          \ar[d]^{\reltwo}|-*=0@{|}
          \\
          Y   
          \ar[r]_{g}   &  
          \setfour  } }
          \)
           \end{center}
           The diagram acts as a graphical representation of the expression 
           $g \comp \relone \subseteq \reltwo \comp f$, which, by \eqref{adj-1},
           is equivalent to $\relone \subseteq \dual{g} \comp \relone \comp f$.

          }

          \longv{
                $\Worlds$-relations form another useful category, namely the 
                category of binary  $\worlds$-(endo)relations,
                which we denote by $\brel_{\worlds}$. Objects of $\brel_{\worlds}$
                are pairs 
                $(X, \relone)$ consisting of a set $X$ and a
                \emph{reflexive} and \emph{transitive} $\Worlds$-relation
                over it. Arrows 
                $(f,g): (X, \relone) \to 
                (Y, \reltwo)$
                are pairs of maps $f,g: X \to Y$ satisfying the following 
                law:
                $$
                \wrel{\relone}{x}{x'}{\wone}
                \implies \wrel{\reltwo}{f(x)}{g(x')}{\wone}.
                $$
                The latter is noting but the inequality 
                $\relone \subseteq \dual{g} \comp \relone \comp f$ 
                that we have just seen. 

                The monoidal structure of $\worlds$ extends to $\brel_{\worlds}$, this way endowing 
                the latter with the structure of  
                a (symmetric) monoidal closed category. 
                \begin{proposition}
                \label{prop:brel-is-monoidal-closed}
                $\brel_{\worlds}$ is a (symmetric)
                monoidal closed category.
                \end{proposition}

                \begin{proof}
                The tensor product $(X, \relone) \tensor (Y, \reltwo)$ of two 
                objects $(X, \relone)$ and
                $(Y, \reltwo)$ in $\brel_{\worlds}$ is the triple 
                $(X \times Y, \relone \tensor \reltwo)$ where
                \begin{align*}
                \wrel{(\relone \tensor \reltwo)}{(x,y)}{(x',y')}{\wone} 
                &\iff \exists \wtwo, \wthree \in \worlds.\ 
                \rone \geq \rtwo \rplus \rthree \text{ and }
                \wrel{\relone}{x}{x'}{\rtwo} \text{ and }\wrel{\reltwo}{y}{y'}{\rthree}
                \end{align*}
                Similarly, we define $(X, \relone) \Rightarrow (Y, \reltwo)$ 
                as the triple 
                $(Y^X, \relone \Rightarrow \reltwo)$
                where
                $$
                \wrel{(\relone \Rightarrow \reltwo)}{f}{g}{\rone} 
                \iff \forall \rtwo, \rthree \in \worlds. \forall x, x'.\
                \text{ if } \rthree \geq \rone \rplus \rtwo \text{ then }
                \wrel{\relone}{x}{x'}{\rtwo} \text{ implies } 
                \wrel{\reltwo}{f(x)}{g(x')}{\rthree}.
                $$
                Straightforward calculation show that  
                $\relone \tensor \reltwo$ and $\relone \Rightarrow \reltwo$ 
                are monotone, provided $\relone$ and $\reltwo$ are, and 
                that the defining axioms of a symmetric monoidal closed category 
                holds. 
                \end{proof}


                Actually, $\brel_{\worlds}$ also enjoys several properties of cartesian categories.
                For instance, we define the cartesian product $(X, \relone) \times 
                (Y, \reltwo)$ of two objects 
                $(X, \relone)$ and $(Y, \reltwo)$ as 
                $(X \times Y, \relone \mathbin{\&} \reltwo)$ where
                $$
                \wrel{(\relone \mathbin{\&} \reltwo)}{(x,y)}{(x',y')}{\rone}
                \iff 
                \wrel{\relone}{x}{x'}{\rone}
                \text{ and } \wrel{\reltwo}{y}{y'}{\rone}.
                $$
                Notice that contrary to the $\Worlds$-relation $\relone \tensor \reltwo$, 
                the $\Worlds$-relation 
                $\relone \mathbin{\&} \reltwo$ uses the \emph{same} resource/possible world to 
                relate different elements, this way internalising 
                duplicability of resources. Notice also how the rich structure 
                of $\brel_{\worlds}$ allows one  
                define suitable intensional logical relations akin to Kripke logical relations 
                \cite{abel}.\footnote{Although we focus 
                on bisimulation-based equivalence, it is a straightforward exercise 
                to define other equivalences such as logical relations, or CIU-like 
                equivalences.}

          \begin{remark}
          $\Worlds$-relations form another useful category,
                denoted by $\brel_{\worlds}$, whose objects 
                are pairs 
                $(X, \relone)$ consisting of a set $X$ and a
                \emph{reflexive} and \emph{transitive} $\Worlds$-relation
                over it, and whose arrows 
                $(f,g): (X, \relone) \to 
                (Y, \reltwo)$
                are pairs of functions $f,g: X \to Y$ such that
                $\wrel{\relone}{x}{x'}{\wone}$
                implies $\wrel{\reltwo}{f(x)}{g(x')}{\wone}$.
                The monoidal structure of $\worlds$ extends to $\brel_{\worlds}$, this way endowing 
                the latter with the structure of  
                a (symmetric) monoidal closed category, which also has 
                cartesian products.
                For this reason, one can use
                $\brel_{\worlds}$ to define what we may call Kripke monoidal logical relations
                \cite{DBLP:journals/pacmpl/AbelB20}. 
            \end{remark}
}
\begin{example}
\label{ex:world-relations}
\begin{varenumerate}
  \item For the grade algebra 
    $([0,\infty], \leq, +,\cdot, 0,1)$, let $\Worlds$ be $([0,\infty], \leq, +, 0)$. 
    A $\Worlds$-relation $\relone$ relate terms 
    with respect to non-negative extended real numbers with the intended meaning that if
    $\wrel{\relone}{\termone}{\termtwo}{a}$ holds, then the 
    $\relone$-distance between $\termone$ and $\termtwo$ is bounded by $a$. 
    Obviously, if $a \leq b$ and $\wrel{\relone}{\termone}{\termtwo}{a}$, 
    we also have $\wrel{\relone}{\termone}{\termtwo}{b}$ 
    (if $\termone$ and $\termtwo$ are at most $a$ far, then they also are 
    at most $b$ far).
    Reed and Pierce \cite{Pierce/DistanceMakesTypesGrowStronger/2010} 
    use $\Worlds$-relations to define a logical relation 
    $\sim_{\typeone}: [0,\infty] \to \Lambda_{\typeone} \times \Lambda_{\typeone}$ 
    to characterise program distance. In fact,
    it is easy to see that $\Worlds$-relations over sets $X$ and $Y$ 
    correspond to metric-like functions $
    X \times Y \to [0,\infty]$ through the maps 
    $\Phi: \wrelcat(X,Y) \to [0,\infty]^{X \times Y}$ and 
    $\Psi: [0,\infty]^{X \times Y} \to \wrelcat(X,Y)$ thus defined:
    \begin{align*}
    \Phi(\relone)(x,y) &= 
    \inf \{a \mid 
    \wrel{\relone}{\termone}{\termtwo}{a}\}
    \\
    \Psi(\delta)(a) &= \{(x,y) \mid \delta(x,y) \leq a\}.
    \end{align*}
  \item Consider the grade algebra $\mathcal{L}$ of security levels, 
  and let $\Worlds$ be $(\mathcal{L}, \geq, \wedge, \bot)$. 
    A $\Worlds$-relation relates terms 
    at specific confidentiality levels, and we read 
    $\wrel{\relone}{\termone}{\termtwo}{\ell}$ as stating that 
    a user with confidentiality level $\ell$ regards
    the terms $\termone$ and $\termtwo$ as $\relone$-related. 
    In particular, for an equivalence $\Worlds$-relation $\sim_{\typeone}$, 
    we read $\termone \sim^{\ell}_{\typeone} \termtwo$ as stating that 
    $\termone$ and $\termtwo$ are indistinguishable for a user with 
    permission level $\ell$.
  \item Given a MKF $\Worlds$, the set 
    $\mathsf{End}(\worldset)$ of $\Worlds$-endomorphisms 
    forms a grade algebra with semiring multiplication 
    given by function composition, unit element given by 
    the identity function, and all other operations defined pointwise. 
    In particular, if $\Worlds$ is 
    $([0,\infty], \leq, +, 0)$, then 
    elements in $\mathsf{End}(\worldset)$ 
    give the so-called \emph{$f$-sensitivities} 
    \cite{DBLP:journals/pacmpl/BartheEGHS18}. 
    \end{varenumerate}
    \end{example}

}

\section{Relational Extensions: Monads and Comonads}
\label{section:corelators-and-relators}

As a follow-up to the previous section, 
we define notions of program equivalence 
as suitable $\Worlds$-equivalences. 
In this work, we study a specific notion of equivalence, namely 
\emph{applicative bisimilarity} \cite{Abramsky/RTFP/1990}. 

Applicative bisimilarity is a coinductively-defined notion of equivalence based on the 
function extensionality principle. 
Although other notions of equivalence 
can be defined and studied in our framework, applicative 
bisimilarity is arguably one of the most interesting ones. In fact, 
being coinductively-defined applicative bisimilarity comes with 
powerful proof techniques which make it well-suited for dealing with 
infinitary behaviours (viz. divergence and recursive types); 
moreover, applicative bisimilarity being extensional, 
it is a lightweight\footnote{In the sense that proof obligations 
to prove relations to be bisimulations have a lower logical complexity 
than proof obligations of, e.g., contextual equivalence or logical relations.}
(and thus practically usable) notion of equivalence. Finally, contrary to other notions of equivalence,
proving applicative bisimilarity to be a congruence is highly non-trivial 
and requires a sophisticated relational technique known as Howe's method 
\cite{Howe/IC/1996,Pitts/ATBC/2011}: we can thus view our general congruence 
theorem for applicative bisimilarity as witnessing the strength and robustness of 
our framework.

Our starting point is the work by Dal Lago, Gavazzo, and collaborators 
\cite{DalLagoGavazzoLevy/LICS/2017,dal-lago/gavazzo-mfps-2019,DBLP:conf/esop/LagoG19,Gavazzo/LICS/2018,DBLP:phd/basesearch/Gavazzo19},
who defined several coinductively-defined notions of equivalence and 
prove general congruence theorems for them relying on the 
notion of a 
\emph{lax extension of a monad}
\cite{Barr/LMM/1970,Hoffman-Seal-Tholem/monoidal-topology/2014,Thijs/PhDThesis/1996}.\footnote{Lax extensions have also been used to prove congruence properties of
logic-based equivalences \cite{Simpson-Niels/Modalities/2018,DBLP:conf/fossacs/MatacheS19}.}
A lax extension of a functor $F$ on the category $\set$ of sets and functions 
is a way to extend $F$ to a \emph{lax functor} $\hat{F}$ on the category 
of sets and \emph{relations}, and thus provides an axiomatisation of the 
notion of a (lax) relational extension.\footnote{Similar notions have been used 
\cite{GoubaultLasotaNowak/MSCS/2008}
in the context of relational semantics \emph{\`a la} Reynolds \cite{Reynolds/Logical-relations/1983}.}

Lax extensions as they are, however, \emph{do not fit} our framework, which is based on 
$\Worlds$-relations rather than on relations. So the first thing we should do 
is to define what a lax extension in our setting is. 
The main feature of $\Lambda_{\gradealg}$ is to be a modal (and coeffectful) 
calculus. To handle modal (and coeffectful) behaviours, 
we introduce the novel notion of a lax extension of a graded (monoidal) comonad, 
which is to comonads what a lax extension of a monad is to a monad. 
The fact that lax extensions of a comonad work 
on $\Worlds$-relations (and not just on relations) is of paramount importance: in fact, 
equivalence at modal types is actually defined by means of 
lax extension of the \emph{identity comonad}. Such lax extensions, 
which we dub \emph{comonadic lax extensions}, essentially 
leave programs unchanged but modify possible worlds, this 
way reflecting
our intuition that modalities do not act on the computational behaviour 
of programs (hence the identity comonad), but on the external observer.

\longv{ 
      \begin{definition}
      \label{def:relator}
      Let $F$ be an endofunctor on $\set$. A \emph{lax extension} 
      $\relatorsymbol$ of $F$ associates to any $\Worlds$-relation 
      $\relone: X \torel Y$ a $\Worlds$-relation 
      $\relatorsymbol{\relone}: F(X) \torel F(Y)$ in such a way that the 
      following hold:
      \begin{align}
      \idrel &\subseteq \relatorsymbol(\idrel)
      \tag{lax functoriality 1}\label{eq:functor-1}
      \\
      \relatorsymbol(\relone); \relatorsymbol(\reltwo) &\subseteq 
      \relatorsymbol(\relone; \reltwo)
      \tag{lax functorialty 2}\label{eq:functor-2}
      \\
      F(f) &\subseteq \relatorsymbol(f)
      \tag{stability 1}\label{eq:stability-1}
      \\
      \dual{F(f)} &\subseteq \relatorsymbol(\dual{f})
      \tag{stability 2}\label{eq:stability-2}
      \\
      \relone \subseteq \reltwo &\implies 
      \relatorsymbol(\relone) \subseteq \relatorsymbol(\reltwo)
      \tag{monotonicity}\label{eq:monotonicity}
      \end{align}
      \end{definition}
      Definition~\ref{def:relator} is nothing but the extension of the usual definition 
      of a lax extension \cite{Barr/LMM/1970} from ordinary relations 
      to $\Worlds$-relations, and it essentially states that 
      the mapping $X \mapsto F(X)$, $\relone \mapsto 
      \relatorsymbol(\relone)$  
        is a \emph{lax functor}\footnote{Recall that a mapping $F = 
          (X \mapsto F(X) ,\relone \to \relatorsymbol(\relone))$ on $\wrelcat$ 
          is a functor is it satisfies the 
          equalities $\relatorsymbol(\idrel) = \idrel$ and 
          $\relatorsymbol(\reltwo \comp \relone) = \relatorsymbol(\reltwo) \comp 
          \relatorsymbol(\relone)$. 
          We say that $F$ is a lax functor (on 
          $\wrelcat$) if it satisfies the inequalities 
          $\idrel \subseteq \relatorsymbol(\idrel)$ and 
          $\relatorsymbol(\reltwo) \comp \relatorsymbol(\relone) \subseteq 
        \relatorsymbol(\reltwo \comp\relone)$.} on $\wrelcat$.
}

\shortv{
    \begin{definition}
    \label{def:relator}
    Let $F$ be a functor on $\set$. A \emph{lax extension} of $F$
    is (family of) map(s) 
    $\relatorsymbol: \wrelcat(X,Y) \to \wrelcat(F(X), F(Y))$
    satisfying the 
    following laws:
    \begin{align*}
    \idrel &\subseteq \relatorsymbol(\idrel)
    &
    \relatorsymbol(\relone); \relatorsymbol(\reltwo) 
    &\subseteq 
    \relatorsymbol(\relone; \reltwo)
    &
    F(f) &\subseteq \relatorsymbol(f)
    \end{align*}
    \vspace{-0.8cm}
    \begin{align*}
    \relone \subseteq \reltwo &\implies \relatorsymbol(\relone) 
    \subseteq \relatorsymbol(\reltwo)
    &
    \dual{F(f)} &\subseteq \relatorsymbol(\dual{f})
    \end{align*}
    \end{definition}
    Definition~\ref{def:relator} is nothing but the usual definition 
      of a lax extension of a functor \cite{Barr/LMM/1970} properly 
      modified to the setting of 
      $\Worlds$-relations: it states that 
      the mapping $X \mapsto F(X)$, $\relone \mapsto 
      \relatorsymbol(\relone)$  
      is a \emph{lax $2$-functor} on $\wrelcat$ that 
      agrees with $F$ on functions. 
}
\longv{
        \begin{lemma}[Stability]
        \label{lemma:stability}
        Lax commutative diagrams in $\wrelcat$ are preserved by the mapping 
        $X \mapsto F(X), \relone \mapsto \relatorsymbol(\relone)$.
        \end{lemma}

        \begin{proof}
        Let us consider a lax commutative diagram in $\wrelcat$ 
        expressed in linear notation: $g \comp \relone \subseteq \reltwo \comp f$.  
        Standard algebraic calculation show \cite{Hoffman-Seal-Tholem/monoidal-topology/2014} 
        that the latter is equivalent to 
        $g \comp \relone \comp \dual{f} \subseteq \reltwo$. By monotonicity of 
        $\relatorsymbol$, we thus obtain 
        $\relatorsymbol(g \comp \relone \comp \dual{f}) \subseteq \relatorsymbol(\reltwo)$, 
        and thus 
        $\relatorsymbol(g) \comp \relatorsymbol(\relone) \comp \relatorsymbol(\dual{f}) 
        \subseteq \relatorsymbol(\reltwo)$, by lax functoriality. 
        We now apply stability on $\relatorsymbol(g)$ and $\relatorsymbol(\dual{f})$, 
        this way obtaining (by monotonicity) 
        $F(g) \comp \relatorsymbol(\relone) \comp \dual{F(f)} 
        \subseteq \relatorsymbol(\reltwo)$, which means 
        $F(g) \comp \relatorsymbol(\relone)
        \subseteq \relatorsymbol(\reltwo) \comp F(f)$.
        \end{proof}
}
We now introduce the notion of a comonadic lax extension 
(i.e. of a lax extension of the identity comonad). 
It is straightforward to generalise 
Definition \ref{definition:corelator} to arbitrary comonads.

\begin{definition}
\label{definition:corelator}
Given a grade algebra $(\gradealg, \gleq, \gplus, 
\gstar, \gzero, \gunit, \gtop)$, a \emph{comonadic lax extension} 
$\corelatorsymbol$ 
associates to any $\Worlds$-relation $\relone: X \torel Y$ 
a $\gradealg$-indexed family of $\Worlds$-relations $\corelatorsymbol_{\gradeone}{\relone}: 
X \torel Y$ in such a way that each $\corelatorsymbol_{\gradeone}{\relone}$ 
is a lax extension of the identity functor and that the following hold, 
where $\duplicate: X \to X \times X$ denotes the 
duplication (or contraction) map sending $x$ to $(x,x)$:
\longv{
    \footnote{
      We label laws with deadjectival nouns. Starting with the adjective objective 
      (such as \emph{comonadic}, we add the suffix \emph{ity}, 
      this way obtaining a noun.
      }
}
\begin{align}
\corelator{\munit}{\relone} &\subseteq {\relone}
\tag{$\mathit{Com}_1$}{\label{eq:comonad-1}}
\\
\corelator{\gradeone \mstar \gradetwo} {\relone} 
&\subseteq \corelator{\gradeone}{\corelator{\gradetwo}{\relone}}
\tag{$\mathit{Com}_2$}{\label{eq:comonad-2}}
\\
\corelator{\gradeone}{\relone} \tensor \corelator{\gradeone}{\reltwo} 
&\subseteq \corelator{\gradeone}{\relone \tensor \reltwo}
\tag{$\mathit{Mon}_1$}{\label{eq:monoidal-1}}
\\
\corelator{\gradeone \mplus \gradetwo}{\relone} 
&\subseteq 
\duplicate ; (\corelator{\gradeone}{\relone} \otimes 
\corelator{\gradetwo}{\relone}); \dual{\duplicate}
\tag{$\mathit{Com}_2$}{\label{eq:monoidal-2}}
\\
\gradeone \mleq \gradetwo 
&\implies \corelator{\gradetwo}{\relone} \subseteq \corelator{\gradeone}{\relone}
\tag{$\mathit{Contra}$}{\label{eq:contravariance}}
\end{align}
\end{definition}

Let us comment on Definition~\ref{definition:corelator}. 
Requiring each maps $\corelatorsymbol_{\gradeone}$ to be a lax extension 
for the identity functor amounts to require each mapping 
$X \mapsto X, \relone \mapsto \corelator{\gradeone}{\relone}$ to 
be a lax $2$-functor on $\wrelcat$. 
\longv{
    Algebraically, 
    we are requiring the following laws to hold.
    \begin{align*}
    \idrel &\subseteq \corelator{\gradeone}{\idrel}
    \\
    \corelator{\gradeone}{\reltwo} \comp \corelator{\gradeone}{\relone} &\subseteq 
    \corelator{\gradeone}{\reltwo \comp \relone}
    \\
    f &\subseteq \corelator{\gradeone}{f}
    \\
    \dual{f} &\subseteq \corelator{\gradeone}{\dual{f}}
    \\
    \relone \subseteq \reltwo &\implies 
    \corelator{\gradeone}{\relone} \subseteq \corelator{\gradeone}{\reltwo}
      \end{align*}
}
The real novelty of Definition~\ref{definition:corelator} 
is that it requires the counit and comultipilication 
of the (identity) comonad to be lax natural transformations 
on $\wrelcat$.
Since both counit and comultuplication are the identity function, 
they are not visible in laws \eqref{eq:comonad-1} and \eqref{eq:comonad-1}.
A similar reading explains laws 
\eqref{eq:monoidal-1} and \eqref{eq:monoidal-2}, where one exploits the 
fact that the identity comonad is \emph{monoidal}. All of this, 
is adapted to a graded setting.
\longv{
      Nonetheless, our definition of a co-relator can be generalised to 
      arbitrary (graded) monoidal comonads $(C, \varepsilon, \delta)$. 
      For we simply consider  
      maps $\corelatorsymbol_{\gradeone}: 
      \wrelcat(X,Y) \to \wrelcat(C_{\gradeone}(X), C_{\gradeone}(Y))$ 
      and modify the conditions in Definition~\ref{definition:corelator} 
      by taking into account the maps $\varepsilon, \delta$ (as well as the other monoidal arrows). 
      For instance, lax comonadicity would be given as the diagrams:
      \[
        \vcenter{\diagramrel{\varepsilon}{\varepsilon}{C_{\munit}(X)}{C_{\munit}(Y)}
        {\corelator{\munit}{\relone}}{X}{Y}{\relone}}
        \qquad
        \vcenter{
        \diagramrel{\delta}{\delta}{C_{\gradeone \mstar \gradetwo}(X)}
        {C_{\gradeone \mstar \gradetwo}(Y)}{\corelator{\gradeone \mstar \gradetwo}{\relone}}
        {C_{\gradeone}(C_{\gradetwo}(X))}{C_{\gradeone}(C_{\gradetwo}(Y))}
        {\corelator{\gradeone}{\corelator{\gradetwo}{\relone}}}}
      \]
      This way, the mapping
      $X \mapsto C_{\gradeone}(X)$, $\relone \mapsto \corelator{\gradeone}{\relone}$ gives a 
      (lax) graded monoidal comonad on $\wrelcat$.
      Once we restrict to identity-on-object comonads, the above diagrams 
      collapse to simple relational inclusions (for such comonads, both $\varepsilon$ and 
      $\delta$ are given by the identity function), and gives the 
      lax comonadicity laws of Definition~\ref{definition:corelator}. 
}
      The only rule in Definition~\ref{definition:corelator} 
      that involves the presence of functions (although regarded as 
      $\Worlds$-relations) is 
      rule \eqref{eq:monoidal-1}.
      Diagramtically, we express this rule as follows.
      \[
      \diagramrel{\duplicate}{\duplicate}{X}{Y}{\corelator{\gradeone \mplus \gradetwo}\relone}
      {X \times X}{Y \times Y}{\corelator{\gradeone}{\relone} \tensor \corelator{\gradetwo}{\relone}}
      \]
Finally, law \eqref{eq:contravariance} states that what we have is actually a
$\opposite{\gradealg}$-graded monoidal comonad. 
Operationally, we can read antitonicity as stating that if two expressions 
are equivalent when used according to $\gradeone$, then they are also equivalent when 
used ``less'' than $\gradeone$ (e.g. if two expressions are equivalent when used 
an arbitrary number of times, then they must be so when used at most $n$ times).
Notice that \eqref{eq:contravariance} and \eqref{eq:comonad-1} imply 
$\corelator{\gradeone}{\relone} \subseteq \relone$, for any $\gradeone \mgeq \munit$.
\longv{
    Before giving examples of comonadic lax extensions, 
    we recall the notion of a lax extension of a \emph{monad} 
    \cite{Barr/LMM/1970,Hoffman-Seal-Tholem/monoidal-topology/2014,Hoffman/Cottage-industry/2015}. 
    Although $\Lambda_{\gradealg}$ is a \emph{pure} calculus, 
    we handled divergence by giving it a monadic (operational) semantics 
    based on the partiality monad. We thus follow
    Dal Lago et al. \cite{DalLagoGavazzoLevy/LICS/2017} 
    and rely on lax extensions of monads to define applicative bisimilarity.

    \begin{definition}
    \label{def: relator-monad}
    A lax extension of a monad $\Monad = (\monad, \unit, \mu)$ is a 
    lax extension of $\monad$ satisfying the following laws:
    \begin{align*}
    \relone \subseteq \dual{\unit}; \relatorsymbol(\relone); \unit
    \label{eq:lax-monad-unit}\tag{lax monad 1}
    \\
    \relatorsymbol(\relatorsymbol(\relone)) 
    \subseteq \dual{\mu}; \relatorsymbol(\relone); \mu
    \label{eq:lax-monad-mu}\tag{lax monad 2}
    \end{align*}
    \end{definition}

    As before, we can conveniently express laws \eqref{eq:lax-monad-unit} 
    and \eqref{eq:lax-monad-mu} as diagrams:
    \[
      \vcenter{
      \diagramrel
      {\unit}
      {\unit}
      {X}
      {Y}
      {\relone}
      {\monad(X)}
      {\monad(Y)}
      {\relatorsymbol(\relone)}
      }
      ;
      \qquad
      \vcenter{
      \diagramrel
      {\mu}
      {\mu}
      {\monad(\monad(X))}
      {\monad(\monad(Y))}
      {\relatorsymbol(\relatorsymbol(\relone))}
      {\monad(X)}
      {\monad(Y)}
      {\relatorsymbol(\relone)}
      }
    \]

    Without much of a surprise, a lax extension $\relatorsymbol$ of $\Monad$ 
    induces a mapping 
    $(X \mapsto \monad(X), \relone \to \relatorsymbol(\relone))$
    which gives a lax monad on $\wrelcat$.

    Since $\Lambda_{\gradealg}$ does not have monadic effect other than divergence, 
    we do not give examples of relators for concrete monads except than for the 
    maybe/partiality monad (but see Section~\ref{section:kripke-meets-lawvere}). 

    \begin{proposition}
    \label{proposition:relator-partiality}
    Let $\relone: X \torel Y$ be a $\Worlds$-relation. 
    Define $\relone_{\divergence}: X_{\divergence} \to Y_{\divergence}$
    as follows:
    \begin{align*}
    \wrel{\relone_{\divergence}}
      {x}
      {y}
      {\wone}
      &\iff 
     x \neq \divergence \to (y \neq \divergence \;\&\;
      \wrel{\relone}
      {x}
      {y}
      {\wone}).
    \end{align*}
    Then, $(-)_{\divergence}$ is a lax extension of the maybe/partiality monad.
    \end{proposition}

    Intuitively, 
    $\relone_{\divergence}$ gives a generalisation of the usual clause used 
    to define operational preorders between programs. Accordingly, 
    a term $\termone$ approximates the behaviour of a term 
    $\termtwo$ at world $\wone$ if either $\termone$ diverges or 
    both $\termone$ and $\termtwo$ converge
    and the resulting values are related at $\wone$. 
    If we take the MKF $([0,\infty], \leq, +, 0)$
    and read $\wrel{\relone}{\termone}{\termtwo}{\gradeone}$ as stating that the $\relone$-distance 
    between $\termone$ and $\termtwo$ is at most $\gradeone$, 
    then $\wrel{\relone_{\divergence}}{\termone}{\termtwo}{\gradeone}$ 
    tells us that if $\termone$ diverges, 
    then the $\relone_{\divergence}$-distance between $\termone$ and $\termtwo$ 
    is bounded by any $\gradeone$ --- and thus it is bounded by $0$. 
    Otherwise, $\termone$ converges to value $\valone$, and thus $\termtwo$ converges to
    a value $\valtwo$ such that the $\relone_{\divergence}$-distance between 
    $\termone$ and $\termtwo$ is the $\relone$-distance between 
    $\valone$ and $\valtwo$. 
}
\longv{
      We summarise the defining laws of relators and co-relators in 
      Figure~\ref{fig:axioms-relators-corelators}. The reader may wonder whether and
      how relators and co-relators should interact.  Indeed, in 
      Section~\ref{section:compositionality-metric-preservation-and-non-interference} 
      we will require 
      relators and co-relators to satisfy a suitable lax distributive law.
      For the moment, however, we will focus on co-relators 
      as the latter are at the heart of our semantical analysis of modal necessity 
      types.

      \begin{figure*}[htbp]
      \hrule 
      \begin{center}
      \def\arraystretch{2}
      \begin{tabular}{ccc}
      \textbf{Lax Functoriality} & \textbf{Stability} & \textbf{Monotonicity}
      \\
      {$\!\begin{aligned}
          \idrel &\subseteq \relatorsymbol(\idrel)
          \\
          \relatorsymbol(\reltwo) \comp \relatorsymbol(\relone) &\subseteq 
          \relatorsymbol(\reltwo \comp \relone)
        \end{aligned}$}
      & 
      {$\!\begin{aligned}
          F(f) &\subseteq \relatorsymbol(f)
          \\
          \dual{F(f)} &\subseteq \relatorsymbol(\dual{f})
        \end{aligned}$}
      & 
      {$\!\begin{aligned}
          \relone \subseteq \reltwo &\implies 
          \relatorsymbol(\relone) \subseteq \relatorsymbol(\reltwo)
        \end{aligned}$}
      \\
      \textbf{Lax comonadicity} & \textbf{Lax monoidality} & \textbf{Antitonicity} 
      \\
      {$\!\begin{aligned}
          \corelator{\munit}{\relone} &\subseteq {\relone}
      \\
      \corelator{\gradeone \mstar \gradetwo} {\relone} 
      &\subseteq \corelator{\gradeone}{\corelator{\gradetwo}{\relone}}
        \end{aligned}$}
      & 
      {$\!\begin{aligned}
          \corelator{\gradeone}{\relone} \tensor \corelator{\gradeone}{\reltwo} 
      &\subseteq \corelator{\gradeone}{\relone \tensor \reltwo}
      \\
      \corelator{\gradeone \mplus \gradetwo}{\relone} &\subseteq 
      \duplicate \comp (\corelator{\gradeone}{\relone} \otimes 
      \corelator{\gradetwo}{\relone}) \comp \dual{\duplicate}
        \end{aligned}$}
      & 
      $\gradeone \mleq \gradetwo \implies \corelator{\gradetwo}{\relone} \subseteq \corelator{\gradeone}{\relone}$
      \\
       & \textbf{Monadicity} & 
      \\
      &
      {$\!
      \begin{aligned}
      \relone \subseteq &\hspace{0.1cm} \unit \comp \relatorsymbol(\relone) \comp \dual{\unit}
      \\
      \relone \subseteq g \comp \relatorsymbol(\reltwo) \comp \dual{f} 
      &\implies \relatorsymbol(\relone) \subseteq \kleisli{g} \comp \relatorsymbol(\reltwo) \comp 
      \dual{(\kleisli{f})} 
      \end{aligned}$
      }
      & 
      \end{tabular}
      \end{center}

      \hrule
      \caption{Axioms of Relators and Corelators}
      \label{fig:axioms-relators-corelators}
      \end{figure*}
}

\subsection{Examples}
\label{section:examples-corelators}
{
\newcommand{\rone}{\gradeone}
\newcommand{\rtwo}{\gradetwo}
\newcommand{\rthree}{\gradethree}
\newcommand{\rfour}{\gradefour}
\newcommand{\rfive}{\gradefive}
\newcommand{\rzero}{\gzero}
\newcommand{\runit}{\gunit}
\newcommand{\rleq}{\gleq}
\newcommand{\rgeq}{\ggeq}
\newcommand{\rstar}{\gstar}
\newcommand{\rplus}{\gplus}

In this section, we give some examples of comonadic lax extensions
that apply to the modal calculi seen so far.
We leave for future research further applications of 
lax extensions of comonads. 

Our first example is 
an abstract extension which we call \emph{action extension}. 
In order to define it, we need an action 
$\widehat{-}: \gradealgset \times \worlds \to \worlds$ making 
$\Worlds$ a lax $\gradealg$-module.

\longv{
    \begin{definition}
      A \emph{lax action} is a monotone map 
      $\widehat{-}: \gradealg \times \worldset \to \worldset$ 
      satisfying the following laws, where 
      we write $\mactop{\gradeone}{\wone}$ for 
      the action of $\widehat{-}$ on $(\gradeone, \wone)$:
      \begin{align*} 
        \mactop{\gradeone}{\wunit} &\wleq \wunit 
        &
        \mactop{\gradeone}{\wone \wcomp \wtwo} 
        &\mleq \mactop{\gradeone}{\wone} \wcomp \mactop{\gradeone}{\wtwo}
        &
        \mactop{\gradeone}{\mactop{\gradetwo}{\wone}} 
        &\wleq \mactop{(\gradeone \mstar \gradetwo)}{\wone}
        \end{align*}
        \vspace{-0.7cm}
        \begin{align*}
        \mactop{\munit}{\wone} &\wleq \wone
        &
        \mactop{\gradeone}{\wone} \wcomp  \mactop{\gradetwo}{\wone}
        &\wleq \mactop{(\gradeone \mplus \gradetwo)}{\wone}
      \end{align*}
    \end{definition}

    A standard example of a lax action is obtained by taking the MKF 
    $(\gradealg, \mleq, \mplus, \mzero)$ and defining 
    $\mactop{\gradeone}{\gradetwo}$ as $\gradeone \mstar \gradetwo$. This is precisely the structure 
    one considers when studying program metric
    \cite{Pierce/DistanceMakesTypesGrowStronger/2010,GaboardiEtAl/POPL/2017,DBLP:conf/lics/AmorimGHK19}. 

    Another example, which we will analyse in Section~\ref{section:kripke-meets-lawvere}, 
    is obtained by taking as modal signature the collection of lax quantale 
    endomorphisms \cite{Rosenthal/Quantales/1990} (also known as change of 
    base endofunctors \cite{Hoffman-Seal-Tholem/monoidal-topology/2014,Kelly/EnrichedCats}) 
    with order and addition defined pointwise and multiplication defined by 
    function composition. This is the approach followed in quantale-based 
    distances \cite{Gavazzo/LICS/2018,Lawvere/GeneralizedMetricSpaces/1973}.

        \begin{definition}
        \label{definition:canonical-corelator}
        Let $\widehat{-}$ be a lax action. 
        The \emph{action extension} $\bang$ is thud defined:
        $$
          \wrel{\bang_{\gradeone}\relone}{x}{y}{\wone} 
          \iff
          \exists \wtwo.\ \wone \wgeq \mactop{\rone}{\wtwo} \;\&\;
          \wrel{\relone}{x}{y}{\wtwo}.
          $$
        \end{definition}
        As the notation suggests, the action extension has been 
        extensively used to deal with linear-like calculi, where 
        one takes the MKF $(\gradealg, \mleq, \mplus, \mzero)$ 
        and the lax action is given by grade multiplication. 
        In those cases, Definition~\ref{definition:canonical-corelator} gives:
        $
          \wrel{\bang_{\gradeone}\relone}{x}{y}{\rtwo} 
        $  if and only if there exists $\rthree$ such that  
        $\rtwo \rgeq \rone \rstar \rthree$ and
        $\wrel{\relone}{x}{y}{\rthree}$. 
        Notice that any modal calculus comes with this `canonical' 
        comonadic lax extension, to which we refer to as 
        the canonical extension.

        \begin{example}[Program Sensitivity]
        \label{example:comonadic-lifting-sensitivity}
        When instantiated on the modal signatures $([0,\infty], \leq, +, \cdot, 0,1, \infty)$, 
        the canonical co-relator gives
        $$\wrel{\bang_{\rone}\relone}{x}{y}{\rtwo} 
          \iff
          \exists \rthree.\ \rtwo \geq \rone \cdot \rthree \text{ and } 
          \wrel{\relone}{x}{y}{\rthree}.
        $$
        In particular, if 
        $\wrel{\relone}{x}{y}{\rtwo}$, then 
        $\wrel{\bang_{\rone}\relone}{x}{y}{\rone \cdot \rtwo}$.
        This is essentially the modal type clause used by Reed and Piece 
        to define metric logical relations \cite{Pierce/DistanceMakesTypesGrowStronger/2010}: 
        intuitively, it states that if $x$ and $y$ are at most $\rtwo$ far 
        when measured by $\relone$, 
        then they also are at most $\rone \cdot \rtwo$ far when 
        measured by $\bang_{\rone}\relone$. 
        Applying the correspondence between $[0,\infty]$-relations and 
        distance functions, we see that if 
        $\delta(x,y) \leq \rtwo$, then $\bang_{\rone}\delta(x,y) \leq \rone \cdot 
        \rtwo$, so that $\bang_{\rone}$ is used to amplify distances 
        of a factor $\rone$. In fact, if we look at the category $\brel_{[0,\infty]}$, then we see that $\bang_{\rone}$ 
        is the relational counterpart of the scaling comonad 
        mapping a generalised metric space $(X, \delta)$ to 
        $(X, \rone \cdot \delta)$, where $(\rone \cdot \delta)(x,y) = 
        \rone \cdot \delta(x,y)$. 
        Such a comonad has been used to give denotational 
        semantics to calculi for program sensitivity and differential privacy 
        \cite{GaboardiEtAl/POPL/2017,DBLP:conf/lics/AmorimGHK19}. 
        Notice also that maps $(f,f): (X, \bang_{\rone}\relone) \to (Y, \reltwo)$ 
        are the relational counterpart of a $\rone$-Lipschitz maps between 
        generalised metric spaces.

        \end{example}

        \begin{example}[$f$-Sensitivity]
        \label{example:f-sensitivity}
        Given a MKF $\Worlds$, the set 
        $\mathsf{End}(\worldset)$ of $\worlds$-endomorphisms 
        forms a modal signature with semiring multiplication 
        given by function composition, unit element given by 
        the identity function, and all other operations defined pointwise 
        (see Section~\ref{section:kripke-meets-lawvere}). 
        This way, we obtain an action $\mact: \mathsf{End}(\worldset) \times 
        \worldset \to \worldset$ given by function application: 
        $h \mact \wone = h(\wone)$. In this case, Definition~\ref{definition:canonical-corelator} 
        gives
        $$
          \wrel{\bang_{h}\relone}{x}{y}{\wone} 
          \iff
          \exists \wtwo.\ \wone \wgeq  h(\wtwo) \textnormal{ and } 
          \wrel{\relone}{x}{y}{\wtwo}.
          $$
        In particular, by taking $\worlds$ as 
        $([0,\infty], \leq, +, \cdot, 0, 1, \infty)$ 
        and looking at $\Worlds$-relations as defining distance functions 
        we obtain something close to the so-called \emph{$f$-sensitivity} 
        \cite{DBLP:journals/pacmpl/BartheEGHS18}. 
        Similar instances of the canonical co-relators have been 
        studied in the context of quantale-based program distances 
        \cite{Gavazzo/LICS/2018}. 
        \end{example}

        Easy calculations show that Example~\ref{example:comonadic-lifting-sensitivity} 
        and Example~\ref{example:f-sensitivity}
        can be modified to deal with other quantitative examples of modal calculi 
        (see Section~\ref{section:examples-modal-calculi}). The next result 
        states that the canonical co-relator is indeed a 
        co-relator.

        \begin{proposition}
        The map $\bang$ is a graded co-relator.
        \end{proposition}

        \begin{proof}
        First, notice that if $\relone$ is a $\gradealg$-relation 
        (and thus monotone), then also $\bang_{\rone}\relone$ is.
        Next, we prove the properties required by Definition~\ref{definition:corelator}.
        \begin{description}
          \item[($\idrel \subseteq \bang_{\rone}\idrel$)] 
            Assume $\wrel{\idrel}{x}{x}{\rtwo}$, since 
            $\rtwo \rgeq \rone \rstar \rzero = \rzero$ and $\wrel{\idrel}{x}{x}{\rzero}$, we conclude 
            $\wrel{\bang_{\rone}\idrel}{x}{x}{\rtwo}$. 
          \item[($\bang_{\rone} \reltwo \comp \bang_{\rone}\relone 
              \subseteq \bang_{\rone}(\reltwo \comp \relone)$)] 
            Let $\rfour_0 \rgeq \rtwo \rplus \rthree$ such that
            $\wrel{(\bang_{\rone}\reltwo \comp \bang_{\rone}\relone)}{x}{z}{\rfour}$ 
            with $\wrel{\bang_{\rone}\relone}{x}{y}{\rtwo}$ and 
            $\wrel{\bang_{\rone}\reltwo}{y}{z}{\rthree}$. 
            As a consequence, we have $\rfour, \rfive \in \gradealg$ such that 
            $\rtwo \rgeq \rone \rstar \rfour$ and $\rthree \rgeq \rone \rstar \rfive$. 
            Therefore, $\rfour_0 \geq \rtwo \rplus \rthree \geq \rone \rstar \rfour \rplus 
            \rone \rstar \rfive =  \rone \rstar (\rfour \rplus \rfive)$. 
            We thus conclude $\wrel{\bang_{\rone}(\reltwo \comp \relone)}{x}{z}{\rfour_0}$.
          \item[($f \subseteq \bang_{\rone} f$)] If $\wrel{(f)}{x}{y}{\rtwo}$, then $y = f(x)$.     
            Therefore, we have 
            $\wrel{(\bang_{\rone} f)}{x}{y}{\rtwo}$, since 
            $\rtwo \rgeq \rone \rstar \rzero = \rzero$. 
            The `dual' stability condition is proved similarly.
          \item[($\relone \subseteq \reltwo \implies \bang_{\rone}\relone \subseteq 
          \bang_{\rone} \reltwo$)] Assume $\relone \subseteq \reltwo$ and 
            $\wrel{\bang_{\rone}\relone}{x}{y}{\rtwo}$. 
            Then $\rtwo \rgeq \rone \rstar \rthree$, for some $\rthree$, and 
            $\wrel{\relone}{x}{y}{\rthree}$. As a consequence, we have 
            $\wrel{\reltwo}{x}{y}{\rthree}$ and thus $\wrel{\bang_{\rone} \reltwo}{x}{y}{\rtwo}$.
          \item[($\bang_{\runit} \relone \subseteq \relone$)] 
            Assume $\wrel{\bang_{\runit}\relone}{x}{y}{\rtwo}$. Then, we have 
            $\wrel{\relone}{x}{y}{\rthree}$ for some $\rthree$ such that 
            $\rtwo \rgeq \runit \rstar \rthree = \rthree$. We conclude 
            $\wrel{\relone}{x}{y}{\rtwo}$ by monotonicity of $\rone$. 
          \item[($\bang_{\rone \rstar \rtwo} \relone \subseteq \bang_{\rone} \bang_{\rtwo} 
          \relone$)] 
            Assume $\wrel{\bang_{\rone \rstar \rtwo}\relone}{x}{y}{\rthree}$. 
            Then $\rthree \rgeq \rone \rstar \rtwo \rstar \rfour$ for some 
            $\rfour$ such that $\wrel{\relone}{x}{y}{\rfour}$. As a consequence, 
            we have $\wrel{\bang_{\rtwo}\relone}{x}{y}{\rtwo \rstar \rfour}$, 
            and thus  $\wrel{\bang_{\rone}\bang_{\rtwo}\relone}{x}{y}{\rone \rstar 
            \rtwo \rstar \rfour}$. We conclude 
            $\wrel{\bang_{\rone}\bang_{\rtwo}\relone}{x}{y}{\rthree}$.
          \item[($\bang_{\rone} \relone \tensor \bang_{\rone} \reltwo \subseteq 
              \bang_{\rone}(\relone \tensor \reltwo)$)]
            Assume $\wrel{\bang_{\rone} \relone \tensor \bang_{\rone} \reltwo}
            {(x,y)}{(x',y')}{\rfive_0}$ for some 
            $\rfive_0 \rgeq \rtwo \rplus \rthree$ such that
            $\wrel{\bang_{\rone}\relone}{x}{y}{\rtwo}$ and 
            $\wrel{\bang_{\rone}\reltwo}{x'}{y'}{\rthree}$. 
            Therefore, we have $\rfour, \rfive$ such that 
            $\rtwo \rgeq \rone \rstar \rfour$, $\rthree \rgeq \rone \rstar \rfive$, 
            and $\wrel{\relone}{x}{y}{\rfour}$, $\wrel{\reltwo}{x'}{y'}{\rfive}$. 
            As a consequence, we have $\wrel{(\relone \tensor \reltwo)}{(x,x')}{(y,y')}{\rfour 
            \rplus \rfive}$, and thus 
            $\wrel{\bang_{\rone}(\relone \tensor \reltwo)}{(x,x')}{(y,y')}{\rone \rstar 
            (\rfour \rplus \rfive)}$. We are done since 
            $\rfive_0 
            \rgeq \rtwo \rplus \rthree
            = \rone \rstar \rfour \rplus \rone \rstar \rfive
            = \rone \rstar (\rfour \rplus \rfive)
            $.
          \item[($\bang_{\rone \rplus \rtwo} \relone \subseteq 
          \duplicate \comp (\bang_{\rone} \relone \tensor \bang_{\rtwo} \relone) \comp \dual{\duplicate}$)]
            Suppose $\wrel{\bang_{r+s}\relone}{x}{y}{p}$, so that 
            $p \rgeq (r+s) \rstar q = r \rstar q + s \rstar q$, for some $q$ such that 
            $\wrel{\relone}{x}{y}{q}$. As a consequence, we have 
            $\wrel{\bang_s \relone}{x}{y}{s \rstar q}$ and 
            $\wrel{\bang_r \relone}{x}{y}{r \rstar q}$, and thus 
            $\wrel{(\bang_r \relone \tensor \bang_s \relone)}{(x,x)}{(y,y)}{p}$.
        \item[($\rone \rleq \rtwo \implies \corelator{\rtwo}{\relone} \subseteq \corelator{\rone}{\relone}$)] Suppose $\wrel{\bang_{\rtwo}\relone}{x}{y}{\rthree}$. Then, there exists 
          $\rfour$ such that $\rthree \rgeq \rtwo \rstar \rfour$ and 
          $\wrel{\relone}{x}{y}{\rfour}$. Since $\rone \rleq \rtwo$ implies 
          $\rone \rstar \rfour \rleq \rtwo \rstar \rfour$ and $\rtwo \rstar \rfour \rleq \rthree$, 
          we conclude $\wrel{\bang_{\rone}\relone}{x}{y}{\rthree}$.
        \end{description}
        \end{proof}

        }

\shortv{

\begin{proposition}
\label{prop:bang-is-corelator}
  Recall that a \emph{lax action} is a monotone map 
  $\widehat{-}: \gradealgset \times \worldset \to \worldset$ 
  satisfying the following laws, where 
  we write $\mactop{\gradeone}{\wone}$ for 
  the action of $\widehat{-}$ on $(\gradeone, \wone)$:
  \begin{align*} 
    \mactop{\gradeone}{\wunit} &\wleq \wunit 
    &
    \mactop{\gradeone}{\wone \wcomp \wtwo} 
    &\mleq \mactop{\gradeone}{\wone} \wcomp \mactop{\gradeone}{\wtwo}
    &
    \mactop{\gradeone}{\mactop{\gradetwo}{\wone}} 
    &\wleq \mactop{(\gradeone \mstar \gradetwo)}{\wone}
    \end{align*}
    \vspace{-0.7cm}
    \begin{align*}
    \mactop{\munit}{\wone} &\wleq \wone
    &
    \mactop{\gradeone}{\wone} \wcomp  \mactop{\gradetwo}{\wone}
    &\wleq \mactop{(\gradeone \mplus \gradetwo)}{\wone}
  \end{align*}
Define the \emph{action extension} $\bang$ as follows:
$$
  \wrel{\bang_{\gradeone}\relone}{x}{y}{\wone} 
  \equiv
  \exists \wtwo.\ \wone \wgeq \mactop{\rone}{\wtwo} \;\&\;
  \wrel{\relone}{x}{y}{\wtwo}.
  $$
Then, $\bang$ is a comonadic lax extension.
\end{proposition}

A standard example of a lax action is obtained by taking the MKF 
$(\gradealgset, \mleq, \mplus, \mzero)$ and defining 
$\mactop{\gradeone}{\gradetwo}$ as $\gradeone \mstar \gradetwo$. This is precisely the structure 
one considers when studying program metrics
\cite{Pierce/DistanceMakesTypesGrowStronger/2010,GaboardiEtAl/POPL/2017,DBLP:conf/lics/AmorimGHK19}. 
Moreover, as the notation suggests, these action extensions are
extensively used to deal with linear-like calculi, where one has
$
  \wrel{\bang_{\gradeone}\relone}{x}{y}{\rtwo} 
$  if and only if there exists $\rthree$ such that  
$\rtwo \rgeq \rone \rstar \rthree$ and
$\wrel{\relone}{x}{y}{\rthree}$. 
Notice that any modal calculus comes with this ``canonical'' 
comonadic lax extension, to which we refer to as 
the \emph{canonical extension}.

\begin{example}
\begin{varitemize}
\item When instantiated to the grade algebra $([0,\infty], \leq, +, \cdot, 0,1, \infty)$, 
  the canonical extension gives
  $\wrel{\bang_{\rone}\relone}{x}{y}{\rtwo}$ if and only if
  $\exists \rthree.\ \rtwo \geq \rone \cdot \rthree \text{ and } 
  \wrel{\relone}{x}{y}{\rthree}.
  $
  In particular, if 
  $\wrel{\relone}{x}{y}{\rtwo}$, then 
  $\wrel{\bang_{\rone}\relone}{x}{y}{\rone \cdot \rtwo}$.
  This is essentially the modal type clause used by Reed and Piece 
  to define metric logical relations \cite{Pierce/DistanceMakesTypesGrowStronger/2010}: 
  intuitively, it states that if $x$ and $y$ are at most $\rtwo$ far 
  when measured by $\relone$, 
  then they also are at most $\rone \cdot \rtwo$ far when 
  measured by $\bang_{\rone}\relone$. 
  \item 
  Recall that the set $\mathsf{End}(\worldset)$ of $\Worlds$-endomorphisms 
  on a MKF $\Worlds$  
  forms a grade algebra with semiring multiplication 
  given by function composition. 
  In this case, a lax action is given by function application 
  $\mactop{f}{\wone} = f(\wone)$, so that 
  the canonical extension 
  gives
  $
    \wrel{\bang_{h}\relone}{x}{y}{\wone} 
  $ if and only if 
  $
    \exists \wtwo.\ \wone \wgeq  h(\wtwo) \textnormal{ and } 
    \wrel{\relone}{x}{y}{\wtwo}.
    $
  In particular, by taking $\Worlds$ as 
  $([0,\infty], \leq, +, 0)$ 
  and looking at $\Worlds$-relations as defining distance functions,
  we obtain the so-called \emph{$f$-sensitivity} 
  \cite{DBLP:journals/pacmpl/BartheEGHS18}.  
\end{varitemize}
\end{example}

}

Our second example of a comonadic lax extension 
comes from modal logic, Kripke semantics of intuitionistic logic,  and 
Kripke logical relations \cite{DBLP:books/daglib/0085577}. 

\begin{proposition}
Let $\gradealg$ be the one-element grade algebra and 
consider the (cartesian) MKF $(\worlds, \leq, \vee, \top)$. 
Define the Kripke extension (we do not write the unique grade subscript) 
$\bbox$ by:
$$
\wrel{\bbox{\relone}}{x}{y}{\wone} 
\defiff 
\forall \wtwo \wgeq \wone.\ \wrel{\relone}{x}{y}{\wtwo}.
$$ 
Then, $\bbox$ gives a comonadic lax extension.
\end{proposition}

The map $\bbox$ is nothing but the relational counterpart of the 
propositional construction used in the Kripke semantics of
 the necessity modality. Notice also that $\bbox$ can be used to 
recover the Kripke logical relation semantics of (intuitionistic) 
arrow types (where we encode $\typeone \to \typetwo$ as 
$\bbox \typeone \mmap \typetwo$).

Our last example of a comonadic lax extension
deals with information flow. 
Let us fix a (op-)lattice of security levels 
$(\mathcal{L}, \geq, \wedge, \vee, \bot, \top)$. 

\shortv{
    \begin{proposition}
    \label{prop:masking-corelator}
    Define the \emph{masking extension} $\mask{}{}$ thus:
    $$
    \wrel{\mask{\rone}{\relone}}{x}{y}{\rtwo} 
    \equiv 
    \rone \nleq \rtwo \text{ or } \wrel{\relone}{x}{y}{\rtwo}.
    $$
    Then, $\mask{}{}$ is a comonadic lax extension.
    \end{proposition}

    The action of $\mask{\rone}{}$ is to
    make code invisible to users with permission below 
    $\rone$. Recall that a judgment of the form 
    $\wrel{\relone}{\termone}{\termtwo}{\rtwo}$ has the intended 
    meaning that terms $\termone$ and $\termtwo$ are 
    $\relone$-indistinguishable to users with security permission 
    $\rtwo$. For instance, two \emph{classified} terms $\termone, \termtwo$
    are indistinguishable 
    to a user with \emph{low} confidentiality permission, 
    even if the two terms are actually different. 
}

\longv{
          \begin{definition}
          The \emph{masking extension} $\mask{}{}$ is thus defined:
          $$
          \wrel{\mask{\rone}{\relone}}{x}{y}{\rtwo} 
          \iff \rone \nleq \rtwo \text{ or } \wrel{\relone}{x}{y}{\rtwo}.
          $$
          \end{definition}

          The action of $\mask{\rone}{}$ is to
          make code invisible to users that has permissions below 
          $\rone$. Recall that a judgment of the form 
          $\wrel{\relone}{\termone}{\termtwo}{\rtwo}$ has the intended 
          meaning that terms $\termone$ and $\termtwo$ are 
          $\relone$-indistinguishable to users with security permission 
          $\rtwo$. For instance, two classified terms $\termone, \termtwo$
          are indistinguishable 
          to a user with low confidentiality permission, 
          even if the two terms are actually different. 

        This is because a user with public permission cannot use 
        secret programs, and thus she will not be able to tell them apart. 
        The masking extension $\mask{\rone}{}$ simply masks (and thus identifies) 
        terms to users with permissions that are not as high as 
        $\rone$. For instance, all terms are $\mask{\secret}{\relone}$-related 
        to a user. Instead, to a user with classified permission 
        such expressions are $\mask{\secret}{\relone}$-related only if they 
        are actually $\relone$-related.

        \begin{proposition}
        \label{prop:masking-corelator}
        The masking extension is a comonadic lax extension.
        \end{proposition}
        \shortv{
        As before, the proof of Proposition~\ref{prop:masking-corelator} 
        checks that all conditions in Definition~\ref{definition:corelator} 
        are satisfied by $\mask{}{}$.
        }

        \begin{proof}
        First, let us observe that $\mask{\rone}{\relone}$ is monotone, 
        provided $\relone$ is. Assume $\wrel{\mask{\rone}{\relone}}{x}{y}{\rtwo}$ 
        and $\rthree \leq \rtwo$ (recall that the order is reversed). 
        If $\wrel{\mask{\rone}{\relone}}{x}{y}{\rtwo}$ holds because 
        $\rtwo \ngeq \rone$, the also $\rthree \ngeq \rone$, 
        and thus we have $\wrel{\mask{\rone}{\relone}}{x}{y}{\rthree}$. 
        If, instead, $\wrel{\mask{\rone}{\relone}}{x}{y}{\rtwo}$ holds because 
        $\wrel{\relone}{x}{y}{\rtwo}$ does, then monotonicity of $\relone$ 
        gives $\wrel{\relone}{x}{y}{\rthree}$, and thus 
        $\wrel{\mask{\rone}{\relone}}{x}{y}{\rthree}$. 
        Next, we have to show that $\mask{}{}$ satisfies all the properties 
        required by Definition~\ref{definition:corelator}. 
        This can be seen through a number of standard calculations similar to those 
        exhibited for the canonical co-relator.
        \end{proof}  
        }
}

\section{Term Relations and their Algebra}

In previous sections, we have introduced $\Worlds$-relations and their 
algebra, in the abstract. However, studying $\Lambda_{\gradealg}$ 
we are interested not in general 
$\Worlds$-relations but in relations between $\Lambda_{\gradealg}$-terms. 
Following standard practice, we refer to such relations as 
\emph{term relations} \cite{Pitts/ATBC/2011,Lassen/PhDThesis}.

\begin{definition}
A \emph{term relation} is a $\Worlds$-relation $\relone$ between 
judgments belonging to the same syntactic class 
(that is, relating values with values, and terms with terms)
satisfying the following properties:
\begin{align*}
  \relone(\envone_1 \compimp \termone: \typeone_1, \envone_2 \compimp \termtwo: \typeone_2, 
  \wone)
  &\implies \envone_1 = \envone_2 \text{ and } \typeone_1 = \typeone_2
  \\
 \relone(\envone_1 \valimp \valone: \typeone_1, \envone_2 \valimp \valtwo: \typeone_2, \wone)
  &\implies \envone_1 = \envone_2 \text{ and } \typeone_1 = \typeone_2.
\end{align*}
\longv{
      We also require term relations to be closed under weakening,\footnote{ 
        Our formulation of weakening derived from the typing rule 
        \[
        \infer[\gradeone \mgeq \munit]{\envone, \graded{\varone}{\typeone}{\gradeone}}{}
        \]
        where we allow for weakening with respect to an arbitrary environment $\envone$. 
        Other calculi \cite{Orchard:2019:QPR:3352468.3341714,Gaboradi-et-al/ICFP/2016} 
        perform weakening not with respect to $\envone$, but with respect to 
        $\mzero \mstar \envone$. As already remarked, our results do not depend on 
        this kind of choices. If the reader wishes to consider such a form of weakening, 
        then she also has to consistently apply it to term relations.
      } meaning that 
      $\relone(\envone \compimp \termone: \typeone, \envone \compimp \termtwo: \typeone, 
      \wone)$ implies 
      ${\relone(\envone \mplus \envtwo \compimp \termone: \typeone, \envone \mplus \envtwo \compimp \termtwo: \typeone, 
      \wone)}$ (and similarly for values).
}
\end{definition}

We employ the notation 
$\wrelcomp{\envone}{\relone}{\termone}{\termtwo}{\wone}{\typeone}$ 
in place of 
$\relone(\envone \compimp \termone: \typeone, \envone \compimp \termtwo: \typeone, \wone)$ 
(and similarly, for values) and write 
$\wrelo{\envone}{\relone}{\termone}{\termtwo}{\wone}{\typeone}$ if the distinction 
between values and terms is not relevant. 
\shortv{
    We also require term relations to be closed under weakening, meaning that
    $\wrelo{\envone}{\relone}{\termone}{\termtwo}{\wone}{\typeone}$ implies 
    $\wrelo{\envone \mplus \envtwo}{\relone}{\termone}{\termtwo}{\wone}{\typeone}$ 
}

\begin{definition}
A \emph{closed} term relation is a term relation relating judgments of the form 
$\emptyenv \imp \termone: \typeone$ (we write 
$\wrelt{\relone}{\termone}{\termtwo}{\wone}{\typeone}$ in place 
of $\wrelo{\emptyenv}{\relone}{\termone}{\termtwo}{\wone}{\typeone}$). 
The \emph{open extension} of a closed term 
relation $\relone$ is the term relation $\open{\relone}$ thus defined
$$
\wrelo{\envone}{\open{\relone}}{\termone}{\termtwo}{\wone}{\typeone}
\equiv 
\forall \substmap \in \mathsf{Subst}(\envone).\ 
\wrelt{\relone}{\termone\substmap}{\termtwo\substmap}{\wone}{\typeone},
$$
where $\mathsf{Subst}(\envone)$ denotes the collection of maps $\gamma$ 
sending variables $(\varone:_{\gradeone} \typeone) \in \envone$ to closed values 
of type $\typeone$. 
Dually, the \emph{closed projection} of a term relation $\relone$ is the 
closed term relation $\closed{\relone}$ obtained by restricting 
$\relone$ to closed terms.
\end{definition}

Term relations being specific $\Worlds$-relations, all the constructions 
seen in previous sections apply to term relations as well. 
Additionally, we can extend the algebra of term relations relying on specific 
features of $\Lambda_{\gradealg}$. 
We already did that with the open extension of a term relation, 
and now we do it again by introducing the central notion a \emph{compatible refinement}.

\begin{definition}
\label{def:compatible-refinement}
The compatible refinement of a term relation $\relone$ 
is the term relation $\refine{\relone}$ inductively defined by the rules in 
Figure~\ref{fig:compatible-refinement}.
\end{definition}

\begin{figure*}[htbp]
\hrule
 $\vspace{0.2cm}$
\[
\infer{
  \wrelval
  {\envone, \graded{\varone}{\typeone}{\gradeone}}
  {\refine{\relone}}
  {\varone}
  {\varone}
  {\wone}
  {\typeone}
}
{\gradeone \mgeq \munit}
\quad
\infer{
  \wrelcomp
  {\envone}
  {\refine{\relone}}
  {\valone}
  {\valtwo}
  {\wone}
  {\typeone}
}
{
  \wrelval
  {\envone}
  {\relone}
  {\valone}
  {\valtwo}
  {\wone}
  {\typeone}
}
\quad
\infer{
  \wrelval
  {\envone}
  {\refine{\relone}}
  {\abs{\varone}{\termone}}
  {\abs{\varone}{\termtwo}}
  {\wone}
  {\typeone \to \typetwo}
}
{
  \wrelcomp
  {\envone, \graded{\varone}{\typeone}{\munit}}
  {\relone}
  {\termone}
  {\termtwo}
  {\wone}
  {\typetwo}
}
\quad
\infer{
  \wrelcomp
  {\envone \mplus \envtwo}
  {\refine{\relone}}
  {\valone \valtwo}
  {\valone' \valtwo'}
  {\wthree}
  {\typetwo}
}
{
  \deduce[ ] 
  {
    \wrelval
    {\envone}
    {\relone}
    {\valone}
    {\valone'}
    {\wone}
    {\typeone \to \typetwo}
  }
  {
    \wrelval
    {\envtwo}
    {\relone}
    {\valtwo}
    {\valtwo'}
    {\wtwo}
    {\typeone}
  }
  &
  \wthree \wgeq \wone \wcomp \wtwo
}
\]
$\vspace{-0.1cm}$
\[
\infer{
  \wrelcomp
  {(\gradeone \mvee \munit) \mstar \envone \mplus \envtwo}
  {\refine{\relone}}
  {\seq{\termone}{\termthree}}
  {\seq{\termtwo}{\termfour}}
  {\wthree}
  {\typetwo}
}
{
  \wrelcomp
  {\envone}
  {\corelator{\gradeone \mvee \munit}{\relone}}
  {\termone}
  {\termtwo}
  {\wone}
  {\typeone}
  &
  \wrelcomp
  {\envtwo, \graded{\varone}{\typeone}{\gradeone}}
  {\relone}
  {\termthree}
  {\termfour}
  {\wtwo}
  {\typetwo}
  &
  \wthree \wgeq \wone \wcomp \wtwo
}
\quad
\infer{
  \wrelcomp
  {\envone}
  {\refine{\relone}}
  {\unfold{\valone}}
  {\unfold{\valtwo}}
  {\wone}
  {\substtype{\typeone}{\typevarone}{\rectype{\typevarone}{\typeone}}}
}
{
  \wrelval
  {\envone}
  {\relone}
  {\valone}
  {\valtwo}
  {\wone}
  {\rectype{\typevarone}{\typeone}}
}
\]
$\vspace{-0.1cm}$
\[
\infer{
  \wrelval
  {\envone}
  {\refine{\relone}}
  {\fold{\valone}}
  {\fold{\valtwo}}
  {\wone}
  {\rectype{\typevarone}{\typeone}}
}
{
  \wrelval
  {\envone}
  {\relone}
  {\valone}
  {\valtwo}
  {\wone}
  {\substtype{\typeone}{\typevarone}{\rectype{\typevarone}{\typeone}}}
}
\quad
\infer{
  \wrelval
  {\gradeone \mstar \envone}
  {\refine{\relone}}
  {\tbox{\valone}}
  {\tbox{\valtwo}}
  {\wone}
  {\bbox_{\gradeone}\typeone}
}
{
  \wrelval
  {\envone}
  {\corelator{\gradeone}{\relone}}
  {\valone}
  {\valtwo}
  {\wone}
  {\typeone}
}
\quad
\infer{
  \wrelcomp
  {\gradetwo \mstar \envone \mplus \envtwo}
  {\refine{\relone}}
  {\letbox{\valone}{\termone}}
  {\letbox{\valtwo}{\termtwo}}
  {\wthree}
  {\typetwo}
}
{
  \deduce[ ]
  {
    \wrelcomp
    {\envtwo, \graded{\varone}{\typeone}{\gradetwo \mstar \gradeone}}
    {\relone}
    {\termone}
    {\termtwo}
    {\wtwo}
    {\typetwo}
  }
  {
    \wrelval
    {\envone}
    {\corelator{\gradetwo}\relone}
    {\valone}
    {\valtwo}
    {\wone}
    {\bbox_{\gradeone}\typeone}
  }
  &
  \wthree \wgeq \wone \wcomp \wtwo
}
\]
\hrule
\caption{Compatible Refinement of $\relone$}
\label{fig:compatible-refinement}
\end{figure*}

Intuitively, the compatible refinement of a term relation 
$\relone$ is the relation obtained 
from $\relone$ by closing $\relone$-related expressions under syntactic constructors. 
Notice that $\refine{\relone}$ is indeed a $\Worlds$-relation (viz. it 
is monotone). 
A natural way to understand the defining rules of $\refine{\relone}$ 
is to look at those rules as diagrams in $\wrelcat$.
\longv{
   We consider the following 
    (families of) maps as illustrative examples (it is straightforward to 
    extend such an example to the full language $\Lambda_{\worlds}$):
    \newcommand{\appmap}{\mathtt{app}}
    \newcommand{\absmap}{\mathtt{abs}}
    \newcommand{\seqmap}{\mathtt{seq}}

    \begin{align*}
    \appmap &: \values_{\envone \valimp \typeone \to \typetwo} \times \values_{\envtwo \valimp \typeone} \to \Lambda_{\envone \mplus \envtwo \compimp \typetwo}
    & 
    \appmap(\valone, \valtwo) &= \valone\valtwo
    \\
    \absmap_{\varone} &: \Lambda_{\envone, \graded{\varone}{\typeone}{\munit} \compimp \typetwo} 
    \to \values_{\envone \valimp \typeone \to \typetwo}
    &
    \absmap_{\varone}(\termone) &= \abs{\varone}{\termone}
    \\
    \seqmap_{\varone} &: \Lambda_{\envone \compimp \typeone} \times 
    \Lambda_{\envtwo, \graded{\varone}{\typeone}{\gradeone} \compimp \typeone} \to 
    \Lambda_{\gradeone \mstar \envone \mplus \envtwo \to \typetwo}
    &
    \seqmap_{\varone}(\termone, \termtwo) &= \seq{\termone}{\termtwo}
    \end{align*}
    Then, we see that the clauses in Figure~\ref{fig:compatible-refinement} 
    for abstraction, application, and sequencing are nothing but the pointwise 
    version of the following lax commutative diagrams in $\wrelcat$.
    \[
        \diagramrel
        {\appmap}
        {\appmap}
        {\values_{\envone \valimp \typeone \to \typetwo} \times 
        \values_{\envtwo \valimp \typeone}}
        {\values_{\envone \valimp \typeone \to \typetwo} \times 
        \values_{\envtwo \valimp \typeone}}
        {\relone \otimes \relone}
        {\Lambda_{\envone \mplus \envtwo \compimp \typetwo}} 
        {\Lambda_{\envone \mplus \envtwo \compimp \typetwo}}
        {\refine{\relone}}  
        \qquad
        \diagramrel
        {\absmap_{\varone}}
        {\absmap_{\varone}}
        {\Lambda_{\envone, \graded{\varone}{\typeone}{\munit} \compimp \typetwo}}
        {\Lambda_{\envone, \graded{\varone}{\typeone}{\munit} \compimp \typetwo}}
        {\relone}
        {\values_{\envone \valimp \typeone \to \typetwo}}
        {\values_{\envone \valimp \typeone \to \typetwo}}
        {\refine{\relone}}
    \]

    \[
    \diagramrel
    {\seqmap_{\varone}}
    {\seqmap_{\varone}}
    {\Lambda_{\envone \compimp \typeone} \times 
     \Lambda_{\envtwo, \graded{\varone}{\typeone}{\gradeone} \compimp \typeone}
    } 
    {\Lambda_{\envone \compimp \typeone} \times 
     \Lambda_{\envtwo, \graded{\varone}{\typeone}{\gradeone} \compimp \typeone}
    }
    {\corelator{\gradeone}{\relone} \tensor {\relone}}
    {\Lambda_{\gradeone \mstar \envone \mplus \envtwo \to \typetwo}}
    {\Lambda_{\gradeone \mstar \envone \mplus \envtwo \to \typetwo}}
    {\refine{\relone}}
    \]
}
\shortv{
    We consider the case of sequencing as an illustrative example.
    \newcommand{\seqmap}{\mathtt{seq}}
    Let the function
    \begin{align*}
    \seqmap_{\varone} &: \Lambda^{\envone}_{\typeone} \times 
    \Lambda^{\envtwo, \graded{\varone}{\typeone}{\gradeone}}_{\typetwo} \to 
    \Lambda^{\gradeone \mstar \envone \mplus \envtwo}_{\typetwo}
    \end{align*} 
    map $(\termone, \termtwo)$ to $\seq{\termone}{\termtwo}$.
    Then, we see that the clause in Figure~\ref{fig:compatible-refinement} 
    for sequencing is nothing but the pointwise 
    version of the following lax commutative diagram in $\wrelcat$.
    \[
    \diagramrel
    {\seqmap_{\varone}}
    {\seqmap_{\varone}}
    {\Lambda^{\envone}_{\typeone} \times 
     \Lambda^{\envtwo, \graded{\varone}{\typeone}{\gradeone}}_{\typetwo}
    } 
    {\Lambda^{\envone}_{\typeone} \times 
     \Lambda^{\envtwo, \graded{\varone}{\typeone}{\gradeone}}_{\typetwo}
    }
    {\corelator{\gradeone}{\relone} \tensor {\relone}}
    {\Lambda^{\gradeone \mstar \envone \mplus \envtwo}_{\typetwo}}
    {\Lambda^{\gradeone \mstar \envone \mplus \envtwo}_{\typetwo}}
    {\refine{\relone}}
    \]
}
Notice how we use the tensor product to account for multiple premises of the rule, 
as well as the comonadic lax extension $\corelatorsymbol_{\gradeone}$ 
to account for graded variables. 
For instance, if we take the canonical extension, 
then we see that, e.g., the 
compatible refinement rule for the box introduction specialises to 
the usual rule one finds in (graded) linear-like type systems 
\cite{DBLP:conf/esop/GhicaS14,Pierce/DistanceMakesTypesGrowStronger/2010}:
\[
\infer{
  \wrelval
  {\gradeone \mstar \envone}
  {\refine{\relone}}
  {\tbox{\valone}}
  {\tbox{\valtwo}}
  {\gradetwo}
  {\bbox_{\gradeone}\typeone}
}
{
  \wrelval
  {\envone}
  {\relone}
  {\valone}
  {\valtwo}
  {\gradethree}
  {\typeone}
  &
  \gradetwo \geq \gradeone \mstar \gradethree
}
\]

\longv{
    Definition~\ref{def:compatible-refinement} induces a \emph{monotone} map 
    $\relone \mapsto \refine{\relone}$ on the collection of term relations that satisfies the following identities:
    \begin{align*}
    \refine{\reltwo \comp \relone} &= \refine{\reltwo}\comp \refine{\relone}
    \\
    \refine{\dual{\relone}} &= \dual{\refine{\relone}}
    \end{align*}
    Moreover, we can formalise the notion of a compatible term relation, i.e. of 
    a term relation closed under syntactical constructors of $\Lambda_{\gradealg}$, 
    relying on the notion of a compatible refinement.
}

\shortv{
    Definition~\ref{def:compatible-refinement} induces a \emph{monotone} map 
    $\relone \mapsto \refine{\relone}$ on the collection of term relations 
    that allows us to formalise the notion of a compatible term relation, i.e. of 
    a term relation closed under syntactical constructors of $\Lambda_{\gradealg}$.
}

\begin{definition}
We say that a term relation is compatible if $\refine{\relone} \subseteq \relone$.
\end{definition}

In particular, a term relation is compatible if and only if it is a pre-fixed point of 
$\relone \mapsto \refine{\relone}$. It is not hard to prove that the identity term 
is such a pre-fixed point, and it actually is the least such. As a consequence, any compatible term relation is reflexive. 
Another important operation on term relations is the one of a 
\emph{substitutive refinement}.

\begin{definition}
The \emph{substitutive refinement} of a term relation $\relone$ 
is the term relation $\substrefine{\relone}$ inductively defined 
by the following rule.
\[
\infer{
  \wrelo
  {\envone}
  {\substrefine{\relone}}
  {\subst{\termone}{\varone}{\valone}}
  {\subst{\termtwo}{\varone}{\valtwo}}
  {\wthree}
  {\typetwo}
}
{\wrelo
{\envone, \graded{\varone}{\typeone}{\gradeone}}
{\relone}
{\termone}
{\termtwo}
{\wone}
{\typetwo}
&
\wrelvalclosed
{\corelator{\gradeone}{\relone}}
{\valone}
{\valtwo}
{\wtwo}
{\typeone}
&
\wthree \wgeq \wone \wcomp \wtwo
}
\]
We say that $\relone$ is \emph{substitutive} if 
$\substrefine{\relone} \subseteq \relone$.
\end{definition}

\longv{
    As before, we see the defining rule of $\substrefine{\relone}$ 
    as the pointwise version of the following lax commutative 
    diagram in $\wrelcat$, where 
    $\substarrow: \Lambda_{\envone, \graded{\varone}{\typeone}{\gradeone} \imp \typetwo} 
    \times \values_{\typeone} \to \Lambda_{\envone \imp \typetwo}$.

    \[
        \diagramrel
        {\substarrow}
        {\substarrow}
        {\Lambda_{\envone, \graded{\varone}{\typeone}{\gradeone} \imp \typetwo} 
        \times \values_{\typeone}}
        {\Lambda_{\envone, \graded{\varone}{\typeone}{\gradeone} \imp \typetwo} 
        \times \values_{\typeone}}
        {\relone \tensor \corelator{\gradeone}{\relone}}
        {\Lambda_{\typetwo}} 
        {\Lambda_{\typetwo}}
        {\substrefine{\relone}}  
    \]
}

In non-modal calculi, substitutivity and compatibility are important properties that 
tell us that equivalent expressions can be safely replaced for one another 
inside a more complex expression, this way giving the 
following compositionality law: if $\termone \simeq \termtwo$, then 
$\ctxone[\termone] \simeq \ctxone[\termtwo]$, where $\ctxone$ is 
an arbitrary context that we regard, for the sake of the arguments, 
as a term with a free variable $\varone$ 
(so that $\ctxone[\termone]$ is $\subst{\ctxone}{\varone}{\termone}$).

Compositionality as it is does not hold for modal calculi, as one 
should also account for possible worlds.
In fact, even if two terms $\termone$, $\termtwo$ are equivalent 
\emph{at a given world} $\wone$, then $\ctxone[\termone]$ 
and $\ctxone[\termtwo]$ may differ at $\wone$.
Why? Because 
$\ctxone$ may use $\termone$ (resp. $\termtwo$) in a modal context, so that what we will observe 
is 
the behaviour of $\termone$ (resp. $\termtwo$) not at $\wone$ but at a different 
world $\wtwo$, where $\termone$ and $\termtwo$ may differ.  
Formally, this means that the `naive' compositionality law 
$\termone \mathbin{{\simeq}{(\wone)}} \termtwo \implies 
\ctxone[\termone] \mathbin{{\simeq}{(\wone)}} \ctxone[\termtwo]$ 
is unsound.

Comonadic lax extensions offer a solution to this problem by replacing 
the premise $\termone \mathbin{{\simeq}{(\wone)}} \termtwo$ 
with $\termone \mathbin{{\corelator{\gradeone}{\simeq}}{(\wone)}} \termtwo$, 
this way letting $\corelator{\gradeone}{\simeq}$ to give information 
on the equality of $\termone$ and $\termtwo$ at world $\wone$ but 
when used as prescribed by $\gradeone$. This is 
summarised by the compositionality law:
\longv{
    \[
    \infer{
      \wrelt
      {\relone}
      {\ctxone[\termone]}
      {\ctxone[\termtwo]}
      {\wone}
      {\typetwo}
    }
      {
        \wrelt
        {\corelator{\gradeone}{\relone}}
        {\termone}
        {\termtwo}
        {\wone}
        {\typeone}
        &
        \graded{\varone}{\typeone}{\gradeone} \imp \ctxone: \typetwo
      }
    \]
In light of all these considerations, it is desirable to define 
a substitutive and compatible equivalence term relation. 
Among the many options available, 
we choose for Abramsky's \emph{applicative bisimilarity} \cite{Abramsky/RTFP/1990}. 
}
\shortv{
    $$
    \termone \mathbin{{\corelator{\gradeone}{\simeq}}{(\wone)}} \termtwo 
    \:\&\;
    \graded{\varone}{\typeone}{\gradeone} \imp \ctxone: \typetwo
    \implies
    \ctxone[\termone] \mathbin{{\simeq}{(\wone)}} \ctxone[\termtwo],
    $$
which follows from substitutivity 
by noticing that 
$\graded{\varone}{\typeone}{\gradeone} \imp \ctxone \relone(\wunit) \ctxone: \typetwo$.
In light of all these considerations, what we look for is 
a \emph{substitutive} and \emph{compatible} equivalence term relation. 
Among the many options available, 
we choose Abramsky's \emph{applicative bisimilarity} \cite{Abramsky/RTFP/1990}. 
}

\section{Modal Applicative (Bi)similarity}
\label{section:modal-applicative-bisimilarity}

In this section, we introduce \emph{modal applicative (bi)similarity}, 
the extension of Abramsky's applicative (bi)similarity \cite{Abramsky/RTFP/1990} 
to a modal and coeffectful setting, and prove that what we obtain is indeed a compatible 
and substitutive term relation.

\longv{
    \begin{definition}
    \label{definition:modal-applicative-bisimulation}
    Recall the definition of the relator $\relatorsymbol^{\divergence}$ 
    for the partiality monad given in Proposition~\ref{proposition:relator-partiality}.
    Define the mapping $\relone \mapsto [\relone]$ on 
    \emph{closed} term relations as follows:
    \begin{align*}
    \wrelcompclosed{[\relone]}{\termone}{\termtwo}{\wone}{\typeone}
    &\iff 
    \wrelvalclosed{\relatorsymbol^{\divergence}(\relone)}{\eval{\termone}}
    {\eval{\termtwo}}{\wone}{\typeone}
    \tag{App eval} \label{eq:app-eval}
    \\
    \wrelvalclosed{[\relone]}{\abs{\varone}{\termone}}{\abs{\varone}{\termtwo}}{\wone} 
    {\typeone \to \typetwo }
    &\iff \forall \valone \in \values_{\typeone}.\ 
    \wrelcompclosed{\relone}{\subst{\termone}{\varone}{\valone}}
    {\subst{\termtwo}{\varone}{\valone}}{\wone}{\typetwo}
    \tag{App abs} \label{eq:app-abs}
    \\
    \wrelvalclosed{[\relone]}{\fold{\valone}}{\fold{\valtwo}}{\wone}{
      \rectype{\typevarone}{\typeone}}
    &\iff
    \wrelvalclosed{\relone}{\valone}{\valtwo}{\wone}{
      \substtype{\typeone}{\typevarone}{\rectype{\typevarone}{\typeone}}}
    \tag{App fold} \label{eq:app-fold}
    \\
    \wrelvalclosed{[\relone]}{\tbox{\valone}}{\tbox{\valtwo}}{\wone}{\bbox_{\gradeone} \typeone} 
    &\iff \wrelvalclosed{\corelator{\gradeone}{\relone}}{\valone}{\valtwo}{\wone}{\typeone} 
    \tag{App box} \label{eq:app-box}
    \end{align*}
    We sat that a closed term relation $\relone$ is an \emph{applicative simulation} 
    if $\relone \subseteq [\relone]$, and that $\relone$ is an \emph{applicative bisimulation} 
    if both $\relone$ and $\dual{\relone}$ are applicative simulation.
    \end{definition}
}

\shortv{
    \begin{definition}
    \label{definition:modal-applicative-bisimulation}
    Given $\relone: X \torel Y$, define 
    $\relone_\bot: X_\bot \to Y_\bot$ by $x \relone_\bot(\wone) y$ iff 
    $x\neq \bot$ implies $y \neq \bot$ and $x\relone(\wone)y$.
    Let $\corelatorsymbol$ be a comonadic lax extension. 
    We define the map $\relone \mapsto [\relone]$ on 
    \emph{closed} term relations as follows:
    \begin{align*}
    \wrelcompclosed{[\relone]}{\termone}{\termtwo}{\wone}{\typeone}
    &\equiv 
    \sem{\termone} \mathbin{\relone^{\scriptscriptstyle \values}_{\bot}(\wone)}
    \sem{\termtwo}
    \\
    \wrelvalclosed{[\relone]}{\valone}{\valtwo}{\wone} 
    {\typeone \gmap \typetwo }
    &\equiv \forall \valthree \in \values_{\typeone}.
    \wrelcompclosed{\relone}{\valone \valthree}
    {\valtwo\valthree}{\wone}{\typetwo}
    \\
    \wrelvalclosed{[\relone]}{\fold{\valone}}{\fold{\valtwo}}{\wone}{
      \rectype{\typevarone}{\typeone}}
    &\equiv
    \wrelvalclosed{\relone}{\valone}{\valtwo}{\wone}{
      \substtype{\typeone}{\typevarone}{\rectype{\typevarone}{\typeone}}}
    \\
    \wrelvalclosed{[\relone]}{\tbox{\valone}}{\tbox{\valtwo}}{\wone}{\bbox_{\gradeone} \typeone} 
    &\equiv 
    \wrelvalclosed{\corelator{\gradeone}{\relone}}{\valone}{\valtwo}{\wone}{\typeone} 
    \end{align*}
    We say that a closed term relation $\relone$ is an \emph{applicative simulation} 
    if $\relone \subseteq [\relone]$, and that $\relone$ is an \emph{applicative bisimulation} 
    if both $\relone$ and $\dual{\relone}$ are applicative simulations.
    \end{definition}
}

The first three clauses in Definition~\ref{definition:modal-applicative-bisimulation} 
are the possible-world counterparts 
of the usual defining clause of a (non-modal) applicative simulation. 
\longv{
        Clause \eqref{eq:app-abs} is function extensionality, whereas 
        clause \eqref{eq:app-fold} is a standard syntax-oriented clause 
        requiring values with the same outermost syntactic constructor 
        to have pairwise related arguments.
        Finally, clause \eqref{eq:app-box} is the real source of intensionality 
        and the actual novelty of (modal) applicative (bi)simulations; it
        defines the relational action of applicative simulations at modal 
        necessity types in terms of co-relators. 

        It is easy to see that a closed term relation $\relone$ is 
        an applicative simulation if and only if the following clauses hold:
        \begin{align*}
        \wrelcompclosed{\relone}{\termone}{\termtwo}{\wone}{\typeone}
         &\implies 
         \wrelvalclosed{\relatorsymbol^{\divergence}(\relone)}{\eval{\termone}}
        {\eval{\termtwo}}{\wone}{\typeone}
        \\
        \wrelvalclosed{\relone}{v}{w}{\wone} 
        {\typeone \to \typetwo }
        &\implies \forall v \in \values_{\typeone}.
        \wrelcompclosed{\relone}{vu}
        {wu}{\wone}{\typetwo}
        \\
        \wrelvalclosed{\relone}{\fold{\valone}}{\fold{\valtwo}}{\wone}{
          \rectype{\typevarone}{\typeone}}
        &\implies
        \wrelvalclosed{\relone}{\valone}{\valtwo}{\wone}{
          \substtype{\typeone}{\typevarone}{\rectype{\typevarone}{\typeone}}}
        \\
        \wrelvalclosed{\relone}{\tbox{\valone}}{\tbox{\valtwo}}{\wone}{\bbox_{\gradeone} \typeone} 
        &\implies \wrelvalclosed{\corelator{\gradeone}{\relone}}{\valone}{\valtwo}{\wone}{\typeone}. 
        \end{align*}
        Moreover, since (co-)relators are monotone, an easy calculation shows that the 
        mapping $\relone \mapsto [\relone]$ is monotone, and thus by the Knaster-Tarski 
        fixed point theorem \cite{tarski1955} it has a \emph{greatest fixed point} which we call 
        \emph{applicative similarly} and denote by $\appsimilarity$. 
        \emph{Applicative bisimilarity} is defined as ${\appsimilarity} \cap 
        {\dual{(\appsimilarity)}}$ and denoted by $\appbisimilarity$. Notice that 
        $\appsimilarity$ being defined coinductively (i.e. as the greatest 
        fixed point of a monotone function) it obeys the following laws, which are 
        known as the coinduction proof principle for (bi)similarly.

        \[
        \infer{{\appsimilarity} = {[{\appsimilarity}]}}{}
        \qquad
        \infer{\relone \subseteq {\appsimilarity}}{\relone \subseteq [\relone]}
        \]

        In particular, say we want to prove a statement of the form 
        $\wrelt{\appsimilarity}{\termone}{\termtwo}{\wone}{\typeone}$. Then, it is 
        enough to find an applicative simulation $\relone$ such that 
        $\wrelt{\relone}{\termone}{\termtwo}{\wone}{\typeone}$ holds.
}
\shortv{
      The real novelty of Definition~\ref{definition:modal-applicative-bisimulation} 
      is the last clause, where modal values are related relying on 
      comonadic lax extensions.

      Since lax extensions are monotone, the 
      mapping $\relone \mapsto [\relone]$ is monotone too 
      and thus it has a \emph{greatest fixed point} which we call 
      \emph{applicative similarly} and denote by $\appsimilarity$. 
      \emph{Applicative bisimilarity} is defined as ${\appsimilarity} \cap 
      {\dual{\appsimilarity}}$ and denoted by $\appbisimilarity$. Notice that 
      $\appsimilarity$ being defined coinductively (i.e. as the greatest 
      fixed point of a monotone function) it obeys the 
      coinduction proof principle, whereby to prove  
      $\termone \mathbin{{\appsimilarity}{(\wone)}} \termtwo: \typeone$ it is 
      enough to find an applicative simulation $\relone$ such that 
      $\wrelt{\relone}{\termone}{\termtwo}{\wone}{\typeone}$.
}
\begin{example}
\begin{varenumerate}
  \item
For the canonical extension, we see that we have
$\wrelvalclosed{\appsimilarity}{\tbox{\valone}}{\tbox{\valtwo}}{\gradetwo}{\bbox_{\gradeone} \typeone} 
$ iff 
there exists $\gradethree$ such that
$\gradetwo \mgeq \gradeone \mstar \gradethree$ and 
$\wrelvalclosed{\appsimilarity}{\valone}{\valtwo}{\gradethree}{\typeone}$
Taking non-negative real numbers both as grade algebra and as 
MKF, we see that modal applicative (bi)similarity gives a relational 
presentation of applicative bisimulation 
metrics \cite{Gavazzo/LICS/2018,CrubilleDalLago/LICS/2015} 
and a coinductive counterpart of the metric logical relations 
by Reed and Pierce \cite{Pierce/DistanceMakesTypesGrowStronger/2010}.
\item For the Kripke extension, we have
  $\wrelvalclosed{\appsimilarity}{\tbox{\valone}}{\tbox{\valtwo}}
  {\wone}{\bbox \typeone}$ iff 
  $\wrelvalclosed{\appsimilarity}{\valone}{\valtwo}
  {\wtwo}{\typeone}$, for any $\wtwo \wgeq \wone$. 
  Therefore, we see that modal applicative bisimilarity 
  gives a coinductive counterpart of the well-known 
  Kripke logical relations \cite{mitchell_2002}.
\item
  For the masking extension $\mask{}{}$ we have
  $\wrelvalclosed{\appsimilarity}{\tbox{\valone}}{\tbox{\valtwo}}{\gradetwo}{\bbox_{\gradeone} \typeone}$ 
  iff
  $\gradetwo \ngeq \gradeone$ or
  $\wrelvalclosed{\appsimilarity}{\valone}{\valtwo}{\gradetwo}{\typeone}$.
  In particular, if two $\gradeone$-masked values $\tbox{\valone}$ and $\tbox{\valtwo}$ 
  are equivalent at security level $\gradetwo$, then either $\gradetwo$ is 
  a too low security level ($\gradetwo \ngeq \gradeone$) to access 
  $\valone$ and $\valtwo$, or the non-masked values $\valone$ and $\valtwo$ are  
  actually equivalent at $\gradetwo$. Modal applicative bisimilarity 
  thus gives the coinductive counterpart of the usual logical 
  relations used in the field of information flow \cite{DBLP:journals/lisp/SabelfeldS01,DBLP:conf/icfp/BowmanA15}.
\end{varenumerate}
\end{example}

Finally, we extend applicative (bi)similarly to open expressions by means of its 
open extension (and write $\appsimilarity$ in place of $\open{\appsimilarity}$). 
\longv{
    In particular, we have:
    $$
    \wrelo{\envone}{\appsimilarity}{\termone}{\termtwo}{\wone}{\typeone}
    \iff
    \forall \substmap \in \mathsf{Subst}(\envone).\
    \wrelt
    {\appsimilarity}{\termone\substmap}{\termtwo\substmap}{\wone}{\typeone}.
    $$
}
What remains to be done is to prove that applicative bisimilarity 
is indeed a notion of program equivalence, in the sense that it
is compatible and 
substitutive $\Worlds$-equivalence.

\begin{proposition}
Applicative similarly $\appsimilarity$ is a $\Worlds$-preorder, and 
applicative bisimilarity $\appbisimilarity$ is a $\Worlds$-equivalence.
\end{proposition}

\begin{proof}
By coinduction, showing, e.g., that ${\appsimilarity}; {\appsimilarity}$ 
is an applicative simulation.
\end{proof}


\subsection{Compositionality and Howe's Method}
\label{section:compositionality-metric-preservation-and-non-interference}

\longv{ 
      In this section, we study compositionality and substitutivity properties of 
      modal applicative bisimilarity. Before entering into the details, however, 
      we spend few words discussing the practical consequences of such properties. 

      In previous sections, we have extensively analysied the relationship between 
      compatibility/substitutivity and compositionality. Our motivations, however, were purely 
      semantical, and one may still ask whether having compatible and substitutive 
      relations (and ultimately having a compositional \emph{intensional} semantics) 
      is of some practical relevance. 
      The practical benefits of extensional semantics are well-known. Do such benefits 
      scale to intensional semantics? 
      We try to answer this question in the affirmative by analysing the case 
      of two main results on semantical approaches to differential privacy 
      and information flow, respectively. Such results are the so-called 
      \emph{metric preservation} \cite{Pierce/DistanceMakesTypesGrowStronger/2010,GaboardiEtAl/POPL/2017} and \emph{non-interference} \cite{DBLP:conf/popl/AbadiBHR99} 
      theorems.

      Let us begin with non-interference. In one of its simplest form non-interference 
      can be stated as follows.

      \begin{theorem}[Non-interference]
      If $\graded{\varone}{\typeone}{\public}, \graded{\vartwo}{\typetwo}{\secret} 
      \imp \termone: \typethree$, then for all values 
      $\emptyenv \valimp \valone: \typeone$, $\emptyenv \valimp \valtwo, \valtwo': \typetwo$ 
      we have:
      $$\wrelt{\appbisimilarity}{\termone[\varone:=\valone, \vartwo := \valtwo]}
      {\termone[\varone:=\valone, \vartwo := \valtwo']}{\public}{\typethree}.
      $$
      \end{theorem}

      Intuitively, non-interference states that users with low permission will 
      never observe secret information flowing out 
      of a program. Consider the expression 
      $\graded{\varone}{\typeone}{\public}, \graded{\vartwo}{\typetwo}{\secret} 
      \imp \termone: \typethree$, and suppose to have public permissions only. 
      Then, we should not be able to obtain any information about any two secret inputs 
      $\valtwo, \valtwo'$ passed to 
      $\termone$. If, however, we are able to observe differences between the behaviour of
       $\termone[\varone:=\valone, \vartwo := \valtwo]$ and the one of
      $\termone[\varone:=\valone, \vartwo := \valtwo']$, then it means that 
      some information concerning $\valtwo$ and $\valtwo'$ flowed out of 
      $\termone$, and we may able to recover such information. 

      Our formulation of non-interference makes use of applicative bisimilarity, but 
      it is immediate to realise that any notion of equivalence can do the job. 
      Actually, we can formulate non-interference without mentioning any notion of program 
      equivalence by simply requiring $\termone[\varone:=\valone, \vartwo := \valtwo]$ 
      and $\termone[\varone:=\valone, \vartwo := \valtwo']$ to both converge 
      (obviously, in order to use term relations to reason about non-interference defined 
      in this way, we need relations ensuring that related programs have the same converge 
      behaviour: notice that applicative bisimilarity is indeed such a relation).

      Bu what compositionality, substitutivity, $\hh$ have to do with non-interference? 
      The answer is simple: the former imply the latter. Indeed, assuming 
      applicative bisimilarity to be substitutive we have:
      \[
      \infer{
        \wrelt
        {\appbisimilarity}
        {\termone[\varone:=\valone, \vartwo := \valtwo]}
        {\termone[\varone:=\valone, \vartwo := \valtwo']}
        {\public}
        {\typethree}
      }
      {
        \infer{\wrelt
          {\mask{\secret}{(\appbisimilarity)}}
          {\valtwo}
          {\valtwo'}
          {\public}
          {\typeone}}
          {\public \ngeq \secret}
        &
        \infer{
          \graded{\vartwo}{\typetwo}{\secret} \imp \termone[\varone:=\valone]: \typethree
        }
        {
        \graded{\varone}{\typeone}{\public}, \graded{\vartwo}{\typetwo}{\secret} 
        \imp \termone: \typethree
        &\emptyenv \valimp \valone: \typeone
        }
      }
      \]

      Now that we have seen how non-interference follows from compatibility 
      and substitutivity, we move to metric preservation \cite{Pierce/DistanceMakesTypesGrowStronger/2010}. 
      {
      \newcommand{\rzero}{0}
      \newcommand{\rone}{r}
      \newcommand{\rtwo}{s}
      \newcommand{\rthree}{p}

      \begin{theorem}[Metric Preservation]

      For any expression $\graded{\varone_1}{\typeone_1}{\rtwo_1}, \hh, 
      \graded{\varone_1}{\typeone_1}{\rtwo_1} \imp \termone: \typeone$, and for all 
      values $\valimp \valone_1, \valtwo_1: \typeone_1, \hh, 
      \valimp \valone_n, \valtwo_n: \typeone_n$, if the distance between 
      each pair of values $\valone_i, \valtwo_i$ is bounded by $\rone_i$, then 
      the distance between 
      $\termone[\varone_1 := \valone_1, \hh, \varone_n := \valone_n]$ 
      and $\termone[\varone_1 := \valtwo_1, \hh, \varone_n := \valtwo_n]$ 
      is bounded by $\sum_i \rtwo_1 \rone_i$.
      \end{theorem}

      Expressing distances by means of term relations (with modal signature 
      $([0,\infty], \leq, +, \cdot, 0,1, \infty)$ and possible worlds 
      $([0,\infty], \leq, +, 0)$),
      we can express metric preservation as follows:
      \[
      \infer{
        \wrelt
        {\appbisimilarity}
        {\termone[\varone_1 := \valone_1, \hh, \varone_n := \valone_n]} 
        {\termone[\varone_1 := \valtwo_1, \hh, \varone_n := \valtwo_n]}
        {\sum_i \rtwo_i \rone_i}
        {\typeone}
      }
      {
        \graded{\varone_1}{\typeone_1}{\rtwo_1}, \hh, 
        \graded{\varone_1}{\typeone_1}{\rtwo_1} \imp \termone: \typeone
        &
        \wrelt
        {\appbisimilarity}
        {\valone_1}
        {\valtwo_1}
        {\rone_1}
        {\typeone_1}
        & 
        \cc
        & 
        \wrelt
        {\appbisimilarity}
        {\valone_n}
        {\valtwo_n}
        {\rone_n}
        {\typeone_n}
      }
      \]
      As for non-interference, the following derivation shows that 
      metric preservation is a direct consequence of compatibility/substitutivity.

      \[
      \infer{
        \wrelt
        {\appbisimilarity}
        {\termone[\varone_1 := \valone_1, \hh, \varone_n := \valone_n]} 
        {\termone[\varone_1 := \valtwo_1, \hh, \varone_n := \valtwo_n]}
        {\sum_i \rtwo_i \rone_i}
        {\typeone}
      }
      {
        \wrelo
        {\graded{\varone_1}{\typeone_1}{\rtwo_1}, \hh, \graded{\varone_1}{\typeone_1}{\rtwo_1}}
        {\appbisimilarity}
        {\termone}
        {\termone}
        {\rzero}
        {\typeone}
        &
        \infer
          {
            \wrelt
            {\bang_{\rtwo_i}(\appbisimilarity)}
            {\valone_i}
            {\valtwo_i}
            {\rtwo_i\rone_i}
            {\typeone_1}
          }
          {
            \wrelt
            {\appbisimilarity}
            {\valone_i}
            {\valtwo_i}
            {\rone_i}
            {\typeone_1}
          }
        & 
        \sum_i \rtwo_i\rone_i \geq  \sum_i \rtwo_i\rone_i
      }
      \]
      }

      Summing up, we have seen how two of the main theorems 
      concerning programming-language approaches to differential privacy 
      and information flow directly follow from compositional properties of 
      term relations (in our specific case, applicative bisimilarity). 
      As a consequence, what remains to be done is to prove that applicative bisimilarity 
      is a compatible and substitutive term relation. We do so using
      Howe's precongruence candidate,\footnote{Also known as Howe's method.} suitably adapted to the setting of 
      $\Worlds$-relations. 
}

Proving that applicative (bi)similarly is compatible and substitutive
is highly nontrivial.
In this section, we prove such results by extending Howe's method 
\cite{Howe/IC/1996,Pitts/ATBC/2011} 
to a modal setting. 
Accordingly, we construct a substitutive and compatible relation $\howe{\appsimilarity}$ 
out of $\appsimilarity$ and prove that $\howe{\appsimilarity}$ and 
$\appsimilarity$ coincide (from which it follows that both $\appsimilarity$ 
and $\appbisimilarity$ are substitutive and compatible).
In non-modal calculi, the proof of substitutivity of $\howe{\appsimilarity}$ 
consists of a routine induction, whereas proving that it coincides with 
applicative similarity is more challenging and requires a mixed 
induction-coinduction argument.
Modal calculi present an additional difficulty, as in such calculi also proving 
substitutivity of $\howe{\appsimilarity}$ is nontrivial; and 
in fact, the defining axioms of a comonadic lax extension 
turned out to be precisely what is needed to ensure such a property.

\begin{definition}
Given a closed term relation $\relone$, define its 
Howe extension $\howe{\relone}$ as the least fixed point 
of the mapping $X \mapsto \refine{\open{X}}; \open{\relone}$.
That is, $\howe{\relone}$ is the least term relation satisfying the following inference 
rule \cite{Gordon/FOSSACS/01}.
\[
\infer{
  \wrelo
  {\envone}
  {\howe{\relone}}
  {\termone}
  {\termtwo}
  {\wone}
  {\typeone}
}
{\wrelo
  {\envone}
  {\refine{\howe{\relone}}}
  {\termone}
  {\termthree}
  {\wtwo}
  {\typeone}
  &
  \wrelo
  {\envone}
  {\open{\relone}}
  {\termthree}
  {\termtwo}
  {\wthree}
  {\typeone}
  &
  \wone \wgeq \wtwo \wcomp \wthree
}
\]
\end{definition}

\longv{ 
        It is convenient to give an explicit, syntax-oriented characterisation 
        of the Howe extension of a term relation $\relone$. We do so by means of judgments 
        of the form $\wrelo{\envone}{\howe{\relone}}{\termone}{\termtwo}{\wone}{\typeone}$ 
        (declined, as usual into value and computation judgments) and of the inference 
        rules in Figure~\ref{fig:howe-extension}. 
        Then, two open terms $\envone \imp \termone, \termtwo: \typeone$ are related 
        by $\howe{\relone}$ at world $\wone$ if and only if 
        $\wrelo{\envone}{\howe{\relone}}{\termone}{\termtwo}{\wone}{\typeone}$ is derivable.  
        We also define the relation $\howen{\relone}{n}$, for $n \in \mathbb{N}$, 
        by saying that $\envone \imp \termone, \termtwo: \typeone$ are related 
        by $\howen{\relone}{n}$ at world $\wone$ if and only if 
        $\wrelo{\envone}{\howe{\relone}}{\termone}{\termtwo}{\wone}{\typeone}$ is derivable 
        with a derivation of depth at most $n$. As a consequence, we see that 
        $\howe{\relone} = \bigcup_{n \geq 0} \howen{\relone}{n}$.

        \begin{figure*}[htbp]
        \hrule
         $\vspace{0.2cm}$
        \[
        \infer[(\gradeone \mgeq \munit)]{
          \wrelval
          {\envone, \graded{\varone}{\typeone}{\gradeone}}
          {\howe{\relone}}
          {\varone}
          {\valone}
          {\wone}
          {\typeone}
        }
        { \wrelval
          {\envone, \graded{\varone}{\typeone}{\gradeone}}
          {\open{\relone}}
          {\varone}
          {\valone}
          {\wone}
          {\typeone}}
        \]
        $\vspace{-0.1cm}$
        \[
        \infer{
          \wrelcomp
          {\envone}
          {\howe{\relone}}
          {\valone}
          {\valtwo}
          {\wone}
          {\typeone}
        }
        {
          \wrelval
          {\envone}
          {\howe{\relone}}
          {\valone}
          {\valthree}
          {\wtwo}
          {\typeone}
          &
          \wrelcomp
          {\envone}
          {\open{\relone}}
          {\valthree}
          {\valtwo}
          {\wthree}
          {\typeone}
          &
          \wone \wgeq \wtwo \wcomp \wthree
        }
        \]
        $\vspace{-0.1cm}$
        \[
        \infer{
          \wrelval
          {\envone}
          {\howe{\relone}}
          {\abs{\varone}{\termone}}
          {\valtwo}
          {\wone}
          {\typeone \to \typetwo}
        }
        {
          \wrelcomp
          {\envone, \graded{\varone}{\typeone}{\munit}}
          {\howe{\relone}}
          {\termone}
          {\termtwo}
          {\wtwo}
          {\typetwo}
          & 
          \wrelval
          {\envone}
          {\open{\relone}}
          {\abs{\varone}{\termtwo}}
          {\valone}
          {\wthree}
          {\typeone \to \typetwo}
          & 
          \wone \wgeq \wtwo \wcomp \wthree
        }
        \]
        $\vspace{-0.1cm}$
        \[
        \infer{
          \wrelcomp
          {\envone \mplus \envtwo}
          {\howe{\relone}}
          {\valone \valtwo}
          {\termone}
          {\wone}
          {\typetwo}
        }
        {
          \wrelval
          {\envone}
          {\howe{\relone}}
          {\valone}
          {\valone'}
          {\wtwo}
          {\typeone \to \typetwo}
          &
          \wrelval
          {\envtwo}
          {\howe{\relone}}
          {\valtwo}
          {\valtwo'}
          {\wthree}
          {\typeone}
          & 
          \wrelcomp
          {\envone \mplus \envtwo}
          {\open{\relone}}
          {\valone'\valtwo'}
          {\termone}
          {\wfour}
          {\typetwo} 
          &
          \wone \wgeq \wtwo \wcomp \wthree \wcomp \wfour
        }
        \]
        $\vspace{-0.1cm}$
        \[
        \infer{
          \wrelcomp
          {(\gradeone \mvee \munit) \mstar \envone \mplus \envtwo}
          {\howe{\relone}}
          {\seq{\termone}{\termtwo}}
          {\termthree}
          {\wone}
          {\typetwo}
        }
        {
          \deduce[ ]
          {
          \wrelcomp
          {(\gradeone \mvee \munit) \mstar \envone \mplus \envtwo}
          {\open{\relone}}
          {\seq{\termone'}{\termtwo'}}
          {\termthree}
          {\wfour}
          {\typetwo}
          }
          {\deduce[ ]{
            \wrelcomp
            {\envtwo, \graded{\varone}{\typeone}{\gradeone}}
            {\howe{\relone}}
            {\termtwo}
            {\termtwo'}
            {\wthree}
            {\typetwo}
            }
            {
            \wrelcomp
            {\envone}
            {\corelator{\gradeone \mvee \munit}{\howe{\relone}}}
            {\termone}
            {\termone'}
            {\wtwo}
            {\typeone}
            }}
          &
          \wone \wgeq \wtwo \wcomp \wthree \wcomp \wfour
        }
        \]
        $\vspace{-0.1cm}$
        \[
        \infer{
          \wrelval
          {\envone}
          {\howe{\relone}}
          {\fold{\valone}}
          {\valthree}
          {\wone}
          {\rectype{\typevarone}{\typeone}}
        }
        {
          \wrelval
          {\envone}
          {\howe{\relone}}
          {\valone}
          {\valtwo}
          {\wtwo}
          {\substtype{\typeone}{\typevarone}{\rectype{\typevarone}{\typeone}}}
          &
          \wrelval
          {\envone}
          {\open{\relone}}
          {\fold{\valtwo}}
          {\valthree}
          {\wthree}
          {\rectype{\typevarone}{\typeone}}
          &
          \wone \wgeq \wtwo \wcomp \wthree
        }
        \]
        $\vspace{-0.1cm}$
        \[
        \infer{
          \wrelcomp
          {\gradeone \mstar \envone \mplus \envtwo}
          {\howe{\relone}}
          {\letfold{\valone}{\termone}}
          {\termthree}
          {\wone}
          {\typetwo}
        }
        {
          \deduce[ ]
          {
          \wrelcomp
          {\gradeone \mstar \envone \mplus \envtwo}
          {\open{\relone}}
          {\letfold{\valtwo}{\termtwo}}
          {\termthree}
          {\wfour}
          {\typetwo}
          }
          {
          \deduce[ ]
            {
             \wrelcomp
            {\envtwo, \graded{\varone}
            {\substtype{\typeone}{\typevarone}{\rectype{\typevarone}{\typeone}}}{\gradeone}}
            {\howe{\relone}}
            {\termone}
            {\termtwo}
            {\wthree}
            {\typetwo}
            }
            {
            \wrelval
            {\envone}
            {\corelator{\gradeone}{\howe{\relone}}}
            {\valone}
            {\valtwo}
            {\wtwo}
            {\rectype{\typevarone}{\typeone}}
            }
          }
          &
          \wone \wgeq \wtwo \wcomp \wthree \wcomp \wfour
        }
        \]
        $\vspace{-0.1cm}$
        \[
        \infer{
          \wrelval
          {\gradeone \mstar \envone}
          {\howe{\relone}}
          {\tbox{\valone}}
          {\valthree}
          {\wone}
          {\bbox_{\gradeone}\typeone}
        }
        {
          \wrelval
          {\envone}
          {\corelator{\gradeone}{\howe{\relone}}}
          {\valone}
          {\valtwo}
          {\wtwo}
          {\typeone}
          &
          \wrelval
          {\gradeone \mstar \envone}
          {\open{\relone}}
          {\tbox{\valtwo}}
          {\valthree}
          {\wthree}
          {\bbox_{\gradeone}\typeone}
          &
          \wone \wgeq \wtwo \wcomp \wthree
        }
        \]
        $\vspace{-0.1cm}$
        \[
        \infer{
          \wrelcomp
          {\gradetwo \mstar \envone \mplus \envtwo}
          {\howe{\relone}}
          {\letbox{\valone}{\termone}}
          {\termthree}
          {\wone}
          {\typetwo}
        }
        {
         \deduce[ ]
         {
         \wrelcomp
          {\gradetwo \mstar \envone \mplus \envtwo}
          {\open{\relone}}
          {\letbox{\valtwo}{\termtwo}}
          {\termthree}
          {\wfour}
          {\typetwo}
         }
         { 
         \deduce[ ]
           {
           \wrelcomp
           {\envtwo, \graded{\varone}{\typeone}{\gradetwo \mstar \gradeone}}
           {\howe{\relone}}
           {\termone}
           {\termtwo}
           {\wthree}
           {\typetwo}
           }
           {
           \wrelval
           {\envone}
           {\corelator{\gradetwo}{\howe{\relone}}}
           {\valone}
           {\valtwo}
           {\wtwo}
           {\bbox_{\gradeone}\typeone}
           }
           }
          &
          \wone \wgeq \wtwo \wcomp \wthree \wcomp \wfour
        }
        \]
        \hrule
        \caption{Howe extension of $\relone$}
        \label{fig:howe-extension}
        \end{figure*}

\subsection{Substitutivity and Compatibility}

}

The Howe extension of a term relation enjoys several nice properties, 
which are summarised by the following result.

\longv{
    \begin{lemma}
    Let $\relone$ be a term relation. Then:
    \begin{enumerate}
      \item If $\relone$ is reflexive, then $\howe{\relone}$ 
        is compatible (and thus reflexive). 
      \item If $\relone$ is transitive, then $\howe{\relone}$ is 
        quasi-transitive, i.e. $\open{\relone} \comp \howe{\relone} \subseteq \howe{\relone}$
      \item If $\relone$ is reflexive and transitive, then 
        $\open{\relone} \subseteq \howe{\relone}$.
    \end{enumerate}
    \end{lemma}
}

\shortv{
      \begin{lemma}
      Let $\relone$ be a reflexive and transitive closed term relation. Then
      $\howe{\relone}$ is a compatible term relation such that 
      $\open{\relone} \subseteq \howe{\relone}$ and 
      $\howe{\relone}; \open{\relone} \subseteq \howe{\relone}$.
      \end{lemma}
}
In particular, $\howe{\appsimilarity}$ is a compatible and reflexive 
term relation that extends $\appsimilarity$. 
Our goal now is to prove that $\howe{\appsimilarity}$ is substitutive 
and equal to $\appsimilarity$ (which entails that 
$\appsimilarity$ itself is compatible and substitutive). 
\longv{
    For non-modal calculi, the proof substitutivity of $\howe{(\appsimilarity)}$ 
    consists of a routine induction, whereas proving that the Howe extension 
    of applicative similarity coincides with applicative similarity itself is 
    more challenging and requires a mixed induction-coinduction 
    argument usually referred to as \emph{Key Lemma} \cite{Howe/IC/1996,Pitts/ATBC/2011}. 

    Modal calculi present an additional difficulty, as in such calculi also proving 
    substitutivity of $\howe{(\appsimilarity)}$ is non-trivial, the proof requiring 
    all the defining axioms of a comonadic lax extension. 
    This observation is particularly interesting
    in light of the role played by \emph{lax extension of monads} 
    in calculi with \emph{monadic} 
    effects \cite{DalLagoGavazzoLevy/LICS/2017}. 
}
In order to prove substitutivity of $\howe{(\appsimilarity)}$, we need
the auxiliary notion of a value substitutive term relation.

\begin{definition}
A term relation $\relone$ is \emph{value substitutive} if 
the following rule holds (notice that $\valone$ is a \emph{closed} value).
\[
\infer{\wrelo{\envone}{\relone}
{\subst{\termone}{\varone}{\valone}}{\subst{\termtwo}{\varone}{\valone}}{\wone}
{\typetwo}
}
{\wrelo{\envone, \graded{\varone}{\typeone}{\gradeone}}{\relone}{\termone}{\termtwo}{\wone}{\typetwo}
& 
\emptyenv \valimp \valone: \typeone
}
\]
\end{definition}

Obviously, any substitutive relation is value substitutive.
\longv{
    (for any closed value $\valone$ of type $\typeone$ we have 
    $\wrelt{\relone}{\valone}{\valone}{\wunit}{\typeone}$, and 
      $\wone \wcomp \wunit = \wone$, for any possible world $\wone$)
}
Moreover, since the defining rule of value substitutivity 
involves \emph{closed} values only (so that 
sequential and simultaneous substitution coincide), we see that 
the open extension of a term relation is always value 
substitutive.
We are now ready to prove our substitutivity lemma.

\begin{lemma}[Substitutivity]
\label{lemma:howe-substitutivity}
Let $\relone$ be a reflexive and transitive term relation. 
Then, $\howe{\relone}$ is substitutive.
\end{lemma}
\shortv{
    \begin{proof}[Proof sketch]
    Define the map
    $
    \substarrow: \Lambda^{\envone, \graded{\varone}{\typeone}{\gradeone}}_{\typetwo} 
    \times \values_{\typeone} \to \Lambda^{\envone}_{\typetwo}
    $ by $\substarrow(\termone, \valone) = \subst{\termone}{\varone}{\valone}$.
    We have to prove $\howe{\relone} \tensor \corelator{\gradeone}{\howe{\relone}} 
    \subseteq \substarrow ; \howe{\relone} ; \dual{\substarrow}$.
    Denoting by $\howen{\relone}{n}$ the $n$-th approximation of 
    $\howe{\relone}$, so that $\howe{\relone} = \bigcup_{n \geq 0} \howen{\relone}{n}$, 
    it is sufficient 
    to show 
    $\left(\bigcup_{n \geq 0} \howen{\relone}{n} \right) 
    \tensor \corelator{\gradeone}{\howe{\relone}} 
    \subseteq \substarrow; \howe{\relone}; \dual{\substarrow}$
    which itself follows from
    $
    \bigcup_{n \geq 0} \big(\howen{\relone}{n} 
    \tensor \corelator{\gradeone}{\howe{\relone}} \big) 
    \subseteq \substarrow; \howe{\relone}; \dual{\substarrow}.
    $
    We prove 
    $\forall n \geq 0.\ \howen{\relone}{n} 
    \tensor \corelator{\gradeone}{\howe{\relone}}
    \subseteq \substarrow; \howe{\relone}; \dual{\substarrow}$ 
     by induction on $n$ relying on the properties 
     of $\corelator{\gradeone}{\howe{\relone}}$.
    \end{proof}
}

\longv{    
                \begin{proof}
                We have to prove $\howe{\relone} \tensor \corelator{\gradeone}{\howe{\relone}} 
                \subseteq \substarrow \comp \howe{\relone} \comp \dual{\substarrow}$, 
                i.e. the admissibility of the following rule:
                \[
                \infer{
                  \wrelo{\envone}{\howe{\relone}}{\subst{\termone}{\varone}{\valone}}
                  {\subst{\termtwo}{\varone}{\valtwo}}{\wone}{\typeone}
                }
                {\wrelo{\envone, \graded{\varone}{\typetwo}{\gradeone}}{\howe{\relone}}{\termone}
                {\termtwo}{\wtwo}{\typeone}
                &
                \wrelt{\corelator{\gradeone}{\howe{\relone}}}{\valone}{\valtwo}{\wthree}{\typetwo}
                &
                \wone \wgeq \wtwo \wcomp \wthree}
                \]   
                Since $\howe{\relone} = \bigcup_{n \geq 0} \howen{\relone}{n}$, it is sufficient 
                to prove 
                $\left(\bigcup_{n \geq 0} \howen{\relone}{n} \right) 
                \tensor \corelator{\gradeone}{\howe{\relone}} 
                \subseteq \substarrow \comp \howe{\relone} \comp \dual{\substarrow}$
                which itself follows from
                $$
                \bigcup_{n \geq 0} \big(\howen{\relone}{n} 
                \tensor \corelator{\gradeone}{\howe{\relone}} \big) 
                \subseteq \substarrow \comp \howe{\relone} \comp \dual{\substarrow}.
                $$
                Therefore, to prove substitutivity it is sufficient 
                to show that
                $\howen{\relone}{n} 
                \tensor \corelator{\gradeone}{\howe{\relone}}
                \subseteq \substarrow \comp \howe{\relone} \comp \dual{\substarrow}$ 
                holds, for any $n \in \mathbb{N}$.
                The latter is nothing but the admissibility of the following ($(\mathbb{N}$-) 
                indexed rule(s):
                \[
                \infer{
                  \wrelo{\envone}{\howe{\relone}}{\subst{\termone}{\varone}{\valone}}
                  {\subst{\termtwo}{\varone}{\valtwo}}{\wone}{\typeone}
                }
                {\wrelo{\envone, \graded{\varone}{\typetwo}{\gradeone}}{\howen{\relone}{n}}{\termone}
                {\termtwo}{\wtwo}{\typeone}
                &
                \wrelt{\corelator{\gradeone}{\howe{\relone}}}{\valone}{\valtwo}{\wthree}{\typetwo}
                &
                \wone \wgeq \wtwo \wcomp \wthree}
                \] 
                We proceed by induction on $n$. The case for $n = 0$ is trivial. 
                We prove $\howen{\relone}{n+1} 
                \tensor \corelator{\gradeone}{\howe{\relone}}
                \subseteq \substarrow \comp \howe{\relone} \comp \dual{\substarrow}$, assuming 
                $\howen{\relone}{m} 
                \tensor \corelator{\gradeone}{\howe{\relone}}
                \subseteq \substarrow \comp \howe{\relone} \comp \dual{\substarrow}$ for all $m \leq n$. 
                Notice that, formally, we are universally quantifying over $\gradeone$. 
                \begin{itemize}
                  \item Suppose to be in the following case:
                    \[
                      \infer{
                        \wrelval{\envone}{\howe{\relone}}{\valone}
                        {\subst{\valthree}{\varone}{\valtwo}}{\wthree}{\typeone}
                        }
                      {
                        \infer[(\gradeone \mgeq \munit)]{
                          \wrelval{\envone, \graded{\varone}{\typeone}{\gradeone}}{\howen{\relone}{1}}{\varone}
                          {\valthree}{\wone}{\typeone}
                        }
                        {
                        \wrelval{\envone, \graded{\varone}{\typeone}{\gradeone}}{\open{\relone}}{\varone}
                        {\valthree}{\wone}{\typeone}
                        }
                      &
                      \wrelvalclosed{\corelator{\gradeone}{\howe{\relone}}}{\valone}{\valtwo}{\wtwo}{\typeone}
                      &
                      \wthree \wgeq \wone \wcomp \wtwo
                      }
                    \]
                    Then, since $\open{\relone}$ is value substitutive, 
                    from 
                    $\wrelval{\envone, \graded{\varone}{\typeone}{\gradeone}}{\open{\relone}}{\varone}
                    {\valthree}{\wone}{\typeone}$ 
                    we infer 
                    $\wrelval{\envone}{\open{\relone}}{\valtwo}
                    {\subst{\valthree}{\varone}{\valtwo}}{\wone}{\typeone}$. 
                    Moreover, since $\gradeone \mgeq \munit$ (and thus 
                    $\corelator{\gradeone}{\howe{\relone}} \subseteq \corelator{\munit}{\howe{\relone}} 
                    \subseteq \howe{\relone}$), from 
                    $\wrelvalclosed{\corelator{\gradeone}{\howe{\relone}}}{\valone}{\valtwo}{\wtwo}{\typeone}$
                    we infer 
                    $\wrelvalclosed{\howe{\relone}}{\valone}{\valtwo}{\wtwo}{\typeone}$
                    and thus (by very definition of term relation)
                    $\wrelval{\envone}{\howe{\relone}}{\valone}{\valtwo}{\wtwo}{\typeone}$. 
                    Putting things together, we obtain 
                    $
                    \wrelval{\envone}{(\open{\relone} \comp \howe{\relone})}{\valone}
                    {\subst{\valthree}{\varone}{\valtwo}}{\wthree}{\typeone}$, which 
                    gives $\wrelval{\envone}{\howe{\relone}}{\valone}
                    {\subst{\valthree}{\varone}{\valtwo}}{\wthree}{\typeone}$, thanks 
                    to quasi-transitivity.
                  \item Suppose to be in the following case:
                    \[
                      \infer{
                        \wrelval{\envone, \graded{\vartwo}{\typetwo}{\gradetwo}}{\howe{\relone}}{\vartwo}
                        {\subst{\valthree}{\varone}{\valtwo}}{\wthree}{\typetwo}
                        }
                      {
                        \infer[(\gradetwo \mgeq \munit)]{
                          \wrelval{\envone, \graded{\varone}{\typeone}{\gradeone}, 
                          \graded{\vartwo}{\typetwo}{\gradetwo}}{\howen{\relone}{1}}{\vartwo}
                          {\valthree}{\wone}{\typetwo}
                        }
                        {
                        \wrelval{\envone, \graded{\varone}{\typeone}{\gradeone}, 
                        \graded{\vartwo}{\typetwo}{\gradetwo}}{\open{\relone}}{\vartwo}
                        {\valthree}{\wone}{\typetwo}
                        }
                      &
                      \wrelvalclosed{\corelator{\gradeone}{\howe{\relone}}}{\valone}{\valtwo}{\wtwo}{\typeone}
                      &
                      \wthree \wgeq \wone \wcomp \wtwo
                      }
                    \]
                    Then, since $\open{\relone}$ is value substitutive, 
                    from 
                    $\wrelval{\envone, \graded{\varone}{\typeone}{\gradeone}, 
                            \graded{\vartwo}{\typetwo}{\gradetwo}}{\open{\relone}}{\vartwo}
                            {\valthree}{\wone}{\typetwo}$
                    we infer 
                    $
                    \wrelval{\envone, \graded{\varone}{\typeone}{\gradeone}}{\open{\relone}}{\vartwo}
                        {\subst{\valthree}{\varone}{\valtwo}}{\wone}{\typetwo}$
                    and thus 
                    $
                    \wrelval{\envone, \graded{\varone}{\typeone}{\gradeone}}{\howe{\relone}}{\vartwo}
                        {\subst{\valthree}{\varone}{\valtwo}}{\wone}{\typetwo}$. 
                    We conclude the thesis, since $\wone \wleq \wone \wcomp \wtwo \wleq \wthree$.
                  \item Suppose to be in the following case:
                    \[
                    \infer{
                    \wrelcomp
                    {(\gradetwo \mvee \munit) \mstar \envone \mplus \envtwo}
                    {\howe{\relone}}
                    {\seqy{\subst{\termone}{\varone}{\valone}}{\subst{\termtwo}{\varone}{\valone}}}
                    {\subst{\termthree}{\varone}{\valtwo}}
                    {\wfive}
                    }
                    {
                    \deduce[\mathcal{D}\vspace{0.3cm}]
                    {
                    \wrelcomp
                      {(\gradetwo \mvee \munit) \mstar \envone \mplus \envtwo, 
                        \graded{\varone}{\typeone}{(\gradetwo \mvee \munit) \mstar \gradeone \mplus \gradethree}}
                      {\howen{\relone}{n+1}}
                      {\seqy{\termone}{\termtwo}}
                      {\termthree}
                      {\wone}
                      {\typethree}
                    }
                    {}
                    &
                    \wrelvalclosed
                    {\corelator{(\gradetwo \mvee \munit) \mstar \gradeone \mplus \gradethree}{\howe{\relone}}}
                    {\valone}
                    {\valtwo}
                    {\wsix}
                    {\typeone}
                    &
                    \wfive \wgeq \wone \wcomp \wsix
                    }
                    \]
                    where $\mathcal{D}$ is the following derivation
                    \[
                    \infer{
                      \wrelcomp
                      {(\gradetwo \mvee \munit) \mstar \envone \mplus \envtwo, 
                        \graded{\varone}{\typeone}{(\gradetwo \mvee \munit) \mstar \gradeone \mplus \gradethree}}
                      {\howen{\relone}{n+1}}
                      {\seqy{\termone}{\termtwo}}
                      {\termthree}
                      {\wone}
                      {\typethree}
                    }
                    { \deduce[]
                      {
                      \wrelcomp
                      {(\gradetwo \mvee \munit) \mstar (\envone, \graded{\varone}{\typeone}{\gradeone}) 
                      \mplus (\envtwo, \graded{\varone}{\typeone}{\gradethree})}
                      {\open{\relone}}
                      {\seqy{\termone'}{\termtwo'}}
                      {\termthree}
                      {\wfour}
                      {\typethree}
                      }
                      {
                      \deduce[]
                      {\wrelcomp
                      {\envtwo,\graded{\varone}{\typeone}{\gradethree}, 
                       \graded{\vartwo}{\typetwo}{\gradetwo}}
                      {\howen{\relone}{n}}
                      {\termtwo}
                      {\termtwo'}
                      {\wthree}
                      {\typethree}
                      }
                      {
                      \wrelcomp
                      {\envone,\graded{\varone}{\typeone}{\gradeone}}
                      {\corelator{\gradetwo \mvee \munit}{\howen{\relone}{n}}}
                      {\termone}
                      {\termone'}
                      {\wtwo}
                      {\typetwo}
                      }
                      }
                      & 
                      \wone \wgeq \wtwo \wcomp \wthree \wcomp \wfour
                    }
                \]
                By law \eqref{eq:monoidal-2}, from 
                $\wrelvalclosed
                {\corelator{(\gradetwo \mvee \munit) \mstar \gradeone \mplus \gradethree}{\howe{\relone}}}
                {\valone}
                {\valtwo}
                {\wsix}
                {\typeone}$ we obtain:
                \begin{align}
                &\wrelvalclosed
                {\corelator{(\gradetwo \mvee \munit) \mstar \gradeone}{\howe{\relone}}}
                {\valone}
                {\valtwo}
                {\wsix_1}
                {\typeone}
                \label{auxiliary-subst-seq-1}
                \\
                &\wrelvalclosed
                {\corelator{\gradethree}{\howe{\relone}}}
                {\valone}
                {\valtwo}
                {\wsix_2}
                {\typeone}
                \label{auxiliary-subst-seq-2}
                \\
                &\wsix \wgeq \wsix_1 \wcomp \wsix_2.
                \end{align}
                Moreover, from \eqref{auxiliary-subst-seq-1} we infer 
                $\wrelvalclosed
                {\corelator{(\gradetwo \mvee \munit)}{\corelator{\gradeone}{\howe{\relone}}}}
                {\valone}
                {\valtwo}
                {\wsix_1}
                {\typeone}$ 
                by law \eqref{eq:comonad-2}. We next apply the induction hypothesis, obtaining
                $$
                \howen{\relone}{n} 
                \tensor \corelator{\gradeone}{\howe{\relone}}
                \subseteq \substarrow \comp \howe{\relone} \comp \dual{\substarrow}.
                $$
                which in turn gives
                $$
                \corelator{\gradetwo \mvee \munit}{\howen{\relone}{n} 
                \tensor \corelator{\gradeone}{\howe{\relone}}}
                \subseteq \substarrow \comp \corelator{\gradetwo \mvee \munit}{\howe{\relone}} \comp \dual{\substarrow}
                $$
                by stability (Lemma~\ref{lemma:stability}). 
                Finally, we use law \eqref{eq:monoidal-1} and obtain
                $$
                \corelator{\gradetwo \mvee \munit}{\howen{\relone}{n}}
                \tensor \corelator{\gradetwo \mvee \munit}{\corelator{\gradeone}{\howe{\relone}}}
                \subseteq
                \corelator{\gradetwo \mvee \munit}{\howen{\relone}{n} 
                \tensor \corelator{\gradeone}{\howe{\relone}}}
                \subseteq 
                \substarrow \comp \corelator{\gradetwo \mvee \munit}{\howe{\relone}} \comp \dual{\substarrow}.
                $$
                From the above inclusion, 
                $\wrelcomp
                {\envone,\graded{\varone}{\typeone}{\gradeone}}
                {\corelator{\gradetwo \mvee \munit}{\howen{\relone}{n}}}
                {\termone}
                {\termone'}
                {\wtwo}
                {\typetwo}$, and 
                $\wrelvalclosed
                {\corelator{(\gradetwo \mvee \munit)}{\corelator{\gradeone}{\howe{\relone}}}}
                {\valone}
                {\valtwo}
                {\wsix_1}
                {\typeone}$ 
                we infer 
                $$
                \wrelcomp
                {\envone}
                {\corelator{\gradetwo \mvee \munit}{\howe{\relone}}}
                {\subst{\termone}{\varone}{\valone}}
                {\subst{\termone'}{\varone}{\valtwo}}
                {\wtwo \wcomp \wsix_1}
                {\typetwo}.
                $$
                We then apply the induction hypothesis on \eqref{auxiliary-subst-seq-2}
                and 
                $\wrelcomp
                {\envtwo,\graded{\varone}{\typeone}{\gradethree}, 
                \graded{\vartwo}{\typetwo}{\gradetwo}}
                {\howen{\relone}{n}}
                {\termtwo}
                {\termtwo'}
                {\wthree}
                {\typethree}$, hence inferring 
                $$
                \wrelcomp
                {\envtwo,\graded{\vartwo}{\typetwo}{\gradetwo}}
                {\howe{\relone}}
                {\subst{\termtwo}{\varone}{\valone}}
                {\subst{\termtwo'}{\varone}{\valtwo}}
                {\wthree \wcomp \wsix_2}
                {\typethree}.
                $$
                Finally, since $\open{\relone}$ is value substitutive
                $\wrelcomp
                {(\gradetwo \mvee \munit) \mstar \envone \mplus \envtwo, 
                \graded{\varone}{\typeone}{(\gradetwo \mvee \munit) \mstar \gradeone \mplus \gradethree}}
                {\open{\relone}}
                {\seqy{\termone'}{\termtwo'}}
                {\termthree}
                {\wfour}
                {\typethree}$ 
                implies 
                $\wrelcomp
                {(\gradetwo \mvee \munit) \mstar \envone \mplus \envtwo}
                {\howen{\relone}{n+1}}
                {\seqy{\subst{\termone}{\varone}{\valone}}{\subst{\termtwo}{\varone}{\valone}}}
                {\subst{\termthree}{\varone}{\valthree}}
                {\wthree \wcomp \wsix_2 \wcomp \wtwo \wcomp \wsix_1 \wcomp \wfour}
                {\typethree}$, 
                so that we can conclude the thesis by the very definition of Howe extension 
                of a relation as follows:
                    \[
                    \infer{
                      \wrelcomp
                      {(\gradetwo \mvee \munit) \mstar \envone \mplus \envtwo}
                      {\howen{\relone}{n+1}}
                      {\seqy{\subst{\termone}{\varone}{\valone}}{\subst{\termtwo}{\varone}{\valone}}}
                      {\subst{\termthree}{\varone}{\valthree}}
                      {\wone}
                      {\typethree}
                    }
                    { \deduce[]
                      {
                      \wrelcomp
                      {(\gradetwo \mvee \munit) \mstar \envone \mplus \envtwo}
                      {\open{\relone}}
                      {\seqy{\subst{\termone'}{\varone}{\valtwo}}{\subst{\termtwo'}{\varone}{\valtwo}}}
                      {\subst{\termthree}{\varone}{\valthree}}
                      {\wfour}
                      {\typethree}
                      }
                      {
                      \deduce[]
                      {\wrelcomp
                      {\envtwo,\graded{\vartwo}{\typetwo}{\gradetwo}}
                      {\howe{\relone}}
                      {\subst{\termtwo}{\varone}{\valone}}
                      {\subst{\termtwo'}{\varone}{\valtwo}}
                      {\wthree \wcomp \wsix_2}
                      {\typethree}
                      }
                      {
                      \wrelcomp
                      {\envone}
                      {\corelator{\gradetwo \mvee \munit}{\howe{\relone}}}
                      {\subst{\termone}{\varone}{\valone}}
                      {\subst{\termone'}{\varone}{\valtwo}}
                      {\wtwo \wcomp \wsix_1}
                      {\typetwo}
                      }
                      }
                      &
                      \wone \wgeq \wthree \wcomp \wsix_2 \wcomp \wtwo \wcomp \wsix_1 \wcomp \wfour
                    }
                \]
                \item Suppose to be in the following case:
                  \[
                  \infer{
                  \wrelval
                  {\gradetwo \mstar \envone}
                  {\howe{\relone}}
                  {\tbox{\subst{\valthree}{\varone}{\valone}}}
                  {\subst{\valfour}{\varone}{\valtwo}}
                  {\wone}
                  {\bbox_{\gradetwo}\typetwo}
                  }
                  {
                  \infer
                  {
                  \wrelval
                    {\gradetwo \mstar \envone, \graded{\varone}{\typeone}{\gradetwo \mstar \gradeone}}
                    {\howen{\relone}{n+1}}
                    {\tbox{\valthree}}
                    {\valfour}
                    {\wfour}
                    {\bbox_{\gradetwo}\typetwo}
                  }
                  {
                    \deduce[ ]
                    {\wrelval
                    {\gradetwo \mstar \envone, \graded{\varone}{\typeone}{\gradetwo \mstar \gradeone}}
                    {\open{\relone}}
                    {\tbox{\valthree'}}
                    {\valfour}
                    {\wtwo}
                    {\bbox_{\gradetwo}\typetwo}
                  }
                  {
                    \wrelval
                    {\envone, \graded{\varone}{\typeone}{\gradeone}}
                    {\corelator{\gradetwo}{\howen{\relone}{n}}}
                    {\valthree}
                    {\valthree'}
                    {\wthree}
                    {\typetwo}
                  }
                  & 
                  \wfour \wgeq \wtwo \wcomp \wthree
                  }
                  &
                  \wrelvalclosed
                  {\corelator{\gradetwo \mstar \gradethree}{\howe{\relone}}}
                  {\valone}
                  {\valtwo}
                  {\wfive}
                  {\typeone}
                  &
                  \wone \wgeq \wfive \wcomp \wfour
                  }
                \]
                We proceed as in previous case. The main passages are
                summarised in the following chain of implication:
                \begin{align*}
                IH 
                &\implies 
                \howen{\relone}{n} \tensor \corelator{\gradeone}{\howe{\relone}} 
                \subseteq \substarrow \comp \howe{\relone} \comp \dual{\substarrow}
                \\
                &\implies 
                \corelator{\gradetwo}{\howen{\relone}{n} \tensor \corelator{\gradeone}{\howe{\relone}}}
                \subseteq \substarrow \comp \corelator{\gradetwo}{\howe{\relone}} \comp \dual{\substarrow}
                & & 
                \text{(By Lemma~\ref{lemma:stability})}
                \\
                &\implies 
                \corelator{\gradetwo}{\howen{\relone}{n}} \tensor \corelator{\gradetwo}{\corelator{\gradeone}{\howe{\relone}}}
                \subseteq \substarrow \comp \corelator{\gradetwo}{\howe{\relone}} \comp \dual{\substarrow}
                & & 
                \text{(By law \eqref{eq:monoidal-1})}
                \\
                &\implies 
                \corelator{\gradetwo}{\howen{\relone}{n}} \tensor \corelator{\gradetwo \mstar \gradeone}{\howe{\relone}}
                \subseteq \substarrow \comp \corelator{\gradetwo}{\howe{\relone}} \comp \dual{\substarrow}
                & & 
                \text{(By law \eqref{eq:comonad-2})}
                \end{align*}
                As a consequence, 
                $
                \wrelval
                {\envone, \graded{\varone}{\typeone}{\gradeone}}
                {\corelator{\gradetwo}{\howen{\relone}{n}}}
                {\valthree}
                {\valthree'}
                {\wthree}
                {\typetwo}
                $
                and 
                $
                \wrelvalclosed
                {\corelator{\gradetwo \mstar \gradethree}{\howe{\relone}}}
                {\valone}
                {\valtwo}
                {\wfive}
                {\typeone}
                $
                implies 
                $$
                \wrelval
                {\envone}
                {\corelator{\gradetwo}{\howen{\relone}{n}}}
                {\subst{\valthree}{\varone}{\valone}}
                {\subst{\valthree'}{\varone}{\valtwo}}
                {\wthree \wcomp \wfive}
                {\typetwo}.
                $$
                Moreover, since $\open{\relone}$ is value substitutive, 
                $
                \wrelval
                {\gradetwo \mstar \envone, \graded{\varone}{\typeone}{\gradetwo \mstar \gradeone}}
                {\open{\relone}}
                {\tbox{\valthree'}}
                {\valfour}
                {\wtwo}
                {\bbox_{\gradetwo}\typetwo}
                $
                entails 
                $$
                \wrelval
                {\gradetwo \mstar \envone}
                {\open{\relone}}
                {\tbox{\subst{\valthree'}{\varone}{\valtwo}}}
                {\subst{\valfour}{\varone}{\valtwo}}
                {\wtwo}
                {\bbox_{\gradetwo}\typetwo},
                $$
                from which the thesis follows by very definition of $\howe{\relone}$.
                \end{itemize}
                The remaining cases follow the same pattern of the one seen so far, or are even easier.
                \end{proof}
        
        }

\longv{
Now that we know that the Howe extension of applicative (bi)similarity is compatible 
and substitutive, it remains to prove that it coincides with applicative (bi)similarity 
itself. 
To do so, we first of all notice that it is enough to focus on 
\emph{closed} expressions only. 

      \begin{lemma} 
      If the closed projection of $\howe{(\appsimilarity)}$ is an applicative 
      simulation, then $\howe{(\appsimilarity)}$ coincides with $\open{(\appsimilarity)}$. 
      \end{lemma}

      \begin{proof}
      We already know that $\open{(\appsimilarity)} \subseteq {\howe{(\appsimilarity)}}$, 
      so that it is enough to prove the converse inclusion. 
      First, notice that since $\closed{(\howe{(\appsimilarity)})}$ is an applicative simulation, 
      $\closed{({\howe{(\appsimilarity)}})}$ is contained in ${\appsimilarity}$, and 
      thus $\open{(\closed{({\howe{(\appsimilarity)}})})}$ is contained in 
      $\open{(\appsimilarity)}$. 
      We are done since ${\howe{(\appsimilarity)}} \subseteq 
      {\open{(\closed{({\howe{(\appsimilarity)}})})}}$.
      \end{proof}

      Therefore, what we need to prove is that the closed projection of 
      $\howe{(\appsimilarity)}$ is an applicative simulation. 

          First, let us observe that the value clauses of 
          Definition~\ref{definition:modal-applicative-bisimulation} 
          are satisfied $\closed{(\howe{(\appsimilarity)})}$. 
          In the remaining part of this section, to improve readability we will write 
          $\closedhowe{\relone}$ for the closed projection of 
          the Howe extension of $\relone$.

          \begin{lemma}
          \label{lemma:howe-extension-values}
          Let $\relone$ be a reflexive and transitive applicative simulation. 
          Then, $\closed{(\howe{\relone})}$ satisfies clauses 
          \eqref{eq:app-abs}, \eqref{eq:app-box}, and \eqref{eq:app-fold}.
          \end{lemma}

          \begin{proof}
          The proof is straightforward, and
          thus we just show the case of clause \eqref{eq:app-box} as an illustrative 
          example. Notice that since we deal with 
          closed relations, if $\wrelt{\closedhowe{\relone}}{\valone}{\valtwo}{\wone}{\typeone}$, 
          then $\valone$ and $\valtwo$ have the same syntactic structure, which is determined 
          by $\typeone$ (for instance, if $\typeone = \bbox_{\gradeone}\typetwo$, then 
          $\valone$ and $\valtwo$ must be two boxed values). 
          So suppose to have the following derivation:
          \[
          \infer{
            \wrelvalclosed
            {(\closedhowe{\relone})}
            {\tbox{\valone}}
            {\tbox{\valthree}}
            {\wone}
            {\bbox_{\gradeone}\typeone}
          }
          {
            \wrelvalclosed  
            {\corelator{\gradeone}{(\closedhowe{\relone})}}
            {\valone}
            {\valtwo}
            {\wtwo}
            {\typeone}
            &
            \wrelvalclosed
            {\relone}
            {\tbox{\valtwo}}
            {\tbox{\valthree}}
            {\wthree}
            {\bbox_{\gradeone}\typeone}
            &
            \wone \wgeq \wtwo \wcomp \wthree
          }
          \]
          Since $\relone$ is an applicative simulation, 
          $\wrelvalclosed
            {\relone}
            {\tbox{\valtwo}}
            {\tbox{\valthree}}
            {\wthree}
            {\bbox_{\gradeone}\typeone}
          $
          implies 
          $\wrelvalclosed
            {\corelator{\gradeone}{\relone}}
            {\valtwo}
            {\valthree}
            {\wthree}
            {\typeone}
          $, which, together with 
          $\wrelvalclosed  
            {\corelator{\gradeone}{(\closedhowe{\relone})}}
            {\valone}
            {\valtwo}
            {\wtwo}
            {\typeone}
          $ and $\wone \wgeq \wtwo \wcomp \wthree$ gives
          $
            \wrelvalclosed
            {(\corelator{\gradeone}{\relone} \comp \corelator{\gradeone}{(\closedhowe{\relone})})}
            {\valone}
            {\valthree}
            {\wone}
            {\typeone}
          $
          and thus the desired thesis by quasi-transitivity.
          \end{proof}
}
\shortv{
Now that we know that the Howe extension of applicative (bi)similarity is compatible 
and substitutive, it remains to prove that it coincides with applicative (bi)similarity 
itself. 
    This is the content of the so-called Key Lemma, which states that 
    if $\relone$ is a reflexive and transitive applicative simulation, 
    then $\howe{\relone}$ (restricted to closed terms) 
    is an applicative simulation too (and thus, by coinduction, it is 
    included in $\appsimilarity$).
Our proof of the key lemma follows the abstract Howe's method 
by Dal Lago et al. \cite{DalLagoGavazzoLevy/LICS/2017}.
Let us sketch how it goes. 
The crux of the argument 
is showing that $\howe{\relone}$ satisfies the first clause 
in Definition~\ref{definition:modal-applicative-bisimulation}. 
Assuming $\termone \howe{\relone}(\wone) \termtwo$, 
we proceed by induction on the evaluation of $\termone$ 
with a case analysis on the shape of $\termone$.
The difficult case 
is given by sequencing, where one sees that for the proof to go through,
the underlying comonadic lax extensions need to satisfy the (lax distributive\footnote{
  Notice that the map $(-)_{\bot}$ defines a lax extension of the functor 
  $M(X) = X_{\bot}$. Since each $\corelatorsymbol_{\gradeone}$ is, in particular, 
  a lax extension of the identity functor $1_{\wrelcat}$, the identity function 
  gives a candidate distributive law of type $1_{\wrelcat} M \Rightarrow 
  M 1_{\wrelcat}$. The inclusion $\corelator{\gradeone \vee \gunit}
  {\relone_\bot}
  \subseteq
  (\corelator{\gradeone \vee \gunit}{\relone})_\bot$ 
  then essentially states that, for any $\gradeone \geq \gunit$, the 
  identity map gives a \emph{lax natural transformation} 
  and thus a lax distributive law of type $1_{\wrelcat} M \Rightarrow 
  M 1_{\wrelcat}$.
}) law 
$\corelator{\gradeone \vee \gunit}
{\relone_\bot}
\subseteq
(\corelator{\gradeone \vee \gunit}{\relone})_\bot$, 
which now becomes part of the hypothesis of the Key Lemma. 
Notice that  $\bang$, $\bbox$, and $\mask{}{}$ all satisfy 
the desired law.

\begin{lemma}[Key Lemma]
\label{lemma:key-lemma-short}
Assume the law $\corelator{\gradeone \vee \gunit}
{\relone_\bot}
\subseteq
(\corelator{\gradeone \vee \gunit}{\relone})_\bot$, for any $\relone$.
Then, for any reflexive and transitive 
applicative simulation $\relone$, $\howe{\relone}$ 
(restricted to closed terms) is an applicative 
simulation.
\end{lemma}

\longv{
        \begin{definition}
        \label{definition:relator-corelator-distributive-law}
        Given a lax extension $\relatorsymbol$ of a monad $(\monad, \unit, \mu)$ 
        and a comonadic lax extension $\corelatorsymbol$, 
        we say that 
        $\corelatorsymbol$ \emph{distributes laxly} over $\relatorsymbol$ 
        if the following inclusion holds for 
        any $\gradeone \mgeq \munit$: 
        \begin{align*}
        \corelator{\gradeone}
        {\relatorsymbol(\relone)}
        \subseteq
        \relatorsymbol(\corelator{\gradeone}{\relone}).
        \tag{dist.} \label{eq:distributive-lax}
        \end{align*}
        \end{definition}
}
 \begin{theorem}
\label{theorem:congruence}
Both applicative similarity and applicative bisimilarity are 
substitutive and compatible term relations. 
\end{theorem}



 


We conclude this section by noticing that
we obtain \emph{Metric Preservation} \cite{Pierce/DistanceMakesTypesGrowStronger/2010}, 
\emph{Non-interference} 
\cite{DBLP:conf/popl/AbadiBHR99}, 
as well as other similar results (such as 
\emph{Proof Irrelevance} \cite{DBLP:conf/lics/Pfenning01})
as immediate corollaries of 
Theorem~\ref{theorem:congruence}.

\begin{corollary}
  \begin{varenumerate}
    \item  
     For any term 
     $\graded{\varone}{\typeone}{\public}, \graded{\vartwo}{\typetwo}{\secret} 
    \imp \termone: \typethree$ and all values 
    $\valone$, $\valtwo$, $\valtwo'$ (of the right type),
    we have $\wrelt{\appbisimilarity}{\termone[\valone, \valtwo/\varone, \vartwo]}
    {\termone[\valone, \valtwo'/\varone, \vartwo]}{\public}{\typethree}$. 
    That is, users with low security permissions cannot observe changes in 
    classified values. 
    \item
    For any term $\graded{\varone_1}{\typeone_1}{\gradeone_1}, \hh, 
\graded{\varone_n}{\typeone_n}{\gradeone_n} \imp \termone: \typeone$ and values 
    $\vect{\valone} = \valone_1, \hh, \valone_n$, 
    $\vect{\valtwo} = \valtwo_1, \hh,\valtwo_n$ (of the right type),
    if $\valone_i \mathbin{{\appbisimilarity}{(\gradethree_i)}} \valtwo_i$ ($\forall i \leq n$), 
    then 
    $\termone[\vect{\valone}/ \vect{\varone}] 
    \mathbin{{\appbisimilarity}{(\sum_{i \leq n} \gradeone_i \cdot \gradethree_i)}}
    \termone[\vect{\valtwo}/ \vect{\varone}]$. 
    That is, terms behave as Lipschitz-continuous functions, with 
    Lipschitz constant given by grades and determined by typing.
\end{varenumerate}
\end{corollary}


}

\longv{
        To conclude our argument, we need to show that $\closedhowe{\relone}$ 
        satisfies clause \eqref{eq:app-eval}, where $\relone$ is a reflexive 
        and transitive applicative simulation. That essentially amounts to 
        show the inclusion
        $$
        \closedhowe{\relone} \subseteq 
        \evalsymbol \comp \relatorsymbol^{\divergence}(\closedhowe{\relone}) \comp \dual{\evalsymbol}.
        $$
        We will do that by case analysis on $\closedhowe{\relone}$. However, 
        since $\evalsymbol$ is defined as $\lub_{n \geq 0} \evalsymboln{n}$, 
        we would also like to reason inductively in terms of $\evalsymboln{n}$. 
        We can do so by observing that $\relatorsymbol^{\divergence}$ supports the following 
        reasoning principles:
        \begin{align*}
        &\wrel{\relatorsymbol^{\divergence}(\relone)}{\divergence}{y}{\wone}
        \tag{Induction 1}\label{eq:induction-1}
        \\
        (\forall n \geq 0.\ 
        \wrel{\relatorsymbol^{\divergence}(\relone)}{x_n}{y}{\wone})
        &\implies 
        \wrel{\relatorsymbol^{\divergence}(\relone)}{\lub_{n \geq 0} x_n}{y}{\wone}.
        \tag{Induction 2}\label{eq:induction-2}
        \end{align*}

        As a consequence, to prove that $\closedhowe{\relone}$ satisfies 
        clause \eqref{eq:app-eval} it is enough to show the following statement:
        $$
        \forall n \geq 0.\ 
        \closedhowe{\relone} \subseteq 
        \evalsymboln{n} \comp \relatorsymbol^{\divergence}(\closedhowe{\relone}) \comp \dual{\evalsymbol}.
        $$
        We proceed by induction on $n$. The case for $n = 0$ directly follows from 
        law \eqref{eq:induction-1}. For the inductive step, we proceed 
        by cases on the definition of $\closedhowe{\relone}$. Most cases are 
        standard, but we encounter a further difficulty in the case of sequencing. 
        Suppose to have:
        \[
        \infer{
          \wrelcompclosed
          {(\closedhowe{\relone})}
          {\seq{\termone}{\termtwo}}
          {\termthree}
          {\wone}
          {\typetwo}
        }
        {
          \wrelcompclosed
          {\corelator{\gradeone \mvee \munit}{(\closedhowe{\relone})}}
          {\termone}
          {\termone'}
          {\wtwo}
          {\typeone}
          &
          \wrelcomp
          {\graded{\varone}{\typeone}{\gradeone}}
          {(\closedhowe{\relone})}
          {\termtwo}
          {\termtwo'}
          {\wthree}
          {\typetwo}
          &
          \wrelcompclosed
          {\relone}
          {\seq{\termone'}{\termtwo'}}
          {\termthree}
          {\wfour}
          {\typetwo}
          &
          \wone \wgeq \wtwo \wcomp \wthree \wcomp \wfour
        }
        \]
        We have to prove 
        $\wrelt
          {\relatorsymbol^{\divergence}(\closedhowe{\relone})}
          {\evaln{n+1}{\seq{\termone}{\termtwo}}}
          {\eval{\termthree}}
          {\wone}
          {\typetwo}.
        $
        Since 
        $$
        \evaln{n+1}{\seq{\termone}{\termtwo}} = 
        \kleisli{(\valone \mapsto \evaln{n}{\subst{\termtwo}{\varone}{\valone}})}(\evaln{n}{\termone})
        $$
        we already see that we may want to rely one lax \eqref{eq:lax-monad-bind}. 
        But let us proceed by step by step. First, by induction hypothesis 
        we have 
        $\closedhowe{\relone} 
        \subseteq 
        \evalsymboln{n} \comp 
        \relatorsymbol^{\divergence}(\closedhowe{\relone}) \comp \dual{\evalsymbol}$ 
        and thus, by stability (Lemma~\ref{lemma:stability}), 
        $$
        \corelator{\gradeone \mvee \munit}{\closedhowe{\relone}}
        \subseteq 
        \evalsymboln{n} \comp 
        \corelator{\gradeone \mvee \munit}
        {\relatorsymbol^{\divergence}(\closedhowe{\relone})} 
        \comp \dual{\evalsymbol}.
        $$
        As a consequence, from
        $\wrelcompclosed
          {\corelator{\gradeone \mvee \munit}{\closedhowe{\relone}}}
          {\termone}
          {\termone'}
          {\wtwo}
          {\typeone}$ 
        we infer 
        $\wrel
          {\corelator{\gradeone \mvee \munit}
          {\relatorsymbol^{\divergence}(\closedhowe{\relone})}}
          {\semn{\termone}{n}}
          {\sem{\termone'}}
          {\wtwo}
        $. 
        Let us now move 
        $\wrelcomp
        {\graded{\varone}{\typeone}{\gradeone}}
        {\closedhowe{\relone}}
        {\termtwo}
        {\termtwo'}
        {\wthree}
        {\typetwo}$. 
        Let write $\hat{\termtwo}$, $\hat{\termtwo'}$ 
        for the maps mapping a closed value $\valone$ 
        of type $\typeone$ to 
        $\subst{\termtwo}{\varone}{\valone}$ 
        and $\subst{\termtwo'}{\varone}{\valone}$, respectively.
        By substitutivity of $\closedhowe{\relone}$ and the 
        induction hypothesis
        we obtain the following lax commutative diagram
        \[
        \xymatrix@C=1.5cm{
        \laxcommuterel
        \values_{\typeone}
        \ar[r]^-{\hat{\termtwo}}
        \ar[d]_{\corelator{\gradeone}{\closedhowe{\relone}}}|@{|}  
        & \Lambda_{\typetwo}  
        \ar[d]^{\closedhowe{\relone}}|@{|}
        \ar[r]^-{\semn{-}{n}}
        & 
        (\values_{\typetwo})_{\divergence}
        \ar[d]^{\relatorsymbol^{\divergence}(\closedhowe{\relone})}|@{|}
        \\
        \values_{\typeone}
        \ar[r]_-{\hat{\termtwo'}}   
        & \Lambda_{\typetwo}  
        \ar[r]^-{\sem{-}}
        & (\values_{\typetwo})_{\divergence} 
        }
        \]
        which, by law \eqref{eq:contravariance}, gives:
        \[
        \xymatrix@C=1.5cm{
        \laxcommuterel
        \values_{\typeone}
        \ar[r]^-{\hat{\termtwo}}
        \ar[d]_{\corelator{\gradeone \mvee \munit}{\closedhowe{\relone}}}|@{|}  
        & \Lambda_{\typetwo}  
        \ar[d]^{\closedhowe{\relone}}|@{|}
        \ar[r]^-{\semn{-}{n}}
        & 
        (\values_{\typetwo})_{\divergence}
        \ar[d]^{\relatorsymbol^{\divergence}(\closedhowe{\relone})}|@{|}
        \\
        \values_{\typeone}
        \ar[r]_-{\hat{\termtwo'}}   
        & \Lambda_{\typetwo}  
        \ar[r]^-{\sem{-}}
        & (\values_{\typetwo})_{\divergence} 
        }
        \]
        Next, we now apply law \eqref{eq:lax-monad-bind}, obtaining:
        \[
        \xymatrix@C=2cm{
        \laxcommuterel
        (\values_{\typeone})_{\divergence}
        \ar[r]^-{\kleisli{(\semn{-}{n} \comp \hat{\termtwo})}}
        \ar[d]_{\relatorsymbol^{\divergence}
        (\corelator{\gradeone \mvee \munit}{\closedhowe{\relone}})}|@{|}  
        & 
        (\values_{\typetwo})_{\divergence}
        \ar[d]^{\relatorsymbol^{\divergence}(\closedhowe{\relone})}|@{|}
        \\
        (\values_{\typeone})_{\divergence}
        \ar[r]_-{\kleisli{(\sem{-} \comp \hat{\termtwo'})}}
        & (\values_{\typetwo})_{\divergence} 
        }
        \]
        At this point we may conclude the thesis,\footnote{
          Actually, we would conclude 
          $\wrel{\relatorsymbol^{\divergence}(\closedhowe{\relone})}
          {\semn{\seq{\termone}{\termtwo}}{n+1}} 
          {\sem{\seq{\termone'}{\termtwo'}}}
          {\wtwo \wcomp \wthree}$
          from which we can then infer the thesis, since we have 
          $\wrelt{\relone}{\seq{\termone'}{\termtwo'}}{\termthree}{\wfour}{\typetwo}$
          (which, $\relone$ being an applicative simulation, entails 
          $\wrel{\relatorsymbol^{\divergence}(\relone)}
          {\sem{\seq{\termone'}{\termtwo'}}}
          {\sem{\termthree}}{\wfour}$) and by quasi-transitivity 
          $\relatorsymbol^{\divergence}(\relone) \comp 
          \relatorsymbol^{\divergence}(\closedhowe{\relone}) 
          \subseteq \relatorsymbol^{\divergence}(\closedhowe{\relone})$.
        } provided 
        that we can infer 
        $\wrel
        {\relatorsymbol^{\divergence}(\corelator{\gradeone \mvee \munit}{\closedhowe{\relone})}}
        {\semn{\termone}{n}}
        {\sem{\termone'}}
        {\wtwo}
        $
        from 
        $\wrel
        {\corelator{\gradeone \mvee \munit}
        {\relatorsymbol^{\divergence}(\closedhowe{\relone})}}
        {\semn{\termone}{n}}
        {\sem{\termone'}}
        {\wtwo}
        $; that is, provided that we have the inclusion
        $$
        \corelator{\gradeone \mvee \munit}
        {\relatorsymbol^{\divergence}(\closedhowe{\relone})}
        \subseteq
        \relatorsymbol^{\divergence}(\corelator{\gradeone \mvee \munit}{\closedhowe{\relone})}.
        $$

        Such an inclusion becomes our condition defining when relators and co-relators 
        properly interact. 

        \begin{definition}
        \label{definition:relator-corelator-distributive-law}
        Given a relator $\relatorsymbol$ for a monad $(\monad, \unit, \kleisli{-})$ 
        and a co-relator $\corelatorsymbol$ as defined in 
        Definition~\ref{definition:corelator}, we say that $\relatorsymbol$ and 
        $\corelatorsymbol$ interact if the following inclusion holds for 
        any $\gradeone \mgeq \munit$:
        \begin{align*}
        \corelator{\gradeone}
        {\relatorsymbol(\relone)}
        \subseteq
        \relatorsymbol(\corelator{\gradeone}{\relone}).
        \tag{Distributivity} \label{eq:distributive-lax}
        \end{align*}
        \end{definition}

        Definition~\ref{definition:relator-corelator-distributive-law}
        gives the existence of distributive laws between the 
        lax comonad $(X \mapsto X, \relone \mapsto \corelator{\gradeone}{\relone})$ 
        and the lax monad $(X \mapsto \monad(X), \relone \mapsto \relatorsymbol(\relone))$ 
        for any $\gradeone \mgeq \munit$.

        Equivalently, we can refine lax \eqref{eq:lax-monad-bind} with the 
        following lax commutative diagram, where $\gradeone \mgeq \munit$.
        \[
        \vcenter{
          \diagramrel
          {f}
          {g}
          {X}
          {Y}
          {\corelator{\gradeone}{\relone}}
          {\monad(Z)}
          {\monad(W)}
          {\relatorsymbol(\reltwo)}
          }
          \implies
           \vcenter{
          \diagramrel
          {\kleisli{f}}
          {\kleisli{g}}
          {\monad(X)}
          {\monad(Y)}
          {\corelator{\gradeone}{\relatorsymbol(\relone)}}
          {\monad(Z)}
          {\monad(W)}
          {\relatorsymbol(\reltwo)}
          }
        \]

        A similar conditions has been used to study applicative distances 
        \cite{Gavazzo/LICS/2018} as well as in the context of metric-based denotational 
        semantics \cite{DBLP:conf/lics/AmorimGHK19} (where it goes under the name of 
        \emph{parametrized Kleisli lifting}). Distributive laws between (graded) monads and 
        comonads, instead, have been study in the categorical semantics of effectful-coeffectful 
        languages \cite{Gaboradi-et-al/ICFP/2016}. Such laws, however, focus on grading 
        and thus are more general than our law \eqref{eq:distributive-lax}, which focuses 
        on a single (non-graded) monad and on an identity-on-object comonad. Nonetheless, 
        it is easy to realise that Definition~\ref{definition:relator-corelator-distributive-law} 
        can be generalised to arbitrary monads (as we will do) and to arbitrary comonads. 
        It can also be refined to a full graded setting along the lines of the work of Gaboardi et
        al. \cite{Gaboradi-et-al/ICFP/2016}. We will say more about that later. 

        Another important difference between law \eqref{eq:distributive-lax} and 
        the aforementioned distributive 
        laws is that the co-relator $\corelatorsymbol_{\gradeone}$ has to distribute over the 
        relator $\relatorsymbol$ only for those $\gradeone$ 
        which are grater or equal than $\munit$. Intuitively, the reason is as follows. 
        Since we want to distribute a co-relator over a relator, we are \emph{a fortiori} 
        dealing with the result of a computation. But if we have such a result, then 
        we must have used the computation, meaning that $\gradeone$ could not be $\mzero$. 
        This restriction, which comes from applicative distances \cite{Gavazzo/LICS/2018}, 
        simplifies the meta-theory of applicative (bi)similarity. Moreover, it is 
        easy to see that all the examples seen so far satisfies the inclusion 
        of Definition~\ref{definition:relator-corelator-distributive-law}.

        \begin{example} 
        Let us consider the canonical co-relator of Definition~\ref{definition:canonical-corelator}
        and let $\gradeone \mgeq \munit$. If $\wrel{\bang_{\gradeone}\relatorsymbol^{\divergence}(\relone)}{x}{y}{\wone}$, 
        then there exists $\wtwo$ such that $\wone \wgeq \gradeone \mact \wtwo$ and 
        either $x = \divergence$, or $\wone = \wtop$, or $x, y \neq \divergence$ and 
        $\wrel{\relone}{x}{y}{\wtwo}$. In the former case, we trivially have 
        $\wrel{\relatorsymbol^{\divergence}(\bang_{\gradeone}\relone)}{x}{y}{\wone}$. 
        If $\wtwo = \wtop$, the it is enough to show $\wone \wgeq \wtop$ (and thus 
        $\wone = \wtop$) too. Since $\gradeone \mgeq \munit$, by monotonicity of $\mact$ 
        we have $\wone \wgeq \gradeone \mact \wtop \wgeq \munit \mact \wtop = \wtop$. 
        Finally, if $x,y \neq \divergence$ and $\wrel{\relone}{x}{y}{\wtwo}$, 
        then we also have $\wrel{\bang_{\gradeone}\relone}{x}{y}{\wone}$, and thus
        $\wrel{\relatorsymbol^{\divergence}\bang_{\gradeone}\relone}{x}{y}{\wone}$. 
        \end{example}

        At this point we have all the ingredients to complete the last passage 
        needed to show that applicative similarly is compatible.

        \begin{lemma}[Key Lemma]
        The closed projection of the 
        Howe extension of an applicative simulation is an applicative simulation.
        \end{lemma}

        \begin{proof}
        The proof goes as outlined in previous pages. Most cases are standard, 
        the difficult one being sequencing. This is handled, as already explained, 
        relying on lax \eqref{eq:distributive-lax}.
        \end{proof}

        Summing up, we have thus proved the following result.

        \begin{theorem}
        \label{theorem:congruence}
        Both applicative similarly and applicative bisimilarity are 
        substitutive and compatible term relations. Additionally, the 
        first one is substitutive and compatible preorder term relation 
        (and thus a precongruence), 
        whereas the second one is substitutive and compatible equivalence 
        term relations (and thus a congruence).
        \end{theorem}

        Theorem~\ref{theorem:congruence} concludes our analysis of 
        (pure) modal applicative (bi)similarly. In the next section, we will 
        formally relate modal applicative (bi)similarly with applicative distances, 
        this way showing that not only metric reasoning is a special form of modal 
        reasoning, but that also the vice versa holds and thus that intensional 
        and metric semantics are essentially equivalent. 
}

\shortv{
      \section{Modal Reasoning $=$ Metric Reasoning}
      \label{section:kripke-meets-lawvere}

      \newcommand{\rone}{\gradeone}
      \newcommand{\rtwo}{\gradetwo}
      \newcommand{\rthree}{\gradethree}
      \newcommand{\rfour}{\mfour}
      \newcommand{\rfive}{\mfive}
      \newcommand{\rzero}{\mzero}
      \newcommand{\runit}{\munit}
      \newcommand{\rleq}{\mleq}
      \newcommand{\rgeq}{\mgeq}
      \newcommand{\rstar}{\mstar}
      \newcommand{\rplus}{\mplus}

      In Example~\ref{ex:world-relations}, 
      we have seen how metric reasoning 
      is a specific example of modal reasoning 
      where possible worlds give upper bounds to distances 
      between programs. 
      In this section, we show that modal reasoning and metric reasoning are
      actually one and the same, provided that the latter is formulated in the 
      general setting of quantale-valued distances 
      \cite{Hoffman-Seal-Tholem/monoidal-topology/2014}, as pioneered by Lawvere 
      in his seminal work on on generalised 
      metric spaces as enriched categories\cite{Lawvere/GeneralizedMetricSpaces/1973}. 
      Accordingly, the distance between two objects is not a number, 
      but an element of a quantale \cite{Rosenthal/Quantales/1990} 
      representing an abstract difference. 

      \begin{definition}
       A commutative \emph{quantale} 
      $\Quantale = (\quantale, \leq, \tensor, \qunit)$ 
      is a complete lattice $(\quantale, \leq)$ equipped with a binary 
      commutative multiplication $\tensor$ such that: 
      (i) $\tensor$ has a unit $\qunit$; (ii) 
      join distributes over multiplication, i.e. 
      $\qone \tensor \big(\join_i \qtwo_i\big) = 
      \join_i (\qone \tensor \qtwo_i)$.
     Given a quantale $\Quantale = (\quantale, \leq, \tensor, \qunit)$, 
        a $\Quantale$-matrix $\vrelone: X \torel Y$
        over sets $X,Y$ is a map
        $\vrelone: X \times Y \to \quantale$. 
      \end{definition}

      When the quantale $\Quantale$ is left unspecified, we generically 
      refer to quantale-valued matrices (or, sometimes, quantale-valued 
      distances).
      Before giving examples of quantales and quantale-valued matrices, 
      we observe that fixed a quantale $\Quantale$, we have a category 
      --- called $\vrel$ ---
      with sets as objects and $\Quantale$-matrices as arrows. 
      The identity arrow $\idvrel: X \torel X$ maps a pair $(x,y)$ to 
      $\qunit$ if $x=y$, and to $\qbot$ (the bottom element of $\Quantale$) 
      otherwise. Given $\Quantale$-matrices $\vrelone: X \torel Y$ 
      and $\vreltwo: Y \torel Z$, their composition 
      $\relcomp{\vrelone}{\vreltwo}: X \torel Z$ is given by the so-called 
      matrix multiplication formula \cite{Hoffman-Seal-Tholem/monoidal-topology/2014}:
      $$
      (\relcomp{\vrelone}{\vreltwo})(x,z) 
      \defeq \join_y \vrelone(x,y) \tensor \vreltwo(y,z).
      $$

      Moreover, the complete lattice structure of $\Quantale$ extends to 
      $\Quantale$-matrices pointwise, so that
      we can say that a $\Quantale$-matrix $\vrelone: X \torel X$ is 
      \emph{reflexive} if $\idvrel \leq \vrelone$, \emph{transitive} 
      if $\relcomp{\vrelone}{\vrelone} \leq \vrelone$, and symmetric if 
      $\dual{\vrelone} \leq \vrelone$ (here, 
      the transpose of a $\Quantale$-matrix 
      $\vrelone: X \torel Y$ is the $\Quantale$-matrix $\dual{\vrelone}: Y \torel X$
      defined by $\dual{\vrelone}(y,x) = \vrelone(x,y)$). 
      Altogether, we obtain the notion of a preorder and equivalence 
      $\Quantale$-matrix.

      \begin{example}
      The following are examples of quantales.
      \begin{varenumerate}
        \item The \emph{Boolean quantale} $\Two = (\{\false, \true\}, \leq, \wedge)$. 
          Notice that $\Two$-matrices are just ordinary relations, and 
          that preorder $\Two$-matrices and equivalence $\Two$-matrices 
          coincide with preorder and equivalence relations.
          More generally, every frame \cite{Vickers/Topology-via-logic} with binary meet as 
          multiplication forms a quantale.
        \item The powerset $\powerset(\mathcal{M})$ of a commutative monoid 
          $(\mathcal{M}, \cdot, 1)$ with multiplication defined by 
          $M \tensor N = \{m \cdot n \mid 
          m \in M, n \in N\}$.
        \item The \emph{Lawvere quantale} $\Lawvere = ([0,\infty], \geq, +, 0)$. 
          Notice that we use the opposite of the natural ordering, 
          so that, e.g., $0$ is the top element 
          of $\Lawvere$. Instantiating transitivity on $\Lawvere$, we obtain the 
          usual \emph{triangle inequality} formula, so that 
          preorder $\Lawvere$-matrices are nothing but generalised metrics 
          \cite{Lawvere/GeneralizedMetricSpaces/1973} and 
          equivalence $\Lawvere$-matrices are pseudometrics \cite{steen/CounterexamplesTopology/1995}.
        \item The \emph{Strong Lawvere quantale} 
          $\StrongLawvere = ([0,\infty], \geq, \max, 0)$. Notice that having 
          replaced addition with binary maximum, transitivity gives 
          the \emph{strong triangle inequality} formula, so that, e.g.,
          equivalence $\StrongLawvere$-matrices are ultra-pseudometrics.
        \item The unit interval with natural order and (quantale) multiplication 
          given by a t-norm \cite{hajek1998metamathematics}.
        \item The set $\Delta \defeq \{f \in [0,1]^{[0,\infty]} \mid f \text{ monotone and }
            f(a) = \join_{b < a} f(b)\}$. Equivalence $\Delta$-matrices 
            gives \emph{probabilistic metric spaces} \cite{HOFMANN20131} 
            (the informal reading of a $\Delta$-relation $\vrelone$ is that 
            $\vrelone(x,y)(a)$ gives the probability that $x$ and $y$ are 
            at most $a$-far).
      \end{varenumerate}
      \end{example}
  But what does quantales and quantale-valued matrices have to do with Kripke monoidal relations?  
  First, we notice that for any MKF $\Worlds = (\worlds, \wleq, \wcomp, \wunit)$, 
  we can regard $\Worlds$-relations as quantale-valued matrices. 
  In fact, the set $\{\wpredone \in \{\false, \true\}^{\worlds} \mid 
  \wpredone \text{ monotone}\}$ carries a quantale 
 structure with the complete lattice structure defined pointwise and 
quantale multiplication defined thus:
            \begin{align*}
              (\wpredone \tensor \wpredtwo)(\wone) 
              &\defiff \exists \wtwo, \wthree.\ \wone \wgeq \wtwo \wcomp \wthree 
                \textnormal{ and } \wpredone(\wtwo) \textnormal{ and } 
                \wpredtwo(\wthree).
            \end{align*}
We denote the resulting quantale as $\kripkequantale$.
As a consequence, we see that $\Worlds$-relations are a special case of 
quantale-valued matrices.\footnote{
  For a $\Worlds$-relation, the corresponding $\kripkequantale$-matrix 
  sends a pair $(x,y)$ to the predicate on $\Worlds$ which holds 
  at $\wone$ if and only if $\wrel{\relone}{x}{y}{\wone}$.
} 
But this is only half of the story. In fact, $\Quantale$-matrices can 
be regarded as monoidal Kripke relations of a special kind, namely 
as meet-preserving monoidal Kripke relations on the MKF 
$\opposite{\Quantale}$ (meaning that 
$\relone(\join_i \qone_i) = \bigcap_i \relone(\qone_i)$).
We write $\mathsf{Inf}(\opposite{\Quantale}, \rel(X,Y))$ for the collection 
of such relations.
Altogether, we obtain the following correspondence 
(cf. \cite{Hoffman-Seal-Tholem/monoidal-topology/2014}).
\begin{proposition}
\label{prop:quantale-relations-iso-kripke-relations}
We have 
$
\Wrel{\Worlds}(X,Y) \cong \Vrel{\kripkequantale}(X,Y)
$
and $\vrel(X,Y) \cong \mathsf{Inf}(\opposite{\Quantale}, \rel(X,Y))$ via
the maps
    \begin{alignat*}{2}
      \wrel{\phi(\vrelone)}{x}{y}{\wone} 
      &\defiff \vrelone(x,y)(\wone)
      &&\text{ }\text{ }
      \psi(\relone)(x,y)(\wone) 
      \defiff \wrel{\relone}{x}{y}{\wone}
      \\
      x \mathbin{\Phi(\vrelone)(\qone)} y 
      &\defiff \qone \leq \vrelone(x,y)
      &&\text{ }\text{ }
      \Psi(\relone)(x,y) =  \join \{\qone \mid \wrel{\relone}{x}{y}{\qone}\}.  
    \end{alignat*}
\end{proposition}

Proposition~\ref{prop:quantale-relations-iso-kripke-relations} 
has a clear mathematical meaning. But what is its pragmatic
relevance? 
Quantale-valued matrices have been extensively studied 
as abstract notions of distances \cite{DBLP:journals/fss/ClementinoH17,DBLP:journals/acs/ClementinoHT04,Hoffman-Seal-Tholem/monoidal-topology/2014,DBLP:journals/tcs/FlaggK97,flagg1992completeness,Flagg1997}, meaning that there is a large body of results 
that, thanks to Proposition~\ref{prop:quantale-relations-iso-kripke-relations}, 
we can rely on to improve our theory of modal program equivalence. 
In particular, Gavazzo \cite{DBLP:phd/basesearch/Gavazzo19,Gavazzo/LICS/2018} 
developed a theory of quantale-based applicative (bi)simulation distances 
for higher-order languages with 
\emph{algebraic effects} \cite{DBLP:journals/entcs/PlotkinP01,PlotkinPower/FOSSACS/01,Plotkin/algebraic-operations-and-generic-effects/2003}. 
Although such a theory deals with algebraic effects (and it is thus more
general than ours, in this respect), the language it builds upon 
has only a single kind of grade algebras. Additionally, 
lacking the general notion of a comonadic lax extension, 
applicative bisimilarity distance is defined with respect to 
the analogue of our lax actions only, and thus it cannot capture the 
behaviour of more general forms of modality.
We can thus rely on Proposition~\ref{prop:quantale-relations-iso-kripke-relations} 
to extend the notion of an applicative bisimulation distance 
to the general setting of modal types. 
First, observe that by Proposition~\ref{prop:quantale-relations-iso-kripke-relations},
any notion we have defined in terms of $\Worlds$-relations 
has a $\vrel$ counterpart. 
Thus, for instance, we have a notion term matrix (cf. term relation),
as well as a notion of a 
comonadic lax extension to $\vrel$. 
Finally, for a $\Quantale$-matrix $\vrelone: X \torel Y$, we define
$\vrelone_{\divergence}: X_{\divergence} \torel Y_{\divergence}$ by
$\vrelone_{\divergence}(x,y) = \qunit$ if $x = \divergence$, 
$\vrelone_{\divergence}(x,y) = \vrelone(x,y)$ if 
$x,y \neq \divergence$, and to $\vrelone_{\divergence}(x,y) = \qbot$ otherwise. 
We refer to the original paper on applicative bisimilarity distance(s) 
\cite{Gavazzo/LICS/2018} for details.

  \begin{definition}
  \label{def:applicative-distance}
    Fixed a quantale $\Quantale$ and 
    comonadic lax extension $\corelatorsymbol$ to $\vrel$, 
    we define \emph{applicative similarity distance} 
    $\vsim$ as the largest term matrix $\vrelone$ 
    such that:\footnote{
    We write $\vrelone^{\scriptscriptstyle \Lambda}_{\typeone}$, 
    $\vrelone^{\scriptscriptstyle \values}_{\typeone}$ 
    for the closed terms matrix $\vrelone$ on terms and values 
    of type $\typeone$, respectively.
    }
     \begin{align*}
      \vrelone^{\scriptscriptstyle \Lambda}_{\typeone}(\termone, \termtwo) 
      &\leq \vrelone^{\scriptscriptstyle \values}_{\divergence}(\sem{\termone}, \sem{\termtwo})
      \\
      \vrelone^{\scriptscriptstyle \values}_{\typeone \to \typetwo}
      (\abs{\varone}{\termone}, \abs{\varone}{\termtwo}) 
      &\leq \meet_{\valone \in \values_{\typeone}} 
      \vrelone^{\scriptscriptstyle \Lambda}_{\typetwo}(\subst{\termone}{\varone}{\valone}, 
      \subst{\termtwo}{\varone}{\valone})
      \\
      \vrelone^{\scriptscriptstyle \values}_{\rectype{\typevarone}{\typeone}}(\fold{\valone}, \fold{\valtwo}) 
      &\leq  \vrelone^{\scriptscriptstyle \values}_{\substtype{\typeone}{\typevarone}{
      \rectype{\typevarone}{\typeone}}}(\valone, \valtwo)
      \\
      \vrelone^{\scriptscriptstyle \values}_{\bbox_{\gradeone}\typeone}(\tbox{\valone}, \tbox{\valtwo})
      &\leq 
      \corelator{\gradeone}{\vrelone^{\scriptscriptstyle \values}_{\typeone}}(\valone, \valtwo),
    \end{align*}
     \emph{Applicative bisimilarity distance} is defined as $\vsim \wedge \dual{\vsim}$.
    \end{definition}

  The following result states that modal reasoning is indeed equal to metric reasoning, via 
  Lawvere.

\begin{theorem}
\label{theorem:modal-reasoning-equal-metric-reasoning}
Let $\Quantale$ and $\Worlds$ be a quantale and a monoidal Kripke frame, respectively.
Modulo the isomorphisms of Proposition~\ref{prop:quantale-relations-iso-kripke-relations}, 
we have that $\Phi(\vsim) = {\appsimilarity}$, where $\appsimilarity$ 
is defined on $\opposite{\Quantale}$, and that 
 $\psi({\appsimilarity}) = {\vsim}$, where $\vsim$ is defined on $\kripkequantale$.
\end{theorem}

Since modal applicative (bi)similarity is compatible and substitutive, 
by Theorem~\ref{theorem:modal-reasoning-equal-metric-reasoning} we obtain that 
applicative (bi)similarity distance is compatible and substitutive, too. 
From this result, it also follows an abstract metric preservation theorem 
stating that we can reason compositionality about program distances.

  \begin{theorem}
  \label{theorem:abstract-metric-preservation}
  For any term $\graded{\varone_1}{\typeone_1}{\gradeone_1}, \hh, 
  \graded{\varone_n}{\typeone_n}{\gradeone_n} \imp \termone: \typeone$, and for all 
  values $\vect{\valone} \defeq \valone_1, \hh, 
  \valone_n$, $\vect{\valtwo} \defeq \valtwo_1, \hh, \valtwo_n$ 
  of the appropriate type, we have:
  $$
  \bigotimes_{i \leq n} \corelator{\gradeone_i}{\vsim_{\typeone_i}}(\valone_i, \valtwo_i)
  \leq \vsim_{\typeone}(\termone[\vect{\valone}/\vect{\varone}],
  \termone[\vect{\valtwo}/\vect{\varone}]).
  $$
  \end{theorem}
}

\longv{
            \section{Kripke Meets Lawvere}
            \label{section:kripke-meets-lawvere}

            \newcommand{\rone}{\gradeone}
            \newcommand{\rtwo}{\gradetwo}
            \newcommand{\rthree}{\gradethree}
            \newcommand{\rfour}{\mfour}
            \newcommand{\rfive}{\mfive}
            \newcommand{\rzero}{\mzero}
            \newcommand{\runit}{\munit}
            \newcommand{\rleq}{\mleq}
            \newcommand{\rgeq}{\mgeq}
            \newcommand{\rstar}{\mstar}
            \newcommand{\rplus}{\mplus}

            In previous sections we have seen how metric reasoning 
            is an example of modal reasoning. More specifically, we have seen how 
            (operationally-based) metric semantics can be seen as special intensional 
            semantics where possible worlds are given by non-negative extended real 
            numbers bounding distances between expressions. 
            In this sections, we show that modal reasoning and metric reasoning are
            actually equivalent, provided that the latter is formulated in the 
            general setting of quantale-valued distances 
            \cite{Hoffman-Seal-Tholem/monoidal-topology/2014}, as pioneered by Lawvere 
            \cite{Lawvere/GeneralizedMetricSpaces/1973}. 
            As a consequence, we see that not only
            metric reasoning is a special kind of modal reasoning, but
            that the converse inclusion holds as well:
            modal reasoning is a special case of metric reasoning, so that 
            modal and metric reasoning coincide.

            Our starting point is the theory of program distance developed by the second author 
            \cite{Gavazzo/LICS/2018} following the seminal work by  
            Lawvere \cite{Lawvere/GeneralizedMetricSpaces/1973} on generalised 
            metric spaces as enriched categories.
            Accordingly, distances take values in a quantale $(\quantale, \join, \tensor, \qunit)$ 
            whose elements represent abstract quantities. Examples of quantales are the two-element 
            (Boolean) quantale $(\{\false, \true\}, \leq, \wedge)$, the Lawvere quantale 
            $([0,\infty], \geq, +, 0)$, and the ultrametric Lawvere quantale 
            $([0,\infty], \geq, \max, 0)$. Notice that the Lawvere quantales rely 
            on the opposite of the natural ordering, so that, e.g., $0$ is the top element 
            of the quantale and $\infty$ is the bottom one. 
            We denote by $\qtop$ and $\qbot$ the top and bottom element of the quantale. 
            Moreover, as in Section~\ref{section:modal-calculi} 
            we assumed $\mzero$ to be the bottom element of the modal signature, here we 
            assume $\qunit$ to be the top element of the quantale. Quantales satisfying such a 
            property are known as \emph{integral quantales}\footnote{Notice that integral 
            always satisfy the law $\qone \tensor \qtwo \qleq \qone$. For, 
            $\qone \tensor \qtwo \qleq \qone \tensor \qtop = \qone \tensor \qunit = \qone $.} 
            \cite{Hoffman-Seal-Tholem/monoidal-topology/2014}.

            Quantales form a category $\quantalescat$ whose objects are quantales, and 
            whose morphisms are quantale homomorphisms. 
            A map $h \in \quantalescat(\quantale, \quantaletwo)$ is a monotone function
            $h: \quantale \to \quantaletwo$ preserving unit and tensor products. 
            \begin{align*}
            h(\qunit_{\quantale}) &= \qunit_{\quantaletwo};
            &
            h(\qone \tensor \qtwo) &= h(\qone) \tensor h(\qtwo);
            & 
            \qone \qleq \qtwo &\implies h(\qone) \qleq h(\qtwo).
            \end{align*}
            The set $\quantalescat(\quantale, \quantale)$ of quantale endomorphism 
            carries, among others, a preordered semiring structure with semiring addition given by 
            pointwise tensor, semiring multiplication given by composition, and 
            preorder defined pointwise. Formally:
            \begin{align*}
              (f \tensor g)(\qone) &= f(\qone) \tensor g(\qone) 
              &
              (g  \comp f)(\qone) &= g(f(\qone))
              &
              f \leq g & \iff \forall \qone.\ f(\qone) \leq g(\qtwo)
            \end{align*}

            The neutral element for the semiring addition and multiplication are defined 
            by the $\qunit$-constant function $\metafun{x}{\qunit}$ 
            and the identity function $\baseid = \metafun{x}{x}$, respectively.  
            In particular, $\quantalescat(\quantale, \quantale)$ gives a modal signature 
            by taking the opposite order as summarised by the following table.

            \begin{center}
            \begin{tabular}{|c|c|c|c|c|c|}
            \hline
            $\quantalescat(\quantale, \quantale)$ & $\geq$ & $\tensor$ & $\comp$ 
            & $\metafun{x}{\qunit}$ &$\baseid$
            \\
            \hline
            $\gradealg$ & $\mleq$ & $\mplus$ & $\mstar$ & $\mzero$ & $\munit$
            \\
            \hline
            \end{tabular}
            \end{center}

            Notice that the zero element of $\quantalescat(\quantale, \quantale)$, namely 
            the constant function $\metafun{x}{\qunit}$, is indeed the bottom element 
            of the preordered semiring, since $\metafun{x}{\qunit} \geq h$, for any 
            homomorphism $h$.

            Given a quantale $(\quantale, \qleq, \tensor, \qunit)$, program distances are 
            formulated for a modal calculus --- called $\quantale$-\Fuzz --- 
            generalising Reed and Pierce's \Fuzz\ \cite{Pierce/DistanceMakesTypesGrowStronger/2010} 
            to arbitrary quantales (and algebraic effects). This design choice prevents 
            undesired phenomena such as distance trivialisation \cite{CrubilleDalLago/ESOP/2017,Gavazzo/LICS/2018}

            $\quantale$-\Fuzz\ is similar to 
            $\Lambda_{\gradealg}$, and can actually coincides with $\Lambda_{\gradealg}$ 
            instantiated with the modal signature we endowed $\quantalescat(\quantale, \quantale)$ 
            with. It is important to keep in mind that such a signature employs the 
            \emph{opposite} of the extension order of the quantale $\quantale$. This explains 
            $\quantale$-\Fuzz\ typing rules such as
            \[
            \infer{\envone \compimp \valone: \typeone}{\envone \valimp \valone:\typeone}
            \qquad
            \infer{
              (\baseone \wedge \baseid) \comp \envone \mplus \envtwo \compimp \seq{\termone}{\termtwo}: 
              \typetwo
            }
            {\envone \compimp \termone: \typeone
            &
            \envtwo, \graded{\varone}{\typeone}{\gradeone} \compimp \termtwo: \typetwo
            }
            \]
            where we rely on the \emph{meet} of $\quantalescat(\quantale, \quantale)$.

            Keeping this observation in mind, it is immediate to generalise 
            $\quantale$-\Fuzz\ from $\quantalescat(\quantale, \quantale)$ 
            to arbitrary modal signature, this way seeing that 
            $\quantale$-\Fuzz\ and $\Lambda_{\gradealg}$ are essentially the same calculus.

            Next, we need to encode $\Worlds$-relations over $X \times Y$ as quantale-valued distances, 
            i.e. as functions $\vrel: X \times Y \to \quantale$. We do so by considering the 
            quantale $\monotone{\worlds}{\two}$ of \emph{monotone} predicates on $\worlds$. 
            That is, an element in $\monotone{\worlds}{\two}$ is a function 
            $\wpredone: \worlds \to \two$ such that if $\wone \wleq \wtwo$ and 
            $\wpredone(\wone) = \true$, then $\wpredone(\wtwo) = \true$ as well.

            \begin{proposition}
            The set $\monotone{\worlds}{\two}$ carries a quantale 
            structure with the complete lattice structure defined pointwise and 
            the monoidal structure defined thus:
            \begin{align*}
              (\wpredone \tensor \wpredtwo)(\wone) = \true 
              &\iff \exists \wtwo, \wthree.\ \wone \wgeq \wtwo \wcomp \wthree 
                \textnormal{ and } \wpredone(\wtwo) = \true \textnormal{ and } 
                \wpredtwo(\wthree) = \true;
              \\
              \qunit(\wone) &= \true.
            \end{align*}
            \end{proposition}


            \begin{proof}
            We show that 
            $
            \wpredone \tensor (\join_i \wpredtwo_i) = \join_i (\wpredone \tensor \wpredone_i).
            $
            We have $\left(\wpredone \tensor (\join_i \wpredtwo_i)\right)(\wone)$ 
            if and only if there exist $\wtwo, \wthree$ such that 
            $\wone \wgeq \wtwo \wcomp \wthree$ and both 
            $\wpredone(\wtwo)$ and $(\join_i \wpredtwo_i)(\wthree)$ hold. The latter holds 
            if and only there exists an index $i$ such that $\wpredtwo_{i}(\wthree)$ holds. 
            But that exactly means that 
            $\join_i (\wpredone \tensor \wpredtwo_i)(\wone)$ holds. 
            \end{proof}

            Having defined the quantale $\monotone{\worlds}{\two}$, 
            we now study $(\monotone{\worlds}{\two})$-distances. Given sets 
            $X,Y$, we write $(\monotone{\worlds}{\two})$-$\rel(X,Y)$ for the 
            collection of such distances, and use letters $\vrelone, \vreltwo, \hh$ 
            to range over such distances. As it is customary, given a quantale 
            $\quantale$, the composition between 
            $\quantale$-distances $\vrelone \in \quantale\text{-}\rel(X,Y)$ 
            and $\vreltwo \in \quantale\text{-}\rel(Y,Z)$ as the 
            $\quantale$-distance
            $\vreltwo \comp \vrelone \in \quantale\text{-}\rel(X,Z)$ 
            defined as:
            $$
            (\vreltwo \comp \vrelone)(x,z) = \join_y \vrelone(x,y) \tensor \vreltwo(y,z).
            $$
            Moreover, we extend the the monoid and lattice structure of 
            $\quantale$ to $\quantale$-distances pointwise, and write 
            $\vidrel$ for the identity $\quantale$-distance\footnote{Where 
            $\vidrel(x,y) = \qunit$, if $x = y$, and $\qbot$, otherwise.}/

            Our first observation is that $\monotone{\worlds}{\two}$-distances 
            and $\Worlds$-relations coincide.

            \begin{proposition}
            For all sets $X,Y$, the sets 
            $(\monotone{\worlds}{\two})$-$\rel(X,Y)$ and 
            $\wrelcat(X,Y)$ are isomorphic through the following maps:
            \begin{align*}
            \Phi(\vrelone) &= \{(x,y, \wone) \mid \vrelone(x,y)(\wone) = \true\}
            \\
            \Psi(\relone) &= \metafun{(x,y)}{\metafun{\wone}{
            \begin{cases}
              \true & \textnormal{ if } \wrel{\relone}{x}{y}{\wone} 
              \\
              \false & \textnormal{ otherwise.}
            \end{cases}
            }}
            \end{align*}
            Moreover, both $\Phi$ and $\Psi$ are monotone, and satisfy the following 
            algebraic laws:
            \begin{align*}
            \Phi(\vreltwo \comp \vrelone) &= \Phi(\vreltwo) \comp \Phi(\vrelone)
            &
            \Psi(\reltwo \comp \relone) &= \Psi(\reltwo) \comp \Psi(\relone)
            \\
            \Phi(\vrelone \tensor \vreltwo) &= \Phi(\vrelone) \tensor \Phi(\vreltwo)
            &
            \Psi(\relone \tensor \reltwo) &= \Psi(\relone) \tensor \Psi(\reltwo)
            \\
            \Phi(\vidrel) &= \idrel
            &
            \Psi(\idrel) &= \vidrel.
            \end{align*}
            \end{proposition} 

            As a consequence, it is easy to see that there is an exact correspondence between 
            categories of $\Worlds$-relations and categories of $(\monotone{\worlds}{\two})$-distances. 
            In particular, the so-called category of $(\monotone{\worlds}{\two})$-categories 
            and functors (which generalise non-expansive maps) corresponds to 
            the category $\brel_{\worlds}$, and the category $(\monotone{\worlds}{\two})\text{-}\rel$ 
            corresponds to $\wrelcat$. That means that we can instantiate all results on 
            quantale-valued distances to the quantale $\monotone{\worlds}{\two}$, this way
            obtaining notions and results on $\Worlds$-relations and, ultimately, on 
            intensional semantics.

            We now move the analysis of applicative distances. The latter have not been studied for 
            arbitrary co-relator but for the canonical co-relator only (recall that $\quantale$-\Fuzz\ 
            has $\mathsf{Quant}(\quantale, \quantale)$ as modal signature). Nonetheless, it is an 
            easy exercise to generalise such a result to arbitrary \emph{quantale-valued co-relators}, 
            i.e. co-relator like maps acting on $\quantale$-relations.

            In particular, when working with the quantale 
            $\monotone{\worlds}{\two}$ any co-relator gives a quantale-valued co-relator
            throughout the map $\Psi$, and vice versa any quantale-valued co-relator gives
            a co-relator via the map $\Phi$.
            Let us now recall the definition of the applicative similarity distance 
            properly generalised to a quantale-valued co-relator $\corelatorsymbol$. 
            Given a $\quantale$-relation $\vrelone$ over $X \times Y$, let us define 
            the $\quantale$-relation $\relatorsymbol^{\bot} \vrelone$ over
            on $X_{\bot} \times Y_{\bot}$ by 
            $$
            \relatorsymbol^{\bot}\vrelone(x,y) = 
            \begin{cases}
              \qunit & \text{ if } x = \bot
              \\
              \vrelone(x,y) & \text{ if } x,y \neq \bot
              \\
              \qbot & \text{ otherwise.}
            \end{cases}
            $$

            \begin{definition}
            A (type-indexed) $\quantale$-valued relation on closed expressions is an 
            \emph{applicative simulation} (distance) if it satisfies the 
            following inequalities, where $\corelatorsymbol$ is quantale-valued co-relator.
            \begin{align*}
              \vrelone_{\typeone}(\termone, \termtwo) 
              &\leq \relatorsymbol^{\bot}(\vrelone)(\sem{\termone}, \sem{\termtwo})
              \\
              \vrelone_{\typeone \to \typetwo}(\abs{\varone}{\termone}, \abs{\varone}{\termtwo}) 
              &\leq \meet_{\valone} \vrelone_{\typetwo}(\subst{\termone}{\varone}{\valone}, 
              \subst{\termtwo}{\varone}{\valone})
              \\
              \vrelone_{\rectype{\typevarone}{\typeone}}(\fold{\valone}, \fold{\valtwo}) 
              &\leq  \vrelone_{\substtype{\typeone}{\typevarone}{
              \rectype{\typevarone}{\typeone}}}(\valone, \valtwo)
              \\
              \vrelone_{\bbox_{\rone}\typeone}(\tbox{\valone}, \tbox{\valtwo})
              &\leq 
              \corelator{\rone}{\vrelone_{\typeone}}(\valone, \valtwo)
            \end{align*}
            \end{definition}

            Applicative similarity distance $\vsim$ is defined as the largest applicative 
            simulation distance. In particular, $\vsim$ satisfies the following identities.
            \begin{align*}
              \vsim_{\typeone}(\termone, \termtwo) 
              &= \relatorsymbol^{\bot}(\vsim)(\sem{\termone}, \sem{\termtwo})
              \\
              \vsim_{\typeone \to \typetwo}(\abs{\varone}{\termone}, \abs{\varone}{\termtwo}) 
              &= \meet_{\valone} \vsim_{\typetwo}(\subst{\termone}{\varone}{\valone}, 
              \subst{\termtwo}{\varone}{\valone})
              \\
              \vsim_{\rectype{\typevarone}{\typeone}}(\fold{\valone}, \fold{\valtwo}) 
              &=  \vsim_{\substtype{\typeone}{\typevarone}{
              \rectype{\typevarone}{\typeone}}}(\valone, \valtwo)
              \\
              \vsim_{\bbox_{\rone}\typeone}(\tbox{\valone}, \tbox{\valtwo})
              &= 
              \corelator{\rone}{\vsim_{\typeone}}(\valone, \valtwo)
            \end{align*}

            Relying on the map $\Phi$, we see that 
            $\vsim_{\typeone}(\termone, \termtwo)(\wone) = \true$ 
            means exactly $\wrelt{\appbisimilarity}{\termone}{\termtwo}{\wone}{\typeone}$. 
            This way, we obtain:
            \begin{align*}
              \vsim_{\typeone \to \typetwo}(\abs{\varone}{\termone}, \abs{\varone}{\termtwo})
              (\wone) = \true
              &\iff \forall \valone \in \values_{\typeone}.\ 
              \wrelt{\appbisimilarity}{\subst{\termone}{\varone}{\valone}} 
              {\subst{\termtwo}{\varone}{\valone}}{\wone}{\typetwo}
            \\
              \vsim_{\bbox_{\rone}\typeone}(\tbox{\valone}, \tbox{\valtwo})(\wone) = \true
              &\iff
              \wrelt{\corelator{\rone}{\appbisimilarity}}{\valone}{\valtwo}{\wone}{\typeone}
            \end{align*}

            We can generalise this observations obtaining the following result.

            \begin{theorem}
            \label{theorem:metric-reasoning-equals-modal-reasoning}
            Let $\corelatorsymbol$ be a co-relator. Then $\appsimilarity$ 
            with respect to $\corelatorsymbol$ coincided with $\Phi(\vsim)$, 
            where $\vsim$ is defined with respect to the quantale-valued co-relator 
            $\Psi \comp \corelatorsymbol \comp \Phi$. 
            Vice versa, given a quantale-valued co-relator $\corelatorsymbol$, 
            $\vsim$ with respect to $\corelatorsymbol$ coincides with 
            $\Psi(\appsimilarity)$, where $\appsimilarity$ is defined with 
            respect to the co-relator $\Phi \comp \corelatorsymbol \comp \Psi$.
            \end{theorem}

            Theorem~\ref{theorem:metric-reasoning-equals-modal-reasoning} extends 
            to the symmetric case of modal applicative bisimilarity and applicative 
            bisimilarity distance. For instance, we see how the abstract metric preservation 
            theorem for applicative (bi)similarity distance gives compositionality of 
            modal applicative bisimilarity.

            \begin{theorem}[Abstract Metric Preservation \cite{Gavazzo/LICS/2018}]
            \label{theorem:abstract-metric-preservation}
            For any expression $\graded{\varone_1}{\typeone_1}{\rtwo_1}, \hh, 
            \graded{\varone_1}{\typeone_1}{\rtwo_1} \imp \termone: \typeone$, and for all 
            values $\valimp \valone_1, \valtwo_1: \typeone_1, \hh, 
            \valimp \valone_n, \valtwo_n: \typeone_n$, we have:
            $$
            \bigotimes_i \corelator{\rone_i}{\vsim_{\typeone_i}}(\valone_i, \valtwo_i)
            \leq \vsim_{\typeone}(\termone[\varone_1 := \valone_1, \hh, \varone_n := \valone_n],
            \termone[\varone_1 := \valtwo_1, \hh, \varone_n := \valtwo_n])
            $$
            \end{theorem}

            %
            That exactly corresponds to the compositionality law \textcolor{red}{[REF]}
            \[
            \infer{
              \wrelt
              {\appbisimilarity}
              {\termone[\varone_1 := \valone_1, \hh, \varone_n := \valone_n]} 
              {\termone[\varone_1 := \valtwo_1, \hh, \varone_n := \valtwo_n]}
              {\wone}
              {\typeone}
            }
            {\forall i.\ \wrelt
              {\corelator{\rone_i}{\appbisimilarity}}
              {\valone_i}
              {\valtwo_i}
              {\wone_i}
              {\typeone_i}
              &
              \wone \wgeq \wone_1 \wcomp \cc \wcomp \wone_n
            }
            \]
}

\begin{remark}[On Effects]
\label{rem:effects}
Theorem~\ref{theorem:abstract-metric-preservation} and 
Theorem~\ref{theorem:modal-reasoning-equal-metric-reasoning}
relate modal and metric reasoning. 
Although mathematically pleasant, the reader may wonder what one really gains 
from such a relationship (after all, one could ignore program distance and 
work directly with $\Worlds$-relations). 
The advantage of the correspondence between program distances and modal equivalences 
is that the former comes with a collection of results and techniques 
that are not readily available in a modal setting. 
For instance, 
since applicative bisimilarity distance has been originally 
defined on languages with arbitrary algebraic effects 
\cite{Plotkin/algebraic-operations-and-generic-effects/2003,PlotkinPower/FOSSACS/01}
(such as pure and probabilistic nondeterminism, imperative stores, exceptions, etc), 
Theorem~\ref{theorem:abstract-metric-preservation} can be easily generalised to 
extensions of $\Lambda_{\gradealg}$ with algebraic operations 
\emph{\`a la} Plotkin and Power. As a consequence, we obtain a collection of 
relational and metric-like techniques for reasoning about programs exhibiting both 
\emph{effectful} and \emph{coeffectful} behaviours.
\end{remark}

\section{Conclusion}
In this work, we have developed a relational theory of program equality 
for higher-order languages with graded modal types and coeffects. 
Such a theory builds upon some nontrivial and abstract notions, notably the 
one of a comonadic lax extension, which make the theory 
a robust and unifying framework for the operational analysis of coeffectful 
languages.
Even if new, we have showed that our relational theory is \emph{de facto} 
equivalent to a general theory of program distance, the latter being built 
on the category of quantale-valued matrices using suitable notions of lax extensions. 
This correspondence allows us to improve both theories at once giving, for instance, 
relational techniques for the analysis of languages with both (algebraic) 
effects and coeffects. 

\paragraph{Future Work} 
We have only touched the results obtainable from the aforementioned correspondence 
between modal relational reasoning and metric reasoning. 
In the future, the authors would like to rely on this correspondence 
to develop B\"ohm tree-like distances by means of
a notion of modal B\"ohm tree equivalence. The latter syntactically compares
B\"ohm trees of programs 
with respect to a possible world determining the granularity of the inspection. 
The action of a comonadic lax extension is then to change this granularity 
making, e.g., parts of the tree visible or invisible. 

The authors would also like to investigate whether abstract metric semantics 
can be used to give a \emph{uniform} denotational semantics to languages with modal types. 
In fact, denotational semantics of 
such languages have been 
given in terms of general categories and 
(graded) monads and comonads \cite{Gaboradi-et-al/ICFP/2016,DBLP:conf/esop/GhicaS14}. 
The specific categories such semantics
instantiate to, however, considerably change from case to case (giving, e.g., 
a presheaves semantics in the case of information flow and a metric semantics 
in the case of program sensitivity). It is thus desirable to have a more uniform, 
albeit more concrete semantics. This has been done by
Breuvart and Pagani \cite{DBLP:conf/csl/BreuvartP15}, who gave a denotational
semantics to coeffectful calculi where programs are interpreted as 
suitable relations. 
We believe that it is also possible to give modal calculi a uniform (abstract) 
metric semantics by interpreting types as categories enriched over a quantale and 
programs as enriched functors (which, in such a setting, generalise non-expansive maps).

Finally, the authors would like to explore further applications of 
comonadic lax extensions. For instance, although in this work the focus 
was on lax extensions of the identity comonad, it is a straightforward 
exercise to generalise our notions to arbitrary comonads. That allows us 
to develop relational techniques for truly comonadic calculi, such as 
those based on Uustalu and Vene's comonadic notions of computation
\cite{DBLP:journals/entcs/UustaluV08}.

\paragraph{Related Work}
In recent years, there has been a growing interest for typing disciplines 
regulating how code can be manipulated. 
Specific examples of such disciplines date back at least to the 90s, 
originating from (bounded) linear logic \cite{DBLP:journals/tcs/Girard87,DBLP:journals/tcs/GirardSS92,DBLP:conf/lics/BentonW96}, programming languages-based approaches 
to information flow \cite{DBLP:conf/popl/AbadiBHR99,DBLP:journals/jcs/VolpanoIS96}, 
and investigations into the Curry-Howard correspondence for modal logic(s) 
\cite{DBLP:journals/mscs/PfenningD01,DBLP:journals/entcs/PfenningW95,DBLP:journals/sLogica/BiermanP00}.
More recently, researchers started to design calculi with types governing more 
general notions of resource consumptions \cite{DBLP:conf/esop/GhicaS14,Brunel-et-al/ESOP/2014}, quantitative aspects of code usage \cite{Pierce/DistanceMakesTypesGrowStronger/2010,DBLP:conf/lics/Atkey18,Orchard:2019:QPR:3352468.3341714}, and environmental requirements \cite{Mycroft-et-al/ICFP/2014,DBLP:phd/ethos/Orchard14}, this way obtaining general modal-like type systems \cite{Gaboradi-et-al/ICFP/2016,Orchard:2019:QPR:3352468.3341714,DBLP:journals/pacmpl/AbelB20}. 
From a semantical perspective, such systems have been investigated by means of 
(comonadic) denotational semantics 
\cite{DBLP:conf/csl/BreuvartP15,Gaboradi-et-al/ICFP/2016,DBLP:conf/esop/GhicaS14} 
and (mostly heap-based) resource sensitive
operational semantics \cite{Brunel-et-al/ESOP/2014,DBLP:journals/corr/abs-2011-04070,DBLP:journals/pacmpl/AbelB20,DBLP:journals/pacmpl/BernardyBNJS18,Orchard:2019:QPR:3352468.3341714}.

Concerning (operationally-based) program equivalence, the work closest to 
ours is the one by Abel and Bernardy \cite{DBLP:journals/pacmpl/AbelB20}, 
where logical relations for a (call-by-name) $\lambda$-calculus with 
modal and polymorphic types is introduced. As we do in this work, 
Abel and Bernardy define logical relations as monoidal Kripke relations. 
Their treatment of modalities, however, is different from 
ours, as they lack the notion of a comonadic lax extension. 
Moreover, the language of Abel and Bernardy includes polymorphism 
(which we do not have) and is pure and strongly normalising 
(the calculus does not have neither general recursion nor effects), 
whereas $\Lambda_{\gradeone}$ has recursive types and, as argued in 
Remark~\ref{rem:effects} we can safely add algebraic 
effects to it.




\bibliographystyle{IEEEtran}
\bibliography{main}

\begin{thebibliography}{10}
\providecommand{\url}[1]{#1}
\csname url@samestyle\endcsname
\providecommand{\newblock}{\relax}
\providecommand{\bibinfo}[2]{#2}
\providecommand{\BIBentrySTDinterwordspacing}{\spaceskip=0pt\relax}
\providecommand{\BIBentryALTinterwordstretchfactor}{4}
\providecommand{\BIBentryALTinterwordspacing}{\spaceskip=\fontdimen2\font plus
\BIBentryALTinterwordstretchfactor\fontdimen3\font minus
  \fontdimen4\font\relax}
\providecommand{\BIBforeignlanguage}[2]{{%
\expandafter\ifx\csname l@#1\endcsname\relax
\typeout{** WARNING: IEEEtran.bst: No hyphenation pattern has been}%
\typeout{** loaded for the language `#1'. Using the pattern for}%
\typeout{** the default language instead.}%
\else
\language=\csname l@#1\endcsname
\fi
#2}}
\providecommand{\BIBdecl}{\relax}
\BIBdecl

\bibitem{Morris/PhDThesis}
J.~Morris, ``Lambda calculus models of programming languages,'' Ph.D.
  dissertation, MIT, 1969.

\bibitem{MasonTalcott/1991}
I.~A. Mason and C.~L. Talcott, ``Equivalence in functional languages with
  effects,'' \emph{J. Funct. Program.}, vol.~1, no.~3, pp. 287--327, 1991.

\bibitem{Plotkin-Lambda-definability-logical-relations}
G.~Plotkin, ``Lambda-definability and logical relations,'' 1973, technical
  Report SAI-RM-4, School of A.I., University of Edinburgh.

\bibitem{Reynolds/Logical-relations/1983}
J.~Reynolds, ``Types, abstraction and parametric polymorphism,'' in
  \emph{{IFIP} Congress}, 1983, pp. 513--523.

\bibitem{Abramsky/RTFP/1990}
S.~Abramsky, ``The lazy lambda calculus,'' in \emph{Research Topics in
  Functional Programming}, D.~Turner, Ed.\hskip 1em plus 0.5em minus
  0.4em\relax Addison Wesley, 1990, pp. 65--117.

\bibitem{DBLP:conf/popl/AbadiBHR99}
M.~Abadi, A.~Banerjee, N.~Heintze, and J.~G. Riecke, ``A core calculus of
  dependency,'' in \emph{{POPL} '99, Proceedings of the 26th {ACM}
  {SIGPLAN-SIGACT} Symposium on Principles of Programming Languages, San
  Antonio, TX, USA, January 20-22, 1999}, 1999, pp. 147--160.

\bibitem{Gaboradi-et-al/ICFP/2016}
M.~Gaboardi, S.~Katsumata, D.~A. Orchard, F.~Breuvart, and T.~Uustalu,
  ``Combining effects and coeffects via grading,'' in \emph{Proc. of {ICFP}
  2016}, 2016, pp. 476--489.

\bibitem{Brunel-et-al/ESOP/2014}
A.~Brunel, M.~Gaboardi, D.~Mazza, and S.~Zdancewic, ``A core quantitative
  coeffect calculus,'' in \emph{Proc. of {ESOP} 2014}, 2014, pp. 351--370.

\bibitem{Orchard:2019:QPR:3352468.3341714}
D.~Orchard, V.-B. Liepelt, and H.~Eades~III, ``Quantitative program reasoning
  with graded modal types,'' \emph{Proc. ACM Program. Lang.}, vol.~3, no. ICFP,
  pp. 110:1--110:30, 2019.

\bibitem{DBLP:conf/lics/Atkey18}
R.~Atkey, ``Syntax and semantics of quantitative type theory,'' in
  \emph{Proceedings of the 33rd Annual {ACM/IEEE} Symposium on Logic in
  Computer Science, {LICS} 2018, Oxford, UK, July 09-12, 2018}, 2018, pp.
  56--65.

\bibitem{Mycroft-et-al/ICFP/2014}
T.~Petricek, D.~A. Orchard, and A.~Mycroft, ``Coeffects: a calculus of
  context-dependent computation,'' in \emph{Proc. of {ICFP} 2014}, 2014, pp.
  123--135.

\bibitem{DBLP:journals/pacmpl/BernardyBNJS18}
J.~Bernardy, M.~Boespflug, R.~R. Newton, S.~{Peyton Jones}, and A.~Spiwack,
  ``Linear haskell: practical linearity in a higher-order polymorphic
  language,'' \emph{{PACMPL}}, vol.~2, no. {POPL}, pp. 5:1--5:29, 2018.

\bibitem{DBLP:journals/pacmpl/AbelB20}
A.~Abel and J.~Bernardy, ``A unified view of modalities in type systems,''
  \emph{Proc. {ACM} Program. Lang.}, vol.~4, no. {ICFP}, pp. 90:1--90:28, 2020.

\bibitem{DBLP:journals/tcs/Girard87}
J.~Girard, ``Linear logic,'' \emph{Theor. Comput. Sci.}, vol.~50, pp. 1--102,
  1987.

\bibitem{DBLP:journals/tcs/GirardSS92}
J.~Girard, A.~Scedrov, and P.~Scott, ``Bounded linear logic: {A} modular
  approach to polynomial-time computability,'' \emph{Theor. Comput. Sci.},
  vol.~97, pp. 1--66, 1992.

\bibitem{Pierce/DistanceMakesTypesGrowStronger/2010}
J.~Reed and B.~Pierce, ``Distance makes the types grow stronger: a calculus for
  differential privacy,'' in \emph{Proc. of {ICFP} 2010}, 2010, pp. 157--168.

\bibitem{DBLP:journals/jcs/VolpanoIS96}
D.~M. Volpano, C.~E. Irvine, and G.~Smith, ``A sound type system for secure
  flow analysis,'' \emph{Journal of Computer Security}, vol.~4, no. 2/3, pp.
  167--188, 1996.

\bibitem{DBLP:conf/lics/Pfenning01}
F.~Pfenning, ``Intensionality, extensionality, and proof irrelevance in modal
  type theory,'' in \emph{16th Annual {IEEE} Symposium on Logic in Computer
  Science, Boston, Massachusetts, USA, June 16-19, 2001, Proceedings}, 2001,
  pp. 221--230.

\bibitem{DBLP:journals/mst/Lambek68}
J.~Lambek, ``Deductive systems and categories i. syntactic calculus and
  residuated categories,'' \emph{Math. Syst. Theory}, vol.~2, no.~4, pp.
  287--318, 1968.

\bibitem{ROUTLEY1973199}
R.~Routley and R.~K. Meyer, ``The semantics of entailment,'' in \emph{Truth,
  Syntax and Modality}, ser. Studies in Logic and the Foundations of
  Mathematics, H.~Leblanc, Ed.\hskip 1em plus 0.5em minus 0.4em\relax Elsevier,
  1973, vol.~68, pp. 199 -- 243.

\bibitem{DBLP:journals/jsyml/Urquhart72}
A.~Urquhart, ``Semantics for relevant logics,'' \emph{J. Symb. Log.}, vol.~37,
  no.~1, pp. 159--169, 1972.

\bibitem{DBLP:books/daglib/0006566}
D.~J. Pym, \emph{The semantics and proof theory of the logic of bunched
  implications}, ser. Applied logic series.\hskip 1em plus 0.5em minus
  0.4em\relax Kluwer, 2002, vol.~26.

\bibitem{Kurz/Tutorial-relation-lifting/2016}
A.~Kurz and J.~Velebil, ``Relation lifting, a survey,'' \emph{J. Log. Algebr.
  Meth. Program.}, vol.~85, no.~4, pp. 475--499, 2016.

\bibitem{Hoffman-Seal-Tholem/monoidal-topology/2014}
D.~Hofmann, G.~Seal, and W.~Tholen, Eds., \emph{Monoidal Topology. A
  Categorical Approach to Order, Metric, and Topology}, ser. Encyclopedia of
  Mathematics and its Applications.\hskip 1em plus 0.5em minus 0.4em\relax
  Cambridge University Press, 2014, no. 153.

\bibitem{Hoffman/Cottage-industry/2015}
D.~Hoffman, ``A cottage industry of lax extensions,'' \emph{Categories and
  General Algebraic Structures with Applications}, vol.~3, no.~1, pp. 113--151,
  2015.

\bibitem{Barr/LMM/1970}
M.~Barr, ``Relational algebras,'' \emph{Lect. Notes Math.}, vol. 137, pp.
  39--55, 1970.

\bibitem{DBLP:journals/fss/ClementinoH17}
M.~M. Clementino and D.~Hofmann, ``The rise and fall of v-functors,''
  \emph{Fuzzy Sets Syst.}, vol. 321, pp. 29--49, 2017.

\bibitem{Thijs/PhDThesis/1996}
A.~Thijs, \emph{Simulation and fixpoint semantics}.\hskip 1em plus 0.5em minus
  0.4em\relax Rijksuniversiteit Groningen, 1996.

\bibitem{Venema-Marti}
J.~Marti and Y.~Venema, ``Lax extensions of coalgebra functors and their
  logic,'' \emph{J. Comput. Syst. Sci.}, vol.~81, no.~5, pp. 880--900, 2015.

\bibitem{Katsumata-Sato/FOSSACS/20S13}
S.~Katsumata and T.~Sato, ``Preorders on monads and coalgebraic simulations,''
  in \emph{Proc. of {FOSSACS} 2013}, 2013, pp. 145--160.

\bibitem{DalLagoGavazzoLevy/LICS/2017}
U.~Dal~Lago, F.~Gavazzo, and P.~Levy, ``Effectful applicative bisimilarity:
  Monads, relators, and howe's method,'' in \emph{Proc. of {LICS} 2017}, 2017,
  pp. 1--12.

\bibitem{DBLP:conf/esop/LagoG19}
U.~D. Lago and F.~Gavazzo, ``Effectful normal form bisimulation,'' in
  \emph{Proc. of {ESOP} 2019}, 2019, pp. 263--292.

\bibitem{dal-lago/gavazzo-mfps-2019}
U.~Da~Lago and F.~Gavazzo, ``On bisimilarity in lambda calculi with continuous
  probabilistic choice,'' 2019, to appear.

\bibitem{GoubaultLasotaNowak/MSCS/2008}
J.~Goubault{-}Larrecq, S.~Lasota, and D.~Nowak, ``Logical relations for monadic
  types,'' \emph{Mathematical Structures in Computer Science}, vol.~18, no.~6,
  pp. 1169--1217, 2008.

\bibitem{Simpson-Niels/Modalities/2018}
A.~Simpson and N.~Voorneveld, ``Behavioural equivalence via modalities for
  algebraic effects,'' in \emph{Proc. of {ESOP} 2018}, 2018, pp. 300--326.

\bibitem{DBLP:conf/csl/BreuvartP15}
F.~Breuvart and M.~Pagani, ``Modelling coeffects in the relational semantics of
  linear logic,'' in \emph{Proc. of {CSL} 2015}, 2015, pp. 567--581.

\bibitem{Howe/IC/1996}
D.~Howe, ``Proving congruence of bisimulation in functional programming
  languages,'' \emph{Inf. Comput.}, vol. 124, no.~2, pp. 103--112, 1996.

\bibitem{Pitts/ATBC/2011}
A.~Pitts, ``Howe's method for higher-order languages,'' in \emph{Advanced
  Topics in Bisimulation and Coinduction}, ser. Cambridge Tracts in Theoretical
  Computer Science, D.~Sangiorgi and J.~Rutten, Eds.\hskip 1em plus 0.5em minus
  0.4em\relax Cambridge University Press, 2011, vol.~52, pp. 197--232.

\bibitem{Gavazzo/LICS/2018}
F.~Gavazzo, ``Quantitative behavioural reasoning for higher-order effectful
  programs: Applicative distances,'' in \emph{Proceedings of the 33rd Annual
  {ACM/IEEE} Symposium on Logic in Computer Science, {LICS} 2018, Oxford, UK,
  July 09-12, 2018}, 2018, pp. 452--461.

\bibitem{Lawvere/GeneralizedMetricSpaces/1973}
F.~Lawvere, ``Metric spaces, generalized logic, and closed categories,''
  \emph{Rend. Sem. Mat. Fis. Milano}, vol.~43, pp. 135--166, 1973.

\bibitem{Barendregt/Book/1984}
H.~Barendregt, \emph{The lambda calculus: its syntax and semantics}, ser.
  Studies in logic and the foundations of mathematics.\hskip 1em plus 0.5em
  minus 0.4em\relax North-Holland, 1984.

\bibitem{DBLP:conf/lics/BentonW96}
P.~N. Benton and P.~Wadler, ``Linear logic, monads and the lambda calculus,''
  in \emph{Proceedings, 11th Annual {IEEE} Symposium on Logic in Computer
  Science, New Brunswick, New Jersey, USA, July 27-30, 1996}, 1996, pp.
  420--431.

\bibitem{DBLP:journals/cacm/Denning76}
D.~E. Denning, ``A lattice model of secure information flow,'' \emph{Commun.
  {ACM}}, vol.~19, no.~5, pp. 236--243, 1976.

\bibitem{Levy/InfComp/2003}
P.~Levy, J.~Power, and H.~Thielecke, ``Modelling environments in call-by-value
  programming languages,'' \emph{Inf. Comput.}, vol. 185, no.~2, pp. 182--210,
  2003.

\bibitem{GaboardiEtAl/POPL/2017}
A.~de~Amorim, M.~Gaboardi, J.~Hsu, S.~Katsumata, and I.~Cherigui, ``A semantic
  account of metric preservation,'' in \emph{Proc. of {POPL} 2017}, 2017, pp.
  545--556.

\bibitem{DBLP:journals/mscs/PfenningD01}
F.~Pfenning and R.~Davies, ``A judgmental reconstruction of modal logic,''
  \emph{Math. Struct. Comput. Sci.}, vol.~11, no.~4, pp. 511--540, 2001.

\bibitem{DBLP:journals/jacm/DaviesP01}
R.~Davies and F.~Pfenning, ``A modal analysis of staged computation,'' \emph{J.
  {ACM}}, vol.~48, no.~3, pp. 555--604, 2001.

\bibitem{DBLP:conf/esop/GhicaS14}
D.~R. Ghica and A.~I. Smith, ``Bounded linear types in a resource semiring,''
  in \emph{Programming Languages and Systems - 23rd European Symposium on
  Programming, {ESOP} 2014, Held as Part of the European Joint Conferences on
  Theory and Practice of Software, {ETAPS} 2014, Grenoble, France, April 5-13,
  2014, Proceedings}, 2014, pp. 331--350.

\bibitem{DBLP:journals/corr/abs-2005-02247}
\BIBentryALTinterwordspacing
J.~Wood and R.~Atkey, ``A linear algebra approach to linear metatheory,''
  \emph{CoRR}, vol. abs/2005.02247, 2020. [Online]. Available:
  \url{https://arxiv.org/abs/2005.02247}
\BIBentrySTDinterwordspacing

\bibitem{DBLP:conf/lics/AmorimGHK19}
A.~A. de~Amorim, M.~Gaboardi, J.~Hsu, and S.~Katsumata, ``Probabilistic
  relational reasoning via metrics,'' in \emph{34th Annual {ACM/IEEE} Symposium
  on Logic in Computer Science, {LICS} 2019, Vancouver, BC, Canada, June 24-27,
  2019}, 2019, pp. 1--19.

\bibitem{DBLP:conf/icfp/DAntoniGAHP13}
L.~D'Antoni, M.~Gaboardi, E.~J.~G. Arias, A.~Haeberlen, and B.~C. Pierce,
  ``Sensitivity analysis using type-based constraints,'' in \emph{Proc. of
  FPCDSL@ICFP 2013}, 2013, pp. 43--50.

\bibitem{DBLP:journals/pacmpl/BartheEGHS18}
G.~Barthe, T.~Espitau, B.~Gr{\'{e}}goire, J.~Hsu, and P.~Strub, ``Proving
  expected sensitivity of probabilistic programs,'' \emph{Proc. {ACM} Program.
  Lang.}, vol.~2, no. {POPL}, pp. 57:1--57:29, 2018.

\bibitem{DBLP:phd/basesearch/Gavazzo19}
\BIBentryALTinterwordspacing
F.~Gavazzo, ``Coinductive equivalences and metrics for higher-order languages
  with algebraic effects,'' Ph.D. dissertation, University of Bologna, Italy,
  2019. [Online]. Available: \url{http://amsdottorato.unibo.it/9075/}
\BIBentrySTDinterwordspacing

\bibitem{DBLP:conf/fossacs/MatacheS19}
C.~Matache and S.~Staton, ``A sound and complete logic for algebraic effects,''
  in \emph{Foundations of Software Science and Computation Structures - 22nd
  International Conference, {FOSSACS} 2019, Held as Part of the European Joint
  Conferences on Theory and Practice of Software, {ETAPS} 2019, Prague, Czech
  Republic, April 6-11, 2019, Proceedings}, 2019, pp. 382--399.

\bibitem{DBLP:books/daglib/0085577}
J.~C. Mitchell, \emph{Foundations for programming languages}, ser. Foundation
  of computing series.\hskip 1em plus 0.5em minus 0.4em\relax {MIT} Press,
  1996.

\bibitem{Lassen/PhDThesis}
S.~Lassen, ``Relational reasoning about functions and nondeterminism,'' Ph.D.
  dissertation, Dept. of Computer Science, University of Aarhus, May 1998.

\bibitem{CrubilleDalLago/LICS/2015}
R.~Crubill{\'{e}} and U.~Dal~Lago, ``Metric reasoning about {lambda}-terms: The
  affine case,'' in \emph{Proc. of {LICS} 2015}, 2015, pp. 633--644.

\bibitem{mitchell_2002}
J.~C. Mitchell, \emph{Concepts in Programming Languages}.\hskip 1em plus 0.5em
  minus 0.4em\relax Cambridge University Press, 2002.

\bibitem{DBLP:journals/lisp/SabelfeldS01}
A.~Sabelfeld and D.~Sands, ``A per model of secure information flow in
  sequential programs,'' \emph{High. Order Symb. Comput.}, vol.~14, no.~1, pp.
  59--91, 2001.

\bibitem{DBLP:conf/icfp/BowmanA15}
W.~J. Bowman and A.~Ahmed, ``Noninterference for free,'' in \emph{Proceedings
  of the 20th {ACM} {SIGPLAN} International Conference on Functional
  Programming, {ICFP} 2015, Vancouver, BC, Canada, September 1-3, 2015}, 2015,
  pp. 101--113.

\bibitem{Gordon/FOSSACS/01}
A.~Gordon, ``A tutorial on co-induction and functional programming,'' in
  \emph{Workshops in Computing}.\hskip 1em plus 0.5em minus 0.4em\relax
  Springer London, September 1994, pp. 78--95.

\bibitem{Rosenthal/Quantales/1990}
K.~Rosenthal, \emph{Quantales and their applications}, ser. Pitman research
  notes in mathematics series.\hskip 1em plus 0.5em minus 0.4em\relax Longman
  Scientific \& Technical, 1990.

\bibitem{Vickers/Topology-via-logic}
S.~Vickers, \emph{Topology Via Logic}, ser. Cambridge Tracts in
  Theoretica.\hskip 1em plus 0.5em minus 0.4em\relax Cambridge University
  Press, 1996.

\bibitem{steen/CounterexamplesTopology/1995}
L.~Steen and J.~Seebach, \emph{Counterexamples in Topology}, ser. Dover books
  on mathematics.\hskip 1em plus 0.5em minus 0.4em\relax Dover Publications,
  1995.

\bibitem{hajek1998metamathematics}
P.~H{\'a}jek, \emph{Metamathematics of Fuzzy Logic}, ser. Trends in
  Logic.\hskip 1em plus 0.5em minus 0.4em\relax Springer Netherlands, 1998.

\bibitem{HOFMANN20131}
D.~Hofmann and C.~Reis, ``Probabilistic metric spaces as enriched categories,''
  \emph{Fuzzy Sets and Systems}, vol. 210, pp. 1 -- 21, 2013.

\bibitem{DBLP:journals/acs/ClementinoHT04}
M.~M. Clementino, D.~Hofmann, and W.~Tholen, ``One setting for all: Metric,
  topology, uniformity, approach structure,'' \emph{Appl. Categorical Struct.},
  vol.~12, no.~2, pp. 127--154, 2004.

\bibitem{DBLP:journals/tcs/FlaggK97}
B.~Flagg and R.~Kopperman, ``Continuity spaces: Reconciling domains and metric
  spaces,'' \emph{Theor. Comput. Sci.}, vol. 177, no.~1, pp. 111--138, 1997.

\bibitem{flagg1992completeness}
R.~C. Flagg, ``Completeness in continuity spaces,'' in \emph{AMS Conference
  proceedings}, vol.~13, 1992, pp. 183--199.

\bibitem{Flagg1997}
------, ``Quantales and continuity spaces,'' \emph{algebra universalis},
  vol.~37, no.~3, pp. 257--276, 1997.

\bibitem{DBLP:journals/entcs/PlotkinP01}
G.~D. Plotkin and J.~Power, ``Semantics for algebraic operations,''
  \emph{Electr. Notes Theor. Comput. Sci.}, vol.~45, pp. 332--345, 2001.

\bibitem{PlotkinPower/FOSSACS/01}
------, ``Adequacy for algebraic effects,'' in \emph{Proc. of {FOSSACS} 2001},
  2001, pp. 1--24.

\bibitem{Plotkin/algebraic-operations-and-generic-effects/2003}
------, ``Algebraic operations and generic effects,'' \emph{Applied Categorical
  Structures}, vol.~11, no.~1, pp. 69--94, 2003.

\bibitem{DBLP:journals/entcs/UustaluV08}
T.~Uustalu and V.~Vene, ``Comonadic notions of computation,'' in
  \emph{Proceedings of the Ninth Workshop on Coalgebraic Methods in Computer
  Science, {CMCS} 2008, Budapest, Hungary, April 4-6, 2008}, 2008, pp.
  263--284.

\bibitem{DBLP:journals/entcs/PfenningW95}
F.~Pfenning and H.~Wong, ``On a modal lambda calculus for {S4},'' in
  \emph{Eleventh Annual Conference on Mathematical Foundations of Programming
  Semantics, {MFPS} 1995, Tulane University, New Orleans, LA, USA, March 29 -
  April 1, 1995}, 1995, pp. 515--534.

\bibitem{DBLP:journals/sLogica/BiermanP00}
G.~M. Bierman and V.~de~Paiva, ``On an intuitionistic modal logic,'' \emph{Stud
  Logica}, vol.~65, no.~3, pp. 383--416, 2000.

\bibitem{DBLP:phd/ethos/Orchard14}
D.~A. Orchard, ``Programming contextual computations,'' Ph.D. dissertation,
  University of Cambridge, {UK}, 2014.

\bibitem{DBLP:journals/corr/abs-2011-04070}
\BIBentryALTinterwordspacing
P.~Choudhury, H.~E. III, R.~A. Eisenberg, and S.~C. Weirich, ``A graded
  dependent type system with a usage-aware semantics (extended version),''
  \emph{CoRR}, vol. abs/2011.04070, 2020. [Online]. Available:
  \url{https://arxiv.org/abs/2011.04070}
\BIBentrySTDinterwordspacing

\bibitem{JohannSimpsonVoigtlander/LICS/2010}
P.~Johann, A.~Simpson, and J.~Voigtl{\"{a}}nder, ``A generic operational
  metatheory for algebraic effects,'' in \emph{Proc. of {LICS} 2010}.\hskip 1em
  plus 0.5em minus 0.4em\relax {IEEE} Computer Society, 2010, pp. 209--218.

\bibitem{MacLane/Book/1971}
S.~MacLane, \emph{Categories for the Working Mathematician}.\hskip 1em plus
  0.5em minus 0.4em\relax Springer-Verlag, 1971.

\bibitem{AbramskyJung/DomainTheory/1994}
S.~Abramsky and A.~Jung, ``Domain theory,'' in \emph{Handbook of Logic in
  Computer Science}.\hskip 1em plus 0.5em minus 0.4em\relax Clarendon Press,
  1994, pp. 1--168.

\bibitem{DBLP:books/cu/Schmidt2011}
G.~Schmidt, \emph{Relational Mathematics}, ser. Encyclopedia of Mathematics and
  its Applications.\hskip 1em plus 0.5em minus 0.4em\relax Cambridge University
  Press, 2011, vol. 132.

\bibitem{Goguen/1977}
J.~A. Goguen, J.~W. Thatcher, E.~G. Wagner, and J.~B. Wright, ``Initial algebra
  semantics and continuous algebras,'' \emph{J. {ACM}}, vol.~24, no.~1, pp.
  68--95, 1977.

\bibitem{Kelly/EnrichedCats}
G.~M. Kelly, ``Basic concepts of enriched category theory,'' \emph{Reprints in
  Theory and Applications of Categories}, no.~10, pp. 1--136, 2005.

\bibitem{Villani/optimal-transport/2008}
C.~Villani, \emph{Optimal Transport: Old and New}, ser. Grundlehren der
  mathematischen Wissenschaften.\hskip 1em plus 0.5em minus 0.4em\relax
  Springer Berlin Heidelberg, 2008.

\bibitem{Kortanek/InfiniteTransportationProblems/1995}
K.~Kortanek and M.~Yamasaki, ``Discrete infinite transportation problems,''
  \emph{Discrete Applied Mathematics}, no.~58, pp. 19--33, 1995.

\bibitem{Wasserstein-metric-and-subordination}
P.~Cl\'ement and W.~Desch, ``Wasserstein metric and subordination,'' 2008.

\bibitem{Munkres/Topology/2000}
J.~Munkres, \emph{Topology}, ser. Featured Titles for Topology Series.\hskip
  1em plus 0.5em minus 0.4em\relax Prentice Hall, Incorporated, 2000.

\bibitem{AKHVLEDIANI20101275}
A.~Akhvlediani, M.~M. Clementino, and W.~Tholen, ``On the categorical meaning
  of hausdorff and gromov distances, i,'' \emph{Topology and its Applications},
  vol. 157, no.~8, pp. 1275 -- 1295, 2010.

\bibitem{Hoffman-topological-theories-as-closed-objects}
D.~Hofmann, ``Topological theories and closed objects,'' \emph{Adv. Math.},
  vol. 215, pp. 789--824, 2007.

\end{thebibliography}

\newpage
\appendices

\onecolumn
\newgeometry{left=3cm, right=3cm, bottom=4cm,top=4cm}
\onecolumn

\section{Modal Calculi}
\label{appendix:modal-calculi}

\newcommand{\kleislimaybe}[1]{#1^{\scriptstyle \Maybe}}
\newcommand{\unitmaybe}{\unit^{\scriptstyle \Maybe}}

In this Section, we provide further details on the syntax and 
(operational) semantics of $\Lambda_{\gradealg}$. First, we
extend the semiring structure of $\gradealg$ to environments 
pointwise. More precisely, we first require two
environments $\envone, \envtwo$ to be compatible, meaning that if $\varone$ 
appears both in $\envone$ and $\envtwo$, then it has the same type. 
We tacitly assume environments to be 
pairwise compatible. 

\begin{definition}
For an operation ${\circ} \in \{\gplus, \gstar\}$, 
we then define $\envone \circ \envtwo$ as follows, 
where the last clause handles the case for $\varone$ not appearing among variables in 
$\envtwo$. 
\begin{align*}
  \envone \circ \emptyenv &\defeq \envone 
  & 
   (\graded{\varone}{\typeone}{\gradeone}, \envone)
   \circ 
   (\graded{\varone}{\typeone}{\gradetwo}, \envtwo)
   &\defeq\graded{\varone}{\typeone}{\gradeone \circ \gradetwo}, \envone \circ \envtwo
   \\
   \emptyenv \circ \envone &\defeq \envone 
   & 
   (\graded{\varone}{\typeone}{\gradeone}, \envone)
   \circ 
  \envtwo
   &\defeq \graded{\varone}{\typeone}{\gradeone}, \envone \circ \envtwo
\end{align*}
\end{definition}

We endow $\Lambda_{\gradealg}$ with an inductive 
call-by-value big step semantics given by the 
rules Figure~\ref{fig:operational-semantics-pure}, 
where in a judgment $\termone \Downarrow \valone$, 
$\termone$ is a closed term and $\valone$ is closed values 
(of the same type of $\termone$).

    \begin{figure*}[htbp]
    \hrule 
    \vspace{0.2cm}
    \[
    \infer{(\abs{\varone}{\termone})\valone \To \valtwo}
    {\subst{\termone}{\varone}{\valone} \To \valtwo}
    \qquad 
    \infer{\letfold{(\fold{\valone})}{\termone} \To \valtwo}
    {\subst{\termone}{\varone}{\valone} \To \valtwo}
    \qquad
    \infer{\letbox{(\tbox{\valone})}{\termone} \To \valtwo}
    {\subst{\termone}{\varone}{\valone} \To \valtwo}
    \qquad
    \infer{\valone \To \valone}{}
    \qquad
    \infer{\seq{\termone}{\termtwo} \To \valtwo}
    {\termone \To \valone 
    &\subst{\termtwo}{\varone}{\valone} \To \valtwo}
    \]

    \hrule
    \caption{Call-by-Value Operational Semantics}
    \label{fig:operational-semantics-pure}
    \end{figure*}

To simplify the meta-theory of $\Lambda_{\gradealg}$, 
we also endow $\Lambda_{\gradealg}$ with a more abstract \emph{monadic} operational 
semantics \cite{PlotkinPower/FOSSACS/01,JohannSimpsonVoigtlander/LICS/2010,DalLagoGavazzoLevy/LICS/2017}: this allow us 
to rely on the abstract Howe's method of Dal Lago et al.
\cite{DalLagoGavazzoLevy/LICS/2017} 
to prove congruence properties of applicative bisimilarity 
as well as to smoothly extend $\Lambda_{\gradealg}$ with effectful primitives.
    
\begin{definition}
The \emph{maybe} or \emph{partiality} monad\footnote{Recall that a monad 
  is a triple $\Monad = (\monad, \unit, \mu)$, with $\monad$ 
  an endofunctor (we consider the case of $\set$ monads only), 
  and $\unit_X: X \to \monad(X)$ and $\mu_X: \monad(\monad(X)) \to \monad(X)$ 
  natural transformations, subject to suitable coherence conditions. 
  Oftentimes, we do not work with monads directly but with the 
  equivalent notion of a Kleisli triple $(\monad, \unit, \kleisli{-})$
  \cite{MacLane/Book/1971}.} (on $\set$) is 
  the triple $\Maybe = (\maybe, \unitmaybe, \kleislimaybe{-})$, where
  $\monad (X) = X_{\divergence} = X + \{\divergence\}$ 
  and (where $f: X \to Y_\bot$):
    \begin{align*}
    \unitmaybe(x) &= x 
    & 
    \kleislimaybe{f}(x) = 
    \begin{cases} 
      \divergence & \text{ if } x = \divergence;
      \\ 
      f(x) & \text{ otherwise.}
    \end{cases}
    \end{align*} 
\end{definition}

Sets of the form $X_{\divergence}$ can always en endowed 
with a $\omega$-complete pointed 
partial order ($\omega$-cppo, for short) structure 
\cite{AbramskyJung/DomainTheory/1994} by considering the flat order\footnote{
 Recall that $x \cpoleq y$ iff $x \neq \divergence \text{ implies } x = y.$ 
} $\cpoleq$.
The bottom element of $X_{\divergence}$ is $\divergence$. Moreover, 
any $\omega$-chain $x_0 \cpoleq x_1 \cpoleq \cc$ in $X_{\divergence}$ 
has a least upper bound which we denote by $\lub_{n \geq 0} x_n$.
The monad and $\omega$-cppo structure of the construction $X_{\divergence}$
properly interact, in the sense that 
the following strictness and continuity laws 
hold, where function spaces of the form $\monad(X) \to \monad(Y)$ 
are endowed with the $\omega$-cppo structure inherited from $\monad(Y)$ 
pointwise.
\begin{align*}
  \kleislimaybe{f}(\divergence) &= \divergence;
  &
  \kleislimaybe{f}\big(\lub_{n \geq 0} x_n\big) &=  \lub_{n \geq 0} \kleislimaybe{f}(x_n);
  &
  \big(\lub_{n \geq 0} \kleislimaybe{f}\big)(x) &= \lub_{n \geq 0} \kleislimaybe{f}(x).
\end{align*}

We can now define an evaluation map mapping each closed term $\termone$ 
to an element $\sem{\termone} \in \values_{\divergence}$. Notice that 
the map $\sem{-}: \Lambda \to \values_{\divergence}$ is thus equivalent 
to a deterministic relation.

\begin{definition}
Define the $\mathbb{N}$-indexed family of (type-indexed) evaluation maps\footnote{
  We omit type subscripts.
} 
$\evalsymboln{n}: \Lambda \to \values_{\divergence}$ recursively as follows:
\begin{align*}
\evaln{0}{\termone} &\defeq \divergence
\\
\evaln{n+1}{\valone} &\defeq \unitmaybe(\valone)
\\
\evaln{n+1}{(\lambda x.\termone)\valone} &\defeq \evaln{n}{\subst{\termone}{\varone}{\valone}}
\\
\evaln{n+1}{\unfold{(\fold{\valone})}} 
&\defeq \evaln{n}{\subst{\termone}{\varone}{\valone}}
\\
\evaln{n+1}{\letbox{\tbox{\valone}}{\termone}}
&\defeq \evaln{n}{\subst{\termone}{\varone}{\valone}}
\\
\evaln{n+1}{\seq{\termone}{\termtwo}} 
&\defeq \kleislimaybe{(\valone \mapsto \evaln{n}{\subst{\termtwo}{\varone}{\valone}})} 
(\evaln{n}{\termone}).
\end{align*}
\end{definition}

The function $\evalsymboln{n}$ maps each (closed) computation of type $\typeone$ 
either to a (closed) value of type $\typeone$ or to the divergence symbol $\divergence$. 
Moreover, 
it is straightforward to see 
that for any (closed) computation $\termone$ we have an $\omega$-chain 
$\evaln{0}{\termone} \cpoleq \evaln{1}{\termone} \cpoleq \cc$ 
so that we ca define $\eval{\termone} = \lub_{n \geq 0} \evaln{n}{\termone}$.

\begin{lemma}
Let $\termone$ be an arbitrary closed computation. Then:
$\termone \To \valone \iff \eval{\termone} = \valone$
and
$\termone \not\To \iff \eval{\termone} = \divergence$.
\end{lemma}

Now that we have endowed $\Lambda_{\gradealg}$ with a typing system 
and an operational semantics, we move to the main topic of this work: 
relational reasoning and program equivalence.

\section{Relational Reasoning}
\label{appendix:categories-of-relations}

We give some preliminaries on relational reasoning.
Oftentimes, it will be helpful to reason about 
\emph{monoidal Kripke relations} 
pointfree style. It is thus useful to keep in mind the 
pointwise reading of relations  
of the form $f; \reltwo; \dual{g}$,
for a relation $\reltwo: Z \torel W$ and functions 
$f: X \to Z$, $g: Y \to W$:
$$
x \mathbin{(f; \reltwo; \dual{g})(\wone)} y \iff 
f(x) \mathbin{\reltwo(\wone)} g(y).
$$
Given $\relone: X \torel Y$ we can thus express a generalised  
monotonicity condition in pointfree fashion as:
$$
\relone \subseteq f; \reltwo; \dual{g}.
$$
Indeed, taking $f = g$, we obtain standard monotonicity of $f$. 
We will make extensively use of the following \emph{adjunction rules} \cite{Hoffman-Seal-Tholem/monoidal-topology/2014}
(also knowns as shunting \cite{DBLP:books/cu/Schmidt2011}), 
for $f: X \to Y$, $g: Y \to Z$, $\relone: X \torel Y$, 
$\reltwo: Y \torel Z$, and $\relthree: X \torel Z$:
\begin{align*}
\relone; g \subseteq \relthree 
  &\iff \relone \subseteq  \relthree; \dual{g}
  \label{adj-1}\tag{adj 1}
  \\
  \dual{f}; \relthree \subseteq \reltwo 
  &\iff
  \relthree \subseteq f; \reltwo.
  \label{adj-2}\tag{adj 2}
\end{align*}
Using \eqref{adj-1} and \eqref{adj-2} we see that generalised monotonicity 
$\relone \subseteq f; \reltwo; \dual{g}$ can be equivalently 
expressed via the following lax commutative diagram: 
\begin{center}
\(
\vcenter{
\xymatrix{
\laxcommuterel
X     
\ar[r]^{f} 
\ar[d]_{\relone}|@{|}  &  
\setthree   
\ar[d]^{\reltwo}|-*=0@{|}
\\
Y   
\ar[r]_{g}   &  
\setfour  } }
\)
 \end{center}
 The diagram acts as a graphical representation of the expression 
 $\relone; g \subseteq f; \reltwo$, which, by \eqref{adj-1},
 is equivalent to $\relone \subseteq f; \reltwo; \dual{g}$.

\subsection{Lax Extensions}

    \begin{lemma}
    \label{lemma:stability}
    Lax commutative diagrams in $\wrelcat$ are preserved by the mapping 
    $X \mapsto F(X), \relone \mapsto \relatorsymbol(\relone)$. That is:
    \[
      \vcenter{\diagramrel{f}{g}{X}{Y}{\relone}{Z}{W}{\reltwo}}
      \implies
      \vcenter{\diagramrel{f}{g}{X}{Y}{\relatorsymbol(\relone)}{F(Z)}{F(W)}{\relatorsymbol(\reltwo)}}
    \]
    \end{lemma}

    \begin{proof}
    Let us consider a lax commutative diagram in $\wrelcat$ 
    expressed in linear notation: $\relone; g \subseteq f;\reltwo$.  
    By shunting, the latter is equivalent to 
    $\dual{f};\relone;g \subseteq \reltwo$. By monotonicity of 
    $\relatorsymbol$, we thus obtain 
    $\relatorsymbol(\dual{f};\relone;g) \subseteq \relatorsymbol(\reltwo)$, 
    and thus 
    $\relatorsymbol(\dual{f}); \relatorsymbol(\relone); \relatorsymbol(g) 
    \subseteq \relatorsymbol(\reltwo)$, by lax functoriality. 
    We now apply stability on $\relatorsymbol(g)$ and $\relatorsymbol(\dual{f})$, 
    this way obtaining (by monotonicity) 
    $\dual{F(f)}; \relatorsymbol(\relone); F(g)  
    \subseteq \relatorsymbol(\reltwo)$, and thus the desired thesis, by 
    shunting.
    \end{proof}

Although $\Lambda_{\gradealg}$ is a \emph{pure} calculus, 
we handled divergence by giving it a monadic (operational) semantics 
based on the partiality monad. We thus follow
Dal Lago et al. \cite{DalLagoGavazzoLevy/LICS/2017} 
and rely on lax extensions of monads to define applicative bisimilarity.

\begin{definition}
\label{def: relator-monad}
A lax extension of a monad $\Monad = (\monad, \unit, \mu)$ is a 
lax extension of $\monad$ satisfying the following laws:
\begin{align*}
\relone \subseteq \unit; \relatorsymbol(\relone); \dual{\unit}
\label{eq:lax-monad-unit}\tag{lax monad 1}
\\
\relatorsymbol(\relatorsymbol(\relone)) 
\subseteq \mu; \relatorsymbol(\relone); \dual{\mu}
\label{eq:lax-monad-mu}\tag{lax monad 2}
\end{align*}
\end{definition}

As before, we can conveniently express laws \eqref{eq:lax-monad-unit} 
and \eqref{eq:lax-monad-mu} as diagrams:
\[
  \vcenter{
  \diagramrel
  {\unit}
  {\unit}
  {X}
  {Y}
  {\relone}
  {\monad(X)}
  {\monad(Y)}
  {\relatorsymbol(\relone)}
  }
  ;
  \qquad
  \vcenter{
  \diagramrel
  {\mu}
  {\mu}
  {\monad(\monad(X))}
  {\monad(\monad(Y))}
  {\relatorsymbol(\relatorsymbol(\relone))}
  {\monad(X)}
  {\monad(Y)}
  {\relatorsymbol(\relone)}
  }
\]

Notice that any lax extension $\relatorsymbol$ 
of a monad $\Monad = (\monad, \unit, \mu)$ satisfies the following law:
\begin{align*}
\relone \subseteq \dual{f}; \relatorsymbol(\reltwo); g
& \implies \relatorsymbol(\relone) \subseteq \dual{(\kleisli{f})}; 
\relatorsymbol(\reltwo); \kleisli{g}
\label{eq:lax-monad-bind}\tag{lax monad bind}
\end{align*}
which can be can conveniently expressed diagrammatically as follows:
\[
  \vcenter{
  \diagramrel
  {f}
  {g}
  {X}
  {Y}
  {\relone}
  {\monad(Z)}
  {\monad(W)}
  {\relatorsymbol(\reltwo)}
  }
  \implies
   \vcenter{
  \diagramrel
  {\kleisli{f}}
  {\kleisli{g}}
  {\monad(X)}
  {\monad(Y)}
  {\relatorsymbol(\relone)}
  {\monad(Z)}
  {\monad(W)}
  {\relatorsymbol(\reltwo)}
  }
\]
Actually, in presence of law \eqref{eq:lax-monad-unit} the laws 
\eqref{eq:lax-monad-mu} and
\eqref{eq:lax-monad-bind} are equivalent \cite{DBLP:phd/basesearch/Gavazzo19}.

\begin{proposition}
\label{proposition:relator-partiality}
Let $\relone: X \torel Y$ be a $\Worlds$-relation. 
Define $\relone_{\divergence}: X_{\divergence} \to Y_{\divergence}$
as follows:
\begin{align*}
\wrel{\relone_{\divergence}}
  {x}
  {y}
  {\wone}
  &\iff 
 x \neq \divergence \to (y \neq \divergence \;\&\;
  \wrel{\relone}
  {x}
  {y}
  {\wone}).
\end{align*}
Then, $(-)_{\divergence}$ is a lax extension of the maybe/partiality monad.
\end{proposition}

Intuitively, 
$\relone_{\divergence}$ gives a generalisation of the usual clause used 
to define operational preorders between programs. Accordingly, 
a term $\termone$ approximates the behaviour of a term 
$\termtwo$ at world $\wone$ if either $\termone$ diverges or 
both $\termone$ and $\termtwo$ converge
and the resulting values are related at $\wone$. 
If we take $([0,\infty], \leq, +, 0)$ as possible worlds structure
and read $\wrel{\relone}{\termone}{\termtwo}{\gradeone}$ as stating that the $\relone$-distance 
between $\termone$ and $\termtwo$ is at most $\gradeone$, 
then $\wrel{\relone_{\divergence}}{\termone}{\termtwo}{\gradeone}$ 
tells us that if $\termone$ diverges, 
then the $\relone_{\divergence}$-distance between $\termone$ and $\termtwo$ 
is bounded by any $\gradeone$ --- and thus it is bounded by $0$. 
Otherwise, $\termone$ converges to value $\valone$, and thus $\termtwo$ converges to
a value $\valtwo$ such that the $\relone_{\divergence}$-distance between 
$\termone$ and $\termtwo$ is the $\relone$-distance between 
$\valone$ and $\valtwo$.

\section{Howe's Method}
\label{appendix:howe-method}

\subsection{Applicative Bisimilarity}
First, recall the definition of a modal applicative (bi)simulation.

 \begin{definition}
    Recall the definition of the relator $\relatorsymbol^{\divergence}$ 
    for the partiality monad given in Proposition~\ref{proposition:relator-partiality}.
    Define the mapping $\relone \mapsto [\relone]$ on 
    \emph{closed} term relations as follows:
    \begin{align*}
    \wrelcompclosed{[\relone]}{\termone}{\termtwo}{\wone}{\typeone}
    &\iff 
    \eval{\termone} \mathbin{\relone^{\scriptscriptstyle \values}_{\bot}(\wone)} \eval{\termtwo}
    \tag{App eval} \label{eq:app-eval}
    \\
    \wrelvalclosed{[\relone]}{\abs{\varone}{\termone}}{\abs{\varone}{\termtwo}}{\wone} 
    {\typeone \to \typetwo }
    &\iff \forall \valone \in \values_{\typeone}.\ 
    \wrelcompclosed{\relone}{\subst{\termone}{\varone}{\valone}}
    {\subst{\termtwo}{\varone}{\valone}}{\wone}{\typetwo}
    \tag{App abs} \label{eq:app-abs}
    \\
    \wrelvalclosed{[\relone]}{\fold{\valone}}{\fold{\valtwo}}{\wone}{
      \rectype{\typevarone}{\typeone}}
    &\iff
    \wrelvalclosed{\relone}{\valone}{\valtwo}{\wone}{
      \substtype{\typeone}{\typevarone}{\rectype{\typevarone}{\typeone}}}
    \tag{App fold} \label{eq:app-fold}
    \\
    \wrelvalclosed{[\relone]}{\tbox{\valone}}{\tbox{\valtwo}}{\wone}{\bbox_{\gradeone} \typeone} 
    &\iff \wrelvalclosed{\corelator{\gradeone}{\relone}}{\valone}{\valtwo}{\wone}{\typeone} 
    \tag{App box} \label{eq:app-box}
    \end{align*}
    We sat that a closed term relation $\relone$ is an \emph{applicative simulation} 
    if $\relone \subseteq [\relone]$, and that $\relone$ is an \emph{applicative bisimulation} 
    if both $\relone$ and $\dual{\relone}$ are applicative simulation.
    \end{definition}
Equivalently, $\relone$ is an applicative simulation if the following hold:
    \begin{align*}
        \wrelcompclosed{\relone}{\termone}{\termtwo}{\wone}{\typeone}
         &\implies 
         \eval{\termone} \mathbin{\relone^{\scriptscriptstyle \values}_{\bot}(\wone)} \eval{\termtwo}
        \\
        \wrelvalclosed{\relone}{v}{w}{\wone} 
        {\typeone \to \typetwo }
        &\implies \forall v \in \values_{\typeone}.
        \wrelcompclosed{\relone}{vu}
        {wu}{\wone}{\typetwo}
        \\
        \wrelvalclosed{\relone}{\fold{\valone}}{\fold{\valtwo}}{\wone}{
          \rectype{\typevarone}{\typeone}}
        &\implies
        \wrelvalclosed{\relone}{\valone}{\valtwo}{\wone}{
          \substtype{\typeone}{\typevarone}{\rectype{\typevarone}{\typeone}}}
        \\
        \wrelvalclosed{\relone}{\tbox{\valone}}{\tbox{\valtwo}}{\wone}{\bbox_{\gradeone} \typeone} 
        &\implies \wrelvalclosed{\corelator{\gradeone}{\relone}}{\valone}{\valtwo}{\wone}{\typeone}. 
        \end{align*}

\begin{proposition}
Applicative similarly $\appsimilarity$ is a preorder term relation, and 
applicative bisimilarity $\appbisimilarity$ is an equivalence term relation.
\end{proposition}

\begin{proof}
By coinduction. For instance, we show that 
${\appsimilarity}; {\appsimilarity}$ 
is an applicative simulation, and thus it is included in $\appsimilarity$. 
We show one case as a paradigmatic example. 
Suppose 
$\wrelcompclosed{({\appsimilarity}; {\appsimilarity})}{\termone}{\termthree}
{\wone}{\typeone}$, so that we have $\wone \wgeq \wtwo \wcomp \wthree$ with 
$\wrelcompclosed{\appsimilarity}{\termone}{\termtwo}
{\wtwo}{\typeone}$ and $\wrelcompclosed{\appsimilarity}{\termtwo}{\termthree}
{\wthree}{\typeone}$, for suitable worlds $\wtwo, \wthree$, and an expression $\termtwo$. 
Say $\termone \To \valone$. Then, 
$\wrelcompclosed{\appsimilarity}{\termone}{\termtwo}{\wtwo}{\typeone}$ 
implies $\termtwo \To \valtwo$ 
for some value $\valtwo$ such that 
$\wrelcompclosed{\appsimilarity}{\valone}{\valtwo}{\wtwo}{\typeone}$. 
From $\termtwo \To \valtwo$ and $\wrelcompclosed{\appsimilarity}{\termtwo}{\termthree}
{\wthree}{\typeone}$
we obtain the existence of a value $\valthree$ such that $\termthree \To \valthree$ and
$\wrelcompclosed{\appsimilarity}{\valtwo}{\valthree}
{\wthree}{\typeone}$. We thus conclude 
$\wrelcompclosed{({\appsimilarity}; {\appsimilarity})}{\valone}{\valthree}
{\wone}{\typeone}$, and thus we are done.
\end{proof}

\subsection{Howe's Method}

It is convenient to give an explicit, syntax-oriented characterisation 
of the Howe extension of a term relation $\relone$. We do so by means of judgments 
of the form $\wrelo{\envone}{\howe{\relone}}{\termone}{\termtwo}{\wone}{\typeone}$ 
(declined, as usual into value and computation judgments) and of the inference 
rules in Figure~\ref{fig:howe-extension}. 
Then, two open terms $\envone \imp \termone, \termtwo: \typeone$ are related 
by $\howe{\relone}$ at world $\wone$ if and only if 
$\wrelo{\envone}{\howe{\relone}}{\termone}{\termtwo}{\wone}{\typeone}$ is derivable.  
We also define the relation $\howen{\relone}{n}$, for $n \in \mathbb{N}$, 
by saying that $\envone \imp \termone, \termtwo: \typeone$ are related 
by $\howen{\relone}{n}$ at world $\wone$ if and only if 
$\wrelo{\envone}{\howe{\relone}}{\termone}{\termtwo}{\wone}{\typeone}$ is derivable 
with a derivation of depth at most $n$. As a consequence, we see that 
$\howe{\relone} = \bigcup_{n \geq 0} \howen{\relone}{n}$.

\begin{figure*}[htbp]
\hrule
 $\vspace{0.2cm}$
\[
\infer[(\gradeone \mgeq \munit)]{
  \wrelval
  {\envone, \graded{\varone}{\typeone}{\gradeone}}
  {\howe{\relone}}
  {\varone}
  {\valone}
  {\wone}
  {\typeone}
}
{ \wrelval
  {\envone, \graded{\varone}{\typeone}{\gradeone}}
  {\open{\relone}}
  {\varone}
  {\valone}
  {\wone}
  {\typeone}}
\]
$\vspace{-0.1cm}$
\[
\infer{
  \wrelcomp
  {\envone}
  {\howe{\relone}}
  {\valone}
  {\valtwo}
  {\wone}
  {\typeone}
}
{
  \wrelval
  {\envone}
  {\howe{\relone}}
  {\valone}
  {\valthree}
  {\wtwo}
  {\typeone}
  &
  \wrelcomp
  {\envone}
  {\open{\relone}}
  {\valthree}
  {\valtwo}
  {\wthree}
  {\typeone}
  &
  \wone \wgeq \wtwo \wcomp \wthree
}
\]
$\vspace{-0.1cm}$
\[
\infer{
  \wrelval
  {\envone}
  {\howe{\relone}}
  {\abs{\varone}{\termone}}
  {\valtwo}
  {\wone}
  {\typeone \to \typetwo}
}
{
  \wrelcomp
  {\envone, \graded{\varone}{\typeone}{\munit}}
  {\howe{\relone}}
  {\termone}
  {\termtwo}
  {\wtwo}
  {\typetwo}
  & 
  \wrelval
  {\envone}
  {\open{\relone}}
  {\abs{\varone}{\termtwo}}
  {\valone}
  {\wthree}
  {\typeone \to \typetwo}
  & 
  \wone \wgeq \wtwo \wcomp \wthree
}
\]
$\vspace{-0.1cm}$
\[
\infer{
  \wrelcomp
  {\envone \mplus \envtwo}
  {\howe{\relone}}
  {\valone \valtwo}
  {\termone}
  {\wone}
  {\typetwo}
}
{
  \wrelval
  {\envone}
  {\howe{\relone}}
  {\valone}
  {\valone'}
  {\wtwo}
  {\typeone \to \typetwo}
  &
  \wrelval
  {\envtwo}
  {\howe{\relone}}
  {\valtwo}
  {\valtwo'}
  {\wthree}
  {\typeone}
  & 
  \wrelcomp
  {\envone \mplus \envtwo}
  {\open{\relone}}
  {\valone'\valtwo'}
  {\termone}
  {\wfour}
  {\typetwo} 
  &
  \wone \wgeq \wtwo \wcomp \wthree \wcomp \wfour
}
\]
$\vspace{-0.1cm}$
\[
\infer{
  \wrelcomp
  {(\gradeone \mvee \munit) \mstar \envone \mplus \envtwo}
  {\howe{\relone}}
  {\seq{\termone}{\termtwo}}
  {\termthree}
  {\wone}
  {\typetwo}
}
{
    \wrelcomp
    {\envone}
    {\corelator{\gradeone \mvee \munit}{\howe{\relone}}}
    {\termone}
    {\termone'}
    {\wtwo}
    {\typeone}
    &
    \wrelcomp
    {\envtwo, \graded{\varone}{\typeone}{\gradeone}}
    {\howe{\relone}}
    {\termtwo}
    {\termtwo'}
    {\wthree}
    {\typetwo}
    &
    \wrelcomp
    {(\gradeone \mvee \munit) \mstar \envone \mplus \envtwo}
    {\open{\relone}}
    {\seq{\termone'}{\termtwo'}}
    {\termthree}
    {\wfour}
    {\typetwo}
    &
  \wone \wgeq \wtwo \wcomp \wthree \wcomp \wfour
}
\]
$\vspace{-0.1cm}$
\[
\infer{
  \wrelval
  {\envone}
  {\howe{\relone}}
  {\fold{\valone}}
  {\valthree}
  {\wone}
  {\rectype{\typevarone}{\typeone}}
}
{
  \wrelval
  {\envone}
  {\howe{\relone}}
  {\valone}
  {\valtwo}
  {\wtwo}
  {\substtype{\typeone}{\typevarone}{\rectype{\typevarone}{\typeone}}}
  &
  \wrelval
  {\envone}
  {\open{\relone}}
  {\fold{\valtwo}}
  {\valthree}
  {\wthree}
  {\rectype{\typevarone}{\typeone}}
  &
  \wone \wgeq \wtwo \wcomp \wthree
}
\]
$\vspace{-0.1cm}$
\[
\infer{
  \wrelcomp
  {\gradeone \mstar \envone \mplus \envtwo}
  {\howe{\relone}}
  {\letfold{\valone}{\termone}}
  {\termthree}
  {\wone}
  {\typetwo}
}
{
\wrelval
    {\envone}
    {\corelator{\gradeone}{\howe{\relone}}}
    {\valone}
    {\valtwo}
    {\wtwo}
    {\rectype{\typevarone}{\typeone}}
&
     \wrelcomp
    {\envtwo, \graded{\varone}
    {\substtype{\typeone}{\typevarone}{\rectype{\typevarone}{\typeone}}}{\gradeone}}
    {\howe{\relone}}
    {\termone}
    {\termtwo}
    {\wthree}
    {\typetwo}
  &
   \wrelcomp
  {\gradeone \mstar \envone \mplus \envtwo}
  {\open{\relone}}
  {\letfold{\valtwo}{\termtwo}}
  {\termthree}
  {\wfour}
  {\typetwo}
  &
  \wone \wgeq \wtwo \wcomp \wthree \wcomp \wfour
}
\]
$\vspace{-0.1cm}$
\[
\infer{
  \wrelval
  {\gradeone \mstar \envone}
  {\howe{\relone}}
  {\tbox{\valone}}
  {\valthree}
  {\wone}
  {\bbox_{\gradeone}\typeone}
}
{
  \wrelval
  {\envone}
  {\corelator{\gradeone}{\howe{\relone}}}
  {\valone}
  {\valtwo}
  {\wtwo}
  {\typeone}
  &
  \wrelval
  {\gradeone \mstar \envone}
  {\open{\relone}}
  {\tbox{\valtwo}}
  {\valthree}
  {\wthree}
  {\bbox_{\gradeone}\typeone}
  &
  \wone \wgeq \wtwo \wcomp \wthree
}
\]
$\vspace{-0.1cm}$
\[
\infer{
  \wrelcomp
  {\gradetwo \mstar \envone \mplus \envtwo}
  {\howe{\relone}}
  {\letbox{\valone}{\termone}}
  {\termthree}
  {\wone}
  {\typetwo}
}
{
   \wrelval
   {\envone}
   {\corelator{\gradetwo}{\howe{\relone}}}
   {\valone}
   {\valtwo}
   {\wtwo}
   {\bbox_{\gradeone}\typeone}
   &
   \wrelcomp
   {\envtwo, \graded{\varone}{\typeone}{\gradetwo \mstar \gradeone}}
   {\howe{\relone}}
   {\termone}
   {\termtwo}
   {\wthree}
   {\typetwo}
  &
  \wrelcomp
  {\gradetwo \mstar \envone \mplus \envtwo}
  {\open{\relone}}
  {\letbox{\valtwo}{\termtwo}}
  {\termthree}
  {\wfour}
  {\typetwo}
  &
  \wone \wgeq \wtwo \wcomp \wthree \wcomp \wfour
}
\]
\hrule
\caption{Howe extension of $\relone$}
\label{fig:howe-extension}
\end{figure*}

\noindent\textbf{Lemma~\ref{lemma:howe-substitutivity}} (Substitutivity)\textbf{.}
\emph{Let $\relone$ be a reflexive and transitive term relation. 
Then, $\howe{\relone}$ is substitutive.
}
{
\renewcommand{\comp}{;}

\begin{proof}[Sketch]
We have to prove $\howe{\relone} \tensor \corelator{\gradeone}{\howe{\relone}} 
\subseteq \substarrow \comp \howe{\relone} \comp \dual{\substarrow}$, 
i.e. the admissibility of the following rule:
\[
\infer{
  \wrelo{\envone}{\howe{\relone}}{\subst{\termone}{\varone}{\valone}}
  {\subst{\termtwo}{\varone}{\valtwo}}{\wone}{\typeone}
}
{\wrelo{\envone, \graded{\varone}{\typetwo}{\gradeone}}{\howe{\relone}}{\termone}
{\termtwo}{\wtwo}{\typeone}
&
\wrelt{\corelator{\gradeone}{\howe{\relone}}}{\valone}{\valtwo}{\wthree}{\typetwo}
&
\wone \wgeq \wtwo \wcomp \wthree}
\]   
Since $\howe{\relone} = \bigcup_{n \geq 0} \howen{\relone}{n}$, it is sufficient 
to prove 
$\left(\bigcup_{n \geq 0} \howen{\relone}{n} \right) 
\tensor \corelator{\gradeone}{\howe{\relone}} 
\subseteq \substarrow \comp \howe{\relone} \comp \dual{\substarrow}$
which itself follows from
$$
\bigcup_{n \geq 0} \big(\howen{\relone}{n} 
\tensor \corelator{\gradeone}{\howe{\relone}} \big) 
\subseteq \substarrow \comp \howe{\relone} \comp \dual{\substarrow}.
$$
Therefore, to prove substitutivity it is sufficient 
to show that
$\howen{\relone}{n} 
\tensor \corelator{\gradeone}{\howe{\relone}}
\subseteq \substarrow \comp \howe{\relone} \comp \dual{\substarrow}$ 
holds, for any $n \in \mathbb{N}$.
The latter is nothing but the admissibility of the following ($(\mathbb{N}$-) 
indexed rule(s):
\[
\infer{
  \wrelo{\envone}{\howe{\relone}}{\subst{\termone}{\varone}{\valone}}
  {\subst{\termtwo}{\varone}{\valtwo}}{\wone}{\typeone}
}
{\wrelo{\envone, \graded{\varone}{\typetwo}{\gradeone}}{\howen{\relone}{n}}{\termone}
{\termtwo}{\wtwo}{\typeone}
&
\wrelt{\corelator{\gradeone}{\howe{\relone}}}{\valone}{\valtwo}{\wthree}{\typetwo}
&
\wone \wgeq \wtwo \wcomp \wthree}
\] 
We proceed by induction on $n$. The case for $n = 0$ is trivial. 
We prove $\howen{\relone}{n+1} 
\tensor \corelator{\gradeone}{\howe{\relone}}
\subseteq \substarrow \comp \howe{\relone} \comp \dual{\substarrow}$, assuming 
$\howen{\relone}{m} 
\tensor \corelator{\gradeone}{\howe{\relone}}
\subseteq \substarrow \comp \howe{\relone} \comp \dual{\substarrow}$ for all $m \leq n$. 
Notice that, formally, we are universally quantifying over $\gradeone$. 
\begin{itemize}
  \item Suppose to be in the following case:
    \[
      \infer{
        \wrelval{\envone}{\howe{\relone}}{\valone}
        {\subst{\valthree}{\varone}{\valtwo}}{\wthree}{\typeone}
        }
      {
        \infer[(\gradeone \mgeq \munit)]{
          \wrelval{\envone, \graded{\varone}{\typeone}{\gradeone}}{\howen{\relone}{1}}{\varone}
          {\valthree}{\wone}{\typeone}
        }
        {
        \wrelval{\envone, \graded{\varone}{\typeone}{\gradeone}}{\open{\relone}}{\varone}
        {\valthree}{\wone}{\typeone}
        }
      &
      \wrelvalclosed{\corelator{\gradeone}{\howe{\relone}}}{\valone}{\valtwo}{\wtwo}{\typeone}
      &
      \wthree \wgeq \wone \wcomp \wtwo
      }
    \]
    Then, since $\open{\relone}$ is value substitutive, 
    from 
    $\wrelval{\envone, \graded{\varone}{\typeone}{\gradeone}}{\open{\relone}}{\varone}
    {\valthree}{\wone}{\typeone}$ 
    we infer 
    $\wrelval{\envone}{\open{\relone}}{\valtwo}
    {\subst{\valthree}{\varone}{\valtwo}}{\wone}{\typeone}$. 
    Moreover, since $\gradeone \mgeq \munit$ (and thus 
    $\corelator{\gradeone}{\howe{\relone}} \subseteq \corelator{\munit}{\howe{\relone}} 
    \subseteq \howe{\relone}$), from 
    $\wrelvalclosed{\corelator{\gradeone}{\howe{\relone}}}{\valone}{\valtwo}{\wtwo}{\typeone}$
    we infer 
    $\wrelvalclosed{\howe{\relone}}{\valone}{\valtwo}{\wtwo}{\typeone}$
    and thus (by very definition of term relation)
    $\wrelval{\envone}{\howe{\relone}}{\valone}{\valtwo}{\wtwo}{\typeone}$. 
    Putting things together, we obtain 
    $
    \wrelval{\envone}{(\open{\relone} \comp \howe{\relone})}{\valone}
    {\subst{\valthree}{\varone}{\valtwo}}{\wthree}{\typeone}$, which 
    gives $\wrelval{\envone}{\howe{\relone}}{\valone}
    {\subst{\valthree}{\varone}{\valtwo}}{\wthree}{\typeone}$, thanks 
    to quasi-transitivity.
  \item Suppose to be in the following case:
    \[
      \infer{
        \wrelval{\envone, \graded{\vartwo}{\typetwo}{\gradetwo}}{\howe{\relone}}{\vartwo}
        {\subst{\valthree}{\varone}{\valtwo}}{\wthree}{\typetwo}
        }
      {
        \infer[(\gradetwo \mgeq \munit)]{
          \wrelval{\envone, \graded{\varone}{\typeone}{\gradeone}, 
          \graded{\vartwo}{\typetwo}{\gradetwo}}{\howen{\relone}{1}}{\vartwo}
          {\valthree}{\wone}{\typetwo}
        }
        {
        \wrelval{\envone, \graded{\varone}{\typeone}{\gradeone}, 
        \graded{\vartwo}{\typetwo}{\gradetwo}}{\open{\relone}}{\vartwo}
        {\valthree}{\wone}{\typetwo}
        }
      &
      \wrelvalclosed{\corelator{\gradeone}{\howe{\relone}}}{\valone}{\valtwo}{\wtwo}{\typeone}
      &
      \wthree \wgeq \wone \wcomp \wtwo
      }
    \]
    Then, since $\open{\relone}$ is value substitutive, 
    from 
    $\wrelval{\envone, \graded{\varone}{\typeone}{\gradeone}, 
            \graded{\vartwo}{\typetwo}{\gradetwo}}{\open{\relone}}{\vartwo}
            {\valthree}{\wone}{\typetwo}$
    we infer 
    $
    \wrelval{\envone, \graded{\varone}{\typeone}{\gradeone}}{\open{\relone}}{\vartwo}
        {\subst{\valthree}{\varone}{\valtwo}}{\wone}{\typetwo}$
    and thus 
    $
    \wrelval{\envone, \graded{\varone}{\typeone}{\gradeone}}{\howe{\relone}}{\vartwo}
        {\subst{\valthree}{\varone}{\valtwo}}{\wone}{\typetwo}$. 
    We conclude the thesis, since $\wone \wleq \wone \wcomp \wtwo \wleq \wthree$.
  \item Suppose to be in the following case:
    \[
    \infer{
    \wrelcomp
    {(\gradetwo \mvee \munit) \mstar \envone \mplus \envtwo}
    {\howe{\relone}}
    {\seqy{\subst{\termone}{\varone}{\valone}}{\subst{\termtwo}{\varone}{\valone}}}
    {\subst{\termthree}{\varone}{\valtwo}}
    {\wfive}
    }
    {
    \deduce[\mathcal{D}\vspace{0.3cm}]
    {
    \wrelcomp
      {(\gradetwo \mvee \munit) \mstar \envone \mplus \envtwo, 
        \graded{\varone}{\typeone}{(\gradetwo \mvee \munit) \mstar \gradeone \mplus \gradethree}}
      {\howen{\relone}{n+1}}
      {\seqy{\termone}{\termtwo}}
      {\termthree}
      {\wone}
      {\typethree}
    }
    {}
    &
    \wrelvalclosed
    {\corelator{(\gradetwo \mvee \munit) \mstar \gradeone \mplus \gradethree}{\howe{\relone}}}
    {\valone}
    {\valtwo}
    {\wsix}
    {\typeone}
    &
    \wfive \wgeq \wone \wcomp \wsix
    }
    \]
    where $\mathcal{D}$ is the following derivation
    \[
    \infer{
      \wrelcomp
      {(\gradetwo \mvee \munit) \mstar \envone \mplus \envtwo, 
        \graded{\varone}{\typeone}{(\gradetwo \mvee \munit) \mstar \gradeone \mplus \gradethree}}
      {\howen{\relone}{n+1}}
      {\seqy{\termone}{\termtwo}}
      {\termthree}
      {\wone}
      {\typethree}
    }
    { \deduce[]
      {
      \wrelcomp
      {(\gradetwo \mvee \munit) \mstar (\envone, \graded{\varone}{\typeone}{\gradeone}) 
      \mplus (\envtwo, \graded{\varone}{\typeone}{\gradethree})}
      {\open{\relone}}
      {\seqy{\termone'}{\termtwo'}}
      {\termthree}
      {\wfour}
      {\typethree}
      }
      {
      \deduce[]
      {\wrelcomp
      {\envtwo,\graded{\varone}{\typeone}{\gradethree}, 
       \graded{\vartwo}{\typetwo}{\gradetwo}}
      {\howen{\relone}{n}}
      {\termtwo}
      {\termtwo'}
      {\wthree}
      {\typethree}
      }
      {
      \wrelcomp
      {\envone,\graded{\varone}{\typeone}{\gradeone}}
      {\corelator{\gradetwo \mvee \munit}{\howen{\relone}{n}}}
      {\termone}
      {\termone'}
      {\wtwo}
      {\typetwo}
      }
      }
      & 
      \wone \wgeq \wtwo \wcomp \wthree \wcomp \wfour
    }
\]
By law \eqref{eq:monoidal-2}, from 
$\wrelvalclosed
{\corelator{(\gradetwo \mvee \munit) \mstar \gradeone \mplus \gradethree}{\howe{\relone}}}
{\valone}
{\valtwo}
{\wsix}
{\typeone}$ we obtain:
\begin{align}
&\wrelvalclosed
{\corelator{(\gradetwo \mvee \munit) \mstar \gradeone}{\howe{\relone}}}
{\valone}
{\valtwo}
{\wsix_1}
{\typeone}
\label{auxiliary-subst-seq-1}
\\
&\wrelvalclosed
{\corelator{\gradethree}{\howe{\relone}}}
{\valone}
{\valtwo}
{\wsix_2}
{\typeone}
\label{auxiliary-subst-seq-2}
\\
&\wsix \wgeq \wsix_1 \wcomp \wsix_2.
\end{align}
Moreover, from \eqref{auxiliary-subst-seq-1} we infer 
$\wrelvalclosed
{\corelator{(\gradetwo \mvee \munit)}{\corelator{\gradeone}{\howe{\relone}}}}
{\valone}
{\valtwo}
{\wsix_1}
{\typeone}$ 
by law \eqref{eq:comonad-2}. We next apply the induction hypothesis, obtaining
$$
\howen{\relone}{n} 
\tensor \corelator{\gradeone}{\howe{\relone}}
\subseteq \substarrow \comp \howe{\relone} \comp \dual{\substarrow}.
$$
which in turn gives
$$
\corelator{\gradetwo \mvee \munit}{\howen{\relone}{n} 
\tensor \corelator{\gradeone}{\howe{\relone}}}
\subseteq \substarrow \comp \corelator{\gradetwo \mvee \munit}{\howe{\relone}} \comp \dual{\substarrow}
$$
by stability (Lemma~\ref{lemma:stability}). 
Finally, we use law \eqref{eq:monoidal-1} and obtain
$$
\corelator{\gradetwo \mvee \munit}{\howen{\relone}{n}}
\tensor \corelator{\gradetwo \mvee \munit}{\corelator{\gradeone}{\howe{\relone}}}
\subseteq
\corelator{\gradetwo \mvee \munit}{\howen{\relone}{n} 
\tensor \corelator{\gradeone}{\howe{\relone}}}
\subseteq 
\substarrow \comp \corelator{\gradetwo \mvee \munit}{\howe{\relone}} \comp \dual{\substarrow}.
$$
From the above inclusion, 
$\wrelcomp
{\envone,\graded{\varone}{\typeone}{\gradeone}}
{\corelator{\gradetwo \mvee \munit}{\howen{\relone}{n}}}
{\termone}
{\termone'}
{\wtwo}
{\typetwo}$, and 
$\wrelvalclosed
{\corelator{(\gradetwo \mvee \munit)}{\corelator{\gradeone}{\howe{\relone}}}}
{\valone}
{\valtwo}
{\wsix_1}
{\typeone}$ 
we infer 
$$
\wrelcomp
{\envone}
{\corelator{\gradetwo \mvee \munit}{\howe{\relone}}}
{\subst{\termone}{\varone}{\valone}}
{\subst{\termone'}{\varone}{\valtwo}}
{\wtwo \wcomp \wsix_1}
{\typetwo}.
$$
We then apply the induction hypothesis on \eqref{auxiliary-subst-seq-2}
and 
$\wrelcomp
{\envtwo,\graded{\varone}{\typeone}{\gradethree}, 
\graded{\vartwo}{\typetwo}{\gradetwo}}
{\howen{\relone}{n}}
{\termtwo}
{\termtwo'}
{\wthree}
{\typethree}$, hence inferring 
$$
\wrelcomp
{\envtwo,\graded{\vartwo}{\typetwo}{\gradetwo}}
{\howe{\relone}}
{\subst{\termtwo}{\varone}{\valone}}
{\subst{\termtwo'}{\varone}{\valtwo}}
{\wthree \wcomp \wsix_2}
{\typethree}.
$$
Finally, since $\open{\relone}$ is value substitutive
$\wrelcomp
{(\gradetwo \mvee \munit) \mstar \envone \mplus \envtwo, 
\graded{\varone}{\typeone}{(\gradetwo \mvee \munit) \mstar \gradeone \mplus \gradethree}}
{\open{\relone}}
{\seqy{\termone'}{\termtwo'}}
{\termthree}
{\wfour}
{\typethree}$ 
implies 
$\wrelcomp
{(\gradetwo \mvee \munit) \mstar \envone \mplus \envtwo}
{\howen{\relone}{n+1}}
{\seqy{\subst{\termone}{\varone}{\valone}}{\subst{\termtwo}{\varone}{\valone}}}
{\subst{\termthree}{\varone}{\valthree}}
{\wthree \wcomp \wsix_2 \wcomp \wtwo \wcomp \wsix_1 \wcomp \wfour}
{\typethree}$, 
so that we can conclude the thesis by the very definition of Howe extension 
of a relation as follows:
    \[
    \infer{
      \wrelcomp
      {(\gradetwo \mvee \munit) \mstar \envone \mplus \envtwo}
      {\howen{\relone}{n+1}}
      {\seqy{\subst{\termone}{\varone}{\valone}}{\subst{\termtwo}{\varone}{\valone}}}
      {\subst{\termthree}{\varone}{\valthree}}
      {\wone}
      {\typethree}
    }
    { \deduce[]
      {
      \wrelcomp
      {(\gradetwo \mvee \munit) \mstar \envone \mplus \envtwo}
      {\open{\relone}}
      {\seqy{\subst{\termone'}{\varone}{\valtwo}}{\subst{\termtwo'}{\varone}{\valtwo}}}
      {\subst{\termthree}{\varone}{\valthree}}
      {\wfour}
      {\typethree}
      }
      {
      \deduce[]
      {\wrelcomp
      {\envtwo,\graded{\vartwo}{\typetwo}{\gradetwo}}
      {\howe{\relone}}
      {\subst{\termtwo}{\varone}{\valone}}
      {\subst{\termtwo'}{\varone}{\valtwo}}
      {\wthree \wcomp \wsix_2}
      {\typethree}
      }
      {
      \wrelcomp
      {\envone}
      {\corelator{\gradetwo \mvee \munit}{\howe{\relone}}}
      {\subst{\termone}{\varone}{\valone}}
      {\subst{\termone'}{\varone}{\valtwo}}
      {\wtwo \wcomp \wsix_1}
      {\typetwo}
      }
      }
      &
      \wone \wgeq \wthree \wcomp \wsix_2 \wcomp \wtwo \wcomp \wsix_1 \wcomp \wfour
    }
\]
\item Suppose to be in the following case:
  \[
  \infer{
  \wrelval
  {\gradetwo \mstar \envone}
  {\howe{\relone}}
  {\tbox{\subst{\valthree}{\varone}{\valone}}}
  {\subst{\valfour}{\varone}{\valtwo}}
  {\wone}
  {\bbox_{\gradetwo}\typetwo}
  }
  {
  \infer
  {
  \wrelval
    {\gradetwo \mstar \envone, \graded{\varone}{\typeone}{\gradetwo \mstar \gradeone}}
    {\howen{\relone}{n+1}}
    {\tbox{\valthree}}
    {\valfour}
    {\wfour}
    {\bbox_{\gradetwo}\typetwo}
  }
  {
    \deduce[ ]
    {\wrelval
    {\gradetwo \mstar \envone, \graded{\varone}{\typeone}{\gradetwo \mstar \gradeone}}
    {\open{\relone}}
    {\tbox{\valthree'}}
    {\valfour}
    {\wtwo}
    {\bbox_{\gradetwo}\typetwo}
  }
  {
    \wrelval
    {\envone, \graded{\varone}{\typeone}{\gradeone}}
    {\corelator{\gradetwo}{\howen{\relone}{n}}}
    {\valthree}
    {\valthree'}
    {\wthree}
    {\typetwo}
  }
  & 
  \wfour \wgeq \wtwo \wcomp \wthree
  }
  &
  \wrelvalclosed
  {\corelator{\gradetwo \mstar \gradethree}{\howe{\relone}}}
  {\valone}
  {\valtwo}
  {\wfive}
  {\typeone}
  &
  \wone \wgeq \wfive \wcomp \wfour
  }
\]
We proceed as in previous case. The main passages are
summarised in the following chain of implication:
\begin{align*}
IH 
&\implies 
\howen{\relone}{n} \tensor \corelator{\gradeone}{\howe{\relone}} 
\subseteq \substarrow \comp \howe{\relone} \comp \dual{\substarrow}
\\
&\implies 
\corelator{\gradetwo}{\howen{\relone}{n} \tensor \corelator{\gradeone}{\howe{\relone}}}
\subseteq \substarrow \comp \corelator{\gradetwo}{\howe{\relone}} \comp \dual{\substarrow}
& & 
\text{(By Lemma~\ref{lemma:stability})}
\\
&\implies 
\corelator{\gradetwo}{\howen{\relone}{n}} \tensor \corelator{\gradetwo}{\corelator{\gradeone}{\howe{\relone}}}
\subseteq \substarrow \comp \corelator{\gradetwo}{\howe{\relone}} \comp \dual{\substarrow}
& & 
\text{(By law \eqref{eq:monoidal-1})}
\\
&\implies 
\corelator{\gradetwo}{\howen{\relone}{n}} \tensor \corelator{\gradetwo \mstar \gradeone}{\howe{\relone}}
\subseteq \substarrow \comp \corelator{\gradetwo}{\howe{\relone}} \comp \dual{\substarrow}
& & 
\text{(By law \eqref{eq:comonad-2})}
\end{align*}
As a consequence, 
$
\wrelval
{\envone, \graded{\varone}{\typeone}{\gradeone}}
{\corelator{\gradetwo}{\howen{\relone}{n}}}
{\valthree}
{\valthree'}
{\wthree}
{\typetwo}
$
and 
$
\wrelvalclosed
{\corelator{\gradetwo \mstar \gradethree}{\howe{\relone}}}
{\valone}
{\valtwo}
{\wfive}
{\typeone}
$
implies 
$$
\wrelval
{\envone}
{\corelator{\gradetwo}{\howen{\relone}{n}}}
{\subst{\valthree}{\varone}{\valone}}
{\subst{\valthree'}{\varone}{\valtwo}}
{\wthree \wcomp \wfive}
{\typetwo}.
$$
Moreover, since $\open{\relone}$ is value substitutive, 
$
\wrelval
{\gradetwo \mstar \envone, \graded{\varone}{\typeone}{\gradetwo \mstar \gradeone}}
{\open{\relone}}
{\tbox{\valthree'}}
{\valfour}
{\wtwo}
{\bbox_{\gradetwo}\typetwo}
$
entails 
$$
\wrelval
{\gradetwo \mstar \envone}
{\open{\relone}}
{\tbox{\subst{\valthree'}{\varone}{\valtwo}}}
{\subst{\valfour}{\varone}{\valtwo}}
{\wtwo}
{\bbox_{\gradetwo}\typetwo},
$$
from which the thesis follows by very definition of $\howe{\relone}$.
\end{itemize}
The remaining cases follow the same pattern of the one seen so far, or are even easier.
\end{proof}

We now prove that applicative (bi)similarity is compatible and substitutive. 

\begin{lemma} 
If the closed projection of $\howe{\appsimilarity}$ is an applicative 
simulation, then $\howe{\appsimilarity}$ coincides with $\open{\appsimilarity}$. 
\end{lemma}

\begin{proof}
We already know that $\open{\appsimilarity} \subseteq {\howe{\appsimilarity}}$, 
so that it is enough to prove the converse inclusion. 
First, notice that since $\closed{(\howe{(\appsimilarity)})}$ is an applicative simulation, 
$\closed{({\howe{\appsimilarity}})}$ is contained in ${\appsimilarity}$, and 
thus $\open{(\closed{({\howe{\appsimilarity}})})}$ is contained in 
$\open{(\appsimilarity)}$. 
We are done since ${\howe{\appsimilarity}} \subseteq 
{\open{(\closed{({\howe{\appsimilarity}})})}}$.
\end{proof}

Therefore, since $\howe{\appsimilarity}$ is compatible and substitutive, 
it is enough to show that the closed projection of $\howe{\appsimilarity}$ 
is an applicative simulation.
First, let us observe that the value clauses of 
Definition~\ref{definition:modal-applicative-bisimulation} 
are satisfied $\closed{(\howe{\appsimilarity})}$. 
In the remaining part of this section, to improve readability we will write 
$\closedhowe{\relone}$ for the closed projection of 
the Howe extension of $\relone$.

\begin{lemma}
\label{lemma:howe-extension-values}
Let $\relone$ be a reflexive and transitive applicative simulation. 
Then, $\closed{(\howe{\relone})}$ satisfies clauses 
\eqref{eq:app-abs}, \eqref{eq:app-box}, and \eqref{eq:app-fold}.
\end{lemma}

\begin{proof}
The proof is straightforward, and
thus we just show the case of clause \eqref{eq:app-box} as an illustrative 
example. Notice that since we deal with 
closed relations, if $\wrelt{\closedhowe{\relone}}{\valone}{\valtwo}{\wone}{\typeone}$, 
then $\valone$ and $\valtwo$ have the same syntactic structure, which is determined 
by $\typeone$ (for instance, if $\typeone = \bbox_{\gradeone}\typetwo$, then 
$\valone$ and $\valtwo$ must be two boxed values). 
So suppose to have the following derivation:
\[
\infer{
  \wrelvalclosed
  {(\closedhowe{\relone})}
  {\tbox{\valone}}
  {\tbox{\valthree}}
  {\wone}
  {\bbox_{\gradeone}\typeone}
}
{
  \wrelvalclosed  
  {\corelator{\gradeone}{(\closedhowe{\relone})}}
  {\valone}
  {\valtwo}
  {\wtwo}
  {\typeone}
  &
  \wrelvalclosed
  {\relone}
  {\tbox{\valtwo}}
  {\tbox{\valthree}}
  {\wthree}
  {\bbox_{\gradeone}\typeone}
  &
  \wone \wgeq \wtwo \wcomp \wthree
}
\]
Since $\relone$ is an applicative simulation, 
$\wrelvalclosed
  {\relone}
  {\tbox{\valtwo}}
  {\tbox{\valthree}}
  {\wthree}
  {\bbox_{\gradeone}\typeone}
$
implies 
$\wrelvalclosed
  {\corelator{\gradeone}{\relone}}
  {\valtwo}
  {\valthree}
  {\wthree}
  {\typeone}
$, which, together with 
$\wrelvalclosed  
  {\corelator{\gradeone}{(\closedhowe{\relone})}}
  {\valone}
  {\valtwo}
  {\wtwo}
  {\typeone}
$ and $\wone \wgeq \wtwo \wcomp \wthree$ gives
$
  \wrelvalclosed
  {(\corelator{\gradeone}{\relone} \comp \corelator{\gradeone}{(\closedhowe{\relone})})}
  {\valone}
  {\valthree}
  {\wone}
  {\typeone}
$
and thus the desired thesis by quasi-transitivity.
\end{proof}

\renewcommand{\closedhowe}[1]{\howe{#1}}
\newcommand{\closedhowebot}[1]{#1^{\scriptscriptstyle H}_{\bot}}

It thus remains to prove that $\closedhowe{\relone}$ 
satisfies clause \eqref{eq:app-eval}, where $\relone$ is a reflexive 
and transitive applicative simulation (we write 
$\howe{\relone}$ for the closed restriction of the Howe extension of 
$\relone$). That essentially amounts to 
show the inclusion
$$
\closedhowe{\relone} \subseteq 
\evalsymbol \comp \closedhowebot{\relone} \comp \dual{\evalsymbol}.
$$
We will do that by case analysis on $\closedhowe{\relone}$. However, 
since $\evalsymbol$ is defined as $\lub_{n \geq 0} \evalsymboln{n}$, 
we would also like to reason inductively in terms of $\evalsymboln{n}$. 
We can do so by observing that 
$(-)_\bot$ supports the following 
reasoning principles:
\begin{align*}
&\wrel{\relone_\bot}{\divergence}{y}{\wone}
\tag{Induction 1}\label{eq:induction-1}
\\
(\forall n \geq 0.\ 
\wrel{\relone_\bot}{x_n}{y}{\wone})
&\implies 
\wrel{\relone_\bot}{\lub_{n \geq 0} x_n}{y}{\wone}.
\tag{Induction 2}\label{eq:induction-2}
\end{align*}

As a consequence, to prove that $\closedhowe{\relone}$ satisfies 
clause \eqref{eq:app-eval} it is enough to show the following statement:
$$
\forall n \geq 0.\ 
\closedhowe{\relone} \subseteq 
\evalsymboln{n} \comp \closedhowebot{\relone}) \comp \dual{\evalsymbol}.
$$
We proceed by induction on $n$. The case for $n = 0$ directly follows from 
law \eqref{eq:induction-1}. For the inductive step, we proceed 
by cases on the definition of $\closedhowe{\relone}$. Most cases are 
standard, but we encounter a further difficulty in the case of sequencing. 
Suppose to have:
\[
\infer{
  \wrelcompclosed
  {(\closedhowe{\relone})}
  {\seq{\termone}{\termtwo}}
  {\termthree}
  {\wone}
  {\typetwo}
}
{
  \wrelcompclosed
  {\corelator{\gradeone \mvee \munit}{(\closedhowe{\relone})}}
  {\termone}
  {\termone'}
  {\wtwo}
  {\typeone}
  &
  \wrelcomp
  {\graded{\varone}{\typeone}{\gradeone}}
  {(\closedhowe{\relone})}
  {\termtwo}
  {\termtwo'}
  {\wthree}
  {\typetwo}
  &
  \wrelcompclosed
  {\relone}
  {\seq{\termone'}{\termtwo'}}
  {\termthree}
  {\wfour}
  {\typetwo}
  &
  \wone \wgeq \wtwo \wcomp \wthree \wcomp \wfour
}
\]
We have to prove 
$\wrelt
  {\closedhowebot{\relone}}
  {\evaln{n+1}{\seq{\termone}{\termtwo}}}
  {\eval{\termthree}}
  {\wone}
  {\typetwo}.
$
Since 
$$
\evaln{n+1}{\seq{\termone}{\termtwo}} = 
\kleisli{(\valone \mapsto \evaln{n}{\subst{\termtwo}{\varone}{\valone}})}(\evaln{n}{\termone})
$$
we already see that we may want to rely one lax \eqref{eq:lax-monad-bind}. 
But let us proceed by step by step. First, by induction hypothesis 
we have 
$\closedhowe{\relone} 
\subseteq 
\evalsymboln{n} \comp 
\closedhowebot{\relone} \comp \dual{\evalsymbol}$ 
and thus, by stability (Lemma~\ref{lemma:stability}), 
$$
\corelator{\gradeone \mvee \munit}{\closedhowe{\relone}}
\subseteq 
\evalsymboln{n} \comp 
\corelator{\gradeone \mvee \munit}
{\closedhowebot{\relone}} 
\comp \dual{\evalsymbol}.
$$
As a consequence, from
$\wrelcompclosed
  {\corelator{\gradeone \mvee \munit}{\closedhowe{\relone}}}
  {\termone}
  {\termone'}
  {\wtwo}
  {\typeone}$ 
we infer 
$\wrel
  {\corelator{\gradeone \mvee \munit}
  {\closedhowebot{\relone}}}
  {\semn{\termone}{n}}
  {\sem{\termone'}}
  {\wtwo}
$. 
Let us now move 
$\wrelcomp
{\graded{\varone}{\typeone}{\gradeone}}
{\closedhowe{\relone}}
{\termtwo}
{\termtwo'}
{\wthree}
{\typetwo}$. 
Let write $\hat{\termtwo}$, $\hat{\termtwo'}$ 
for the maps mapping a closed value $\valone$ 
of type $\typeone$ to 
$\subst{\termtwo}{\varone}{\valone}$ 
and $\subst{\termtwo'}{\varone}{\valone}$, respectively.
By substitutivity of $\closedhowe{\relone}$ and the 
induction hypothesis
we obtain the following lax commutative diagram
\[
\xymatrix@C=1.5cm{
\laxcommuterel
\values_{\typeone}
\ar[r]^-{\hat{\termtwo}}
\ar[d]_{\corelator{\gradeone}{\closedhowe{\relone}}}|@{|}  
& \Lambda_{\typetwo}  
\ar[d]^{\closedhowe{\relone}}|@{|}
\ar[r]^-{\semn{-}{n}}
& 
(\values_{\typetwo})_{\divergence}
\ar[d]^{\closedhowebot{\relone}}|@{|}
\\
\values_{\typeone}
\ar[r]_-{\hat{\termtwo'}}   
& \Lambda_{\typetwo}  
\ar[r]^-{\sem{-}}
& (\values_{\typetwo})_{\divergence} 
}
\]
which, by law \eqref{eq:contravariance}, gives:
\[
\xymatrix@C=1.5cm{
\laxcommuterel
\values_{\typeone}
\ar[r]^-{\hat{\termtwo}}
\ar[d]_{\corelator{\gradeone \mvee \munit}{\closedhowe{\relone}}}|@{|}  
& \Lambda_{\typetwo}  
\ar[d]^{\closedhowe{\relone}}|@{|}
\ar[r]^-{\semn{-}{n}}
& 
(\values_{\typetwo})_{\divergence}
\ar[d]^{\closedhowebot{\relone}}|@{|}
\\
\values_{\typeone}
\ar[r]_-{\hat{\termtwo'}}   
& \Lambda_{\typetwo}  
\ar[r]^-{\sem{-}}
& (\values_{\typetwo})_{\divergence} 
}
\]
Next, we now apply law \eqref{eq:lax-monad-bind}, obtaining:
\[
\xymatrix@C=2cm{
\laxcommuterel
(\values_{\typeone})_{\divergence}
\ar[r]^-{\kleisli{(\semn{-}{n} \comp \hat{\termtwo})}}
\ar[d]_{
(\corelator{\gradeone \mvee \munit}{\closedhowe{\relone}})_{\bot}}|@{|}  
& 
(\values_{\typetwo})_{\divergence}
\ar[d]^{\closedhowebot{\relone}}|@{|}
\\
(\values_{\typeone})_{\divergence}
\ar[r]_-{\kleisli{(\sem{-} \comp \hat{\termtwo'})}}
& (\values_{\typetwo})_{\divergence} 
}
\]
At this point we may conclude the thesis,\footnote{
  Actually, we would conclude 
  $\wrel{\closedhowebot{\relone}}
  {\semn{\seq{\termone}{\termtwo}}{n+1}} 
  {\sem{\seq{\termone'}{\termtwo'}}}
  {\wtwo \wcomp \wthree}$
  from which we can then infer the thesis, since we have 
  $\wrelt{\relone}{\seq{\termone'}{\termtwo'}}{\termthree}{\wfour}{\typetwo}$
  (which, $\relone$ being an applicative simulation, entails 
  $\wrel{\relone_\bot}
  {\sem{\seq{\termone'}{\termtwo'}}}
  {\sem{\termthree}}{\wfour}$) and by quasi-transitivity 
  $\relone_\bot \comp 
  \closedhowebot{\relone}
  \subseteq \closedhowebot{\relone}$.
} provided 
that we can infer 
$\wrel
{(\corelator{\gradeone \mvee \munit}{\closedhowe{\relone})_{\bot}}}
{\semn{\termone}{n}}
{\sem{\termone'}}
{\wtwo}
$
from 
$\wrel
{\corelator{\gradeone \mvee \munit}
{\closedhowebot{\relone}}}
{\semn{\termone}{n}}
{\sem{\termone'}}
{\wtwo}
$; that is, provided that we have the inclusion
$$
\corelator{\gradeone \mvee \munit}
{\closedhowebot{\relone}}
\subseteq
(\corelator{\gradeone \mvee \munit}{\closedhowe{\relone})_{\bot}}.
$$

Such an inclusion is specific instance of a more general 
(lax distributive law) condition defining when 
lax extensions of monads and 
comonadic lax extensions
properly interact. 

\begin{definition}
\label{definition:relator-corelator-distributive-law}
Given a lax extension $\relatorsymbol$ of a monad $(\monad, \unit, \kleisli{-})$ 
and a comonadic lax extension $\corelatorsymbol$ as defined in 
Definition~\ref{definition:corelator}, we say that 
$\corelatorsymbol$ distributes over $\relatorsymbol$ if the following inclusion holds for 
any $\gradeone \mgeq \munit$:
\begin{align*}
\corelator{\gradeone}
{\relatorsymbol(\relone)}
\subseteq
\relatorsymbol(\corelator{\gradeone}{\relone}).
\tag{Distributivity} \label{eq:distributive-lax}
\end{align*}
\end{definition}

Equivalently, we can refine law \eqref{eq:lax-monad-bind} with the 
following lax commutative diagram, where $\gradeone \mgeq \munit$.
\[
\vcenter{
  \diagramrel
  {f}
  {g}
  {X}
  {Y}
  {\corelator{\gradeone}{\relone}}
  {\monad(Z)}
  {\monad(W)}
  {\relatorsymbol(\reltwo)}
  }
  \implies
   \vcenter{
  \diagramrel
  {\kleisli{f}}
  {\kleisli{g}}
  {\monad(X)}
  {\monad(Y)}
  {\corelator{\gradeone}{\relatorsymbol(\relone)}}
  {\monad(Z)}
  {\monad(W)}
  {\relatorsymbol(\reltwo)}
  }
\]
}

We have thus proved the following result.

\noindent \textbf{Lemma~\ref{lemma:key-lemma-short}} (Key Lemma)\textbf{.}
\emph{Assume the law $\corelator{\gradeone \vee \gunit}
{\relone_\bot}
\subseteq
(\corelator{\gradeone \vee \gunit}{\relone})_\bot$, for any $\relone$.
Then, for any reflexive and transitive 
applicative simulation $\relone$, $\howe{\relone}$ 
(restricted to closed terms) is an applicative 
simulation.
}

It then follows that applicative similarity is a preorder and 
that applicative bisimilarity is a congruence.

\section{On Effectful Extensions}

We informally sketch how to extend the main results in this paper 
to an effectful setting. We follow the same structure given by 
Gavazzo \cite{Gavazzo/LICS/2018}, to which we refer for details.

The ingredients are:
\begin{enumerate}
\item A collection $\signature$ of effect-triggering operation symbols
  (such as probabilistic choice operations or primitives for input-output).
\item A monad $\Monad$ to interpret such operations (such as the distribution monad). 
  Formally, that means to any $n$-ary operation symbol in $\signature$ 
  is associated an $n$-ary algebraic operation on 
  $\Monad$ \cite{DBLP:journals/entcs/PlotkinP01,PlotkinPower/FOSSACS/01,Plotkin/algebraic-operations-and-generic-effects/2003}.
\item A lax extension $\relatorsymbol$ of $\Monad$ to $\vrel$. 
\end{enumerate}
First, we extend $\Lambda_{\gradealg}$ with operations in $\signature$: for 
any $n$-ary operation $\op \in \signature$, we stipulate that 
$\op(\termone_1, \hh, \termone_n)$ is a term of the calculus, whenever 
$\termone_1, \hh, \termone_n$ are. The operational semantics of the calculus thus 
obtained, which we denote by $\Lambda_{\gradealg, \signature}$, is defined 
as for $\Lambda_{\gradealg}$: there, in fact, we rely on the monad structure 
of the partiality monad, rather than on partiality itself. However, 
we also rely on the domain structure of $X_{\divergence}$ and some strictness 
and continuity properties of Kleisli extension: we require such properties 
to hold for $\Monad$ too, this way restricting our analysis to 
continuous monads \cite{DalLagoGavazzoLevy/LICS/2017,Goguen/1977}. 
A monad $\Monad = (\monad, \unit, \kleisli{-})$ is continuous\footnote{
  Notice that continuous monads is just an $\omega$-cppo enriched monad 
  \cite{Kelly/EnrichedCats}.
}
if $\monad X$ is an $\omega$-cppo for any set $X$, 
and satisfy the laws
\begin{align*}
  \kleisli{f}(\divergence) &= \divergence;
  &
  \kleisli{f}\big(\lub_{n \geq 0} x_n\big) &=  \lub_{n \geq 0} \kleisli{f}(x_n);
  &
  \big(\lub_{n \geq 0} \kleisli{f}\big)(x) &= \lub_{n \geq 0} \kleisli{f}(x).
\end{align*}

Next, we move to the definition of applicative similarity distance, 
which is defined as in Definition~\ref{def:applicative-distance} 
except for the clause on terms which now involves the lax extension 
$\relatorsymbol$ of $\Monad$:
$$
\toterm{\vsim}(\termone, \termtwo) \leq 
\relatorsymbol(\toval{\vsim})(\sem{\termone}, \sem{\termtwo}).
$$ 
To prove compatibility of applicative similarity distance, we also need 
the lax extension $\relatorsymbol$ of $\Monad$ to satisfy the properties 
we relied on in our proof of the key lemma (Lemma~\ref{lemma:key-lemma-short}), 
namely distributivity of comonadic lax extensions over $\envone$, and 
a suitable induction principle based on continuity of $\Monad$. 
We summarise these properties as follows:
\begin{enumerate}
  \item $\relatorsymbol$ must be a lax extension of $\Monad$ (on $\vrel$). 
  \item $\relatorsymbol$ must be inductive, meaning that the following 
    laws hold:
    \begin{align*}
     \qunit &\leq \relatorsymbol(\vrelone)(\bot, \chi)  
    &
    \meet_n \relatorsymbol(\vrelone)(\chi_n, \phi)
    &\leq  \relatorsymbol(\vrelone)(\lub_n \chi_n, \phi)
    \end{align*}
  \item $\corelatorsymbol$ must distributes over $\relatorsymbol$:
    $$\corelator{\gradeone}
    {\relatorsymbol(\vrelone)}
    \leq
    \relatorsymbol(\corelator{\gradeone}{\vrelone}).
    $$
\end{enumerate}
If these conditions are satisfied, then applicative similarity distance 
is  substitutive and compatible, and the abstract metric preservation theorem 
holds. 

We conclude this overview by mentioning 
a couple of concrete examples of effectful extensions of 
$\Lambda_{\gradealg}$ (see \cite{Gavazzo/LICS/2018} for further examples).

\begin{example}
\begin{enumerate}
\item 
    The discrete subdistribution monad $\distribution$ is continuous and probabilistic choice 
    operations are algebraic on it. A lax extension of $\distribution$ to 
    $\Vrel{\Lawvere}$ is given 
    by the so-called Wasserstein-Kantorovich metric 
    lifting \cite{Villani/optimal-transport/2008}. 
    The duality theorem for countable transportation problems \cite{Kortanek/InfiniteTransportationProblems/1995} 
    stating that the Wasserstein and Kantorovich metric coincide 
    gives us two equivalent formulations of such 
    a lifting, which allow one to prove that we indeed have a lax extension of $\distribution$. 
    Such extension satisfies the required induction principle 
    \cite{Wasserstein-metric-and-subordination} and comonadic lax extension 
    given by multiplication by a constant in $[0,\infty]$ distributes 
    over it. The resulting notion of (probabilistic) applicative similarity distance 
    is thus compatible and substitutive. 
\item
    The powerset monad $\powerset$ is countable and set-theoretic union is algebraic on it.
    A lax extension of $\powerset$ is given the so-called Hausdorff lifting \cite{Munkres/Topology/2000}. The Hausdorff lifting has been extended to arbitrary quantales 
    \cite{AKHVLEDIANI20101275} (and thus to any possible worlds structure), so that 
    it is always possible to inject pure nondeterminism in modal calculi. 
\item 
    More generally, canonical lax extensions of monads $\Monad$ to $\vrel$ can be obtained 
    by algebra maps $\xi: \monad(\quantale) \to \quantale$ respecting 
    the quantale structure of $\Quantale$ \cite{Hoffman-topological-theories-as-closed-objects}.
\end{enumerate}
\end{example}

\restoregeometry

\end{document}